\pdfoutput=1
\documentclass[prd,showpacs,letterpaper,preprintnumbers,10pt,aps,notitlepage,superscriptaddress,nofootinbib]{revtex4-1}

\usepackage{amsfonts,amsmath,graphicx,bm}
\usepackage{mathrsfs}
\usepackage{slashed}

\renewcommand{\ol}[1]{\overline{#1}}

\newcommand{\nb}{\phantom{0}}

\def\mathbi#1{\textbf{\em #1}}

\begin{document}

\preprint{JLAB-THY-11-1459}

\title{Calculation of the heavy-hadron axial couplings $\mathbi{g}_\mathbf{1}$, $\mathbi{g}_\mathbf{2}$, and $\mathbi{g}_\mathbf{3}$ using lattice QCD}

\author{William Detmold}
\email{wdetmold@jlab.org}
\affiliation{Department of Physics, College of William and Mary, Williamsburg, VA 23187, USA}
\affiliation{Jefferson Laboratory, 12000 Jefferson Avenue, Newport News, VA 23606, USA}
\author{C.-J.~David Lin}
\email{dlin@mail.nctu.edu.tw}
\affiliation{Institute of Physics, National Chiao-Tung University, Hsinchu 300, Taiwan}
\affiliation{Physics Division, National Centre for Theoretical Sciences, Hsinchu 300, Taiwan}
\author{Stefan Meinel}
\email{smeinel@wm.edu}
\affiliation{Department of Physics, College of William and Mary, Williamsburg, VA 23187, USA}


\begin{abstract}
In a recent letter [Phys.~Rev.~Lett.~{\bf 108}, 172003 (2012), arXiv:1109.2480] we have reported on a lattice QCD calculation
of the heavy-hadron axial couplings $g_1$, $g_2$, and $g_3$.
These quantities are low-energy constants of heavy-hadron chiral perturbation theory (HH$\chi$PT)
and are related to the $B^* B \,\pi$, $\Sigma_b^*\,\Sigma_b\,\pi$, and $\Sigma_b^{(*)} \Lambda_b \, \pi$ couplings.
In the following, we discuss important details of the calculation and give further
results. To determine the axial couplings, we explicitly match the matrix elements of the axial current in QCD with
the corresponding matrix elements in HH$\chi$PT. We construct the ratios of correlation functions
used to calculate the matrix elements in lattice QCD, and study the contributions from excited states.
We present the complete numerical results and discuss the data analysis in depth. In particular, we demonstrate
the convergence of $SU(4|2)$ HH$\chi$PT for the axial-current matrix elements at pion masses up to about
400 MeV and show the impact of the nonanalytic loop contributions.
Finally, we present additional predictions for strong and radiative decay widths of charm and bottom baryons.
\end{abstract}

\maketitle

\section{Introduction}

Two important symmetries which allow many predictions in nonperturbative QCD
are chiral symmetry \cite{Gasser:1983yg} and heavy-quark symmetry \cite{Shifman:1987rj}.
Both symmetries are a consequence of the large separation of scales in the quark masses, relative to the intrinsic
scale of QCD. In chiral perturbation theory, an expansion is performed around the limit where the light-quark masses vanish,
and the dynamics is determined by derivatively coupled pions, associated with the spontaneous breaking of chiral symmetry.
In heavy-quark effective theory, an expansion is performed around the static limit, where a quark mass is infinitely heavy. In that limit,
the heavy quark acts as a pointlike color source with a fixed velocity, and the spin of the heavy quark decouples.
Corrections to the static limit are suppressed by powers of $\Lambda_{QCD}/m_Q$,
where $m_Q$ is the heavy-quark mass. Because $\Lambda_{QCD}/m_b$ is particularly small,
the static limit is a good approximation to describe the dynamics of hadrons containing a bottom quark. For
charmed hadrons, the static limit is still a reasonable first approximation.

At the hadronic level, the combination of chiral symmetry and heavy-quark symmetry into a single effective theory leads
to a framework known as heavy-hadron chiral perturbation theory (HH$\chi$PT), which describes the interactions of
heavy-light hadrons with pions and kaons \cite{Wise:1992hn, Burdman:1992gh, Yan:1992gz, Cho:1992gg, Cho:1992cf, Cheng:1993kp}.
At leading order, the HH$\chi$PT Lagrangian contains three axial couplings $g_1$, $g_2$ and $g_3$. The coupling $g_1$
determines the strength of the interaction between heavy-light mesons and pions, while $g_2$ and $g_3$
similarly determine the interaction of heavy-light baryons with pions.

In the meson sector, the strong decay $B^* \to B \:\pi$ is kinematically forbidden. However, virtual pion loops
contribute to much of the physics of $B$ mesons, and the coupling $g_1$ appears in calculations of important observables
such as $B$ meson masses, decay constants, bag parameters and form factors within chiral perturbation theory
\cite{Grinstein:1992qt, Goity:1992tp, Boyd:1994pa}. Precise knowledge of these hadronic observables is needed for
flavor physics, both within and beyond the Standard Model. In this context, chiral perturbation theory is needed, in particular, to describe the quark-mass
dependence of lattice QCD results for such observables. Most current lattice QCD calculations are performed at unphysically large values of the
up- and down quark masses to reduce the amount of computer time needed. The results from a range of quark masses must then be extrapolated to
the physical values of the light-quark masses. Chiral perturbation theory predicts the functional form needed for
this extrapolation, but the uncertainty in the final result is influenced significantly by the uncertainty in the value of the
axial coupling $g_1$ \cite{Kronfeld:2002ab}.

While most of heavy-flavor physics has traditionally focused on the $B$ mesons, measurements at the LHC and super-$B$ factories
will extend the knowledge of flavor physics in the bottom-baryon sector, which provides complementary constraints on new physics models because
of the different spin of the baryons. Therefore, lattice QCD calculations of bottom-baryon observables such as form factors are needed,
and as in the meson sector, chiral extrapolations of the data need to be performed. For baryons, the accuracy of such
extrapolations can be improved dramatically if the values of the couplings $g_2$ and $g_3$ are known.
The coupling $g_3$ is related to the strong decays $\Sigma_b^{(*)} \to \Lambda_b\: \pi$, which are kinematically allowed.
The widths of these decays have recently been measured at Fermilab \cite{CDF:2011ac}, but the experimental uncertainty is still large.

The axial couplings $g_1$, $g_2$, and $g_3$ are calculable from the underlying theory, QCD. The only reliable
approach for these nonperturbative observables is lattice QCD. While there are no previous lattice calculations of $g_2$ and $g_3$,
a number of groups have performed lattice computations of the coupling $g_1$, both in the quenched approximation
(i.e. neglecting the vacuum-polarization effects of light quarks) \cite{deDivitiis:1998kj, Abada:2003un, Negishi:2006sc} and with $n_f=2$
dynamical flavors \cite{Ohki:2008py, Becirevic:2009yb, Bulava:2010ej}. In these lattice calculations, one computes matrix elements
of the axial current, and relates these matrix elements to the coupling $g_1$. To fit the data and extract $g_1$, theoretical knowledge
of the light-quark mass dependence of the axial-current matrix elements is required. The previous lattice calculations used fits
that were linear or quadratic in the pion mass, or including logarithms, but the correct chiral perturbation theory predictions were not known.

We have recently derived the expressions for the axial-current matrix elements at next-to-leading order in partially quenched heavy-hadron chiral perturbation theory, both
for the mesons and the baryons \cite{Detmold:2011rb}. We have then performed the first complete lattice QCD calculation of the three axial couplings $g_1$, $g_2$, $g_3$,
controlling all systematic uncertainties \cite{Detmold:2011bp}.
In the following, we discuss important details of the analysis that were omitted for brevity in Ref.~\cite{Detmold:2011bp}, and present some additional results.

Our calculation includes $n_f=2+1$ flavors of dynamical light quarks, and makes use of data at six different values of the quark masses corresponding to (valence)
pion masses as low as 227 MeV. Two different lattice spacings of $a=0.112$ fm and $a=0.085$ fm are used to perform a continuum extrapolation. The spatial volume is $(2.7\:{\rm fm})^3$,
large enough so that finite-size effects are very small and can be removed by using finite-volume heavy-hadron chiral perturbation theory in the $p$ regime.
Because the axial couplings $g_1$, $g_2$, $g_3$ are defined in the static limit, we use the static lattice action of Ref.~\cite{Eichten:1989kb},
modified using smeared gauge links to reduce noise \cite{DellaMorte:2003mn}, for the heavy quark. We implement the light quarks with a domain-wall action
\cite{Kaplan:1992bt, Furman:1994ky, Shamir:1993zy}. This is a five-dimensional formulation that realizes a lattice chiral symmetry
for the four-dimensional theory, which becomes exact, even at finite lattice spacing,
when the extent of the auxiliary fifth dimension is taken to infinity. As a result, the renormalization of operators is simplified and discretization
errors are small. Our calculations make use of gauge field ensembles generated
by the RBC and UKQCD collaborations \cite{Aoki:2010dy}.

This paper is organized as follows: We begin with an introduction to HH$\chi$PT in Sec.~\ref{sec:hhchipt}.
We derive the matrix elements of the axial Noether current at leading order in HH$\chi$PT using canonical quantization in
Sec.~\ref{sec:axialcurrentmatrixeltshhchPT}. The ratios of correlation functions used to calculate the corresponding matrix elements
in lattice QCD are constructed in Sec.~\ref{sec:axialcurrentmeLQCD}, where we also show their spectral decomposition.
The details of the lattice actions and parameters are given in Sec.~\ref{sec:latticeparams}. We present the numerical results for the ratios of correlation
functions in Sec.~\ref{sec:resultsratios}, and explain our method for extracting the ground-state contributions to the matrix elements
in Sec.~\ref{sec:sourcesinkextrap}. In Sec.~\ref{sec:chiralfits}, we then describe the chiral fits using $SU(4|2)$ HH$\chi$PT,
including the effects of finite volume and nonzero lattice spacing. We compare our lattice QCD results for the axial couplings to various
estimates reported in the literature in Sec.~\ref{sec:comparison}. The calculations of heavy-baryon decay widths
are presented in Sec.~\ref{sec:decays}, and we conclude in Sec.~\ref{sec:conclusions}. Appendix \ref{sec:rawdata} contains further plots of numerical data,
and Appendix \ref{sec:summationmethod} contains a comparison of our ratio method with an alternative approach (the summation method).

\section{Axial couplings from ratios of correlation functions}

\subsection{Heavy-hadron chiral perturbation theory}

\label{sec:hhchipt}

For hadrons containing a heavy quark (or antiquark), in the static limit $m_Q \rightarrow \infty$, the spin of the light degrees of freedom, $s_l$, becomes a conserved quantum number.
The lowest-lying static-light mesons have $s_l=1/2$, and are therefore grouped into pseudoscalar mesons, described by a field $P^i$, and vector mesons, described by a field $P^{*i}_{\mu}$. We work with heavy-light mesons containing
a light quark of flavor $i=u,d$ and a heavy antiquark.
The vector meson field satisfies $v^\mu P^{*i}_{\mu}=0$, where $v$ is the four-velocity; this is a parameter of the effective theory, subject to the constraint $v^2=1$.
Because of heavy-quark spin symmetry, the pseudoscalar and vector mesons are degenerate.  To make the heavy-quark symmetry manifest, the pseudoscalar and vector meson
fields can be combined into a single field $H$, which is $4\times4$-matrix-valued and given by \cite{Wise:1992hn, Grinstein:1992qt}
\begin{equation}
 H^i=\left[ - P^i\gamma_5 + P^{*i}_{\mu} \gamma^\mu \right]\frac{1-\slashed{v}}{2}. \label{eq:Hfield}
\end{equation}
This field satisfies the constraint $ H^i \:\frac{1-\slashed{v}}{2} =H^i$. Next, we consider static-light baryons
containing two light quarks of flavors $i$ and $j$ and a heavy quark. We include both the states with $s_l=0$ and $s_l=1$.
The states with $s_l=1$ form two multiplets with $J=1/2$ and $J=3/2$, and are described by a Dirac spinor field $B^{ij}$ and a Rarita-Schwinger spinor field $B^{*ij}_\mu$.
These $s_l=1$ fields are symmetric in flavor: $B^{ij}=B^{ji}$ and $B^{*ij}_\mu=B^{*ji}_\mu$. For two quark flavors one has (using the notation for bottom baryons)
\begin{equation}
B=\left( \begin{array}{cc} \Sigma_b^+ & \frac{1}{\sqrt{2}}\Sigma_b^0 \\ \frac{1}{\sqrt{2}}\Sigma_b^0 & \Sigma_b^- \end{array} \right),
\end{equation}
and similarly for $B^*_\mu$. Again, because of heavy-quark spin symmetry, the $J=1/2$ and $J=3/2$ baryons with $s_l=1$ are degenerate, and the corresponding fields can be combined into a single field \cite{Georgi:1990cx, Cho:1992cf}
\begin{equation}
 S_\mu^{ij}=S_\mu^{ji}=\sqrt{\frac{1}{3}}(\gamma_\mu + v_\mu)\gamma_5 B^{ij} + B^{*ij}_{\mu}, \label{eq:Sfield}
\end{equation}
satisfying $v^\mu S_\mu^{ij}=0$ and $\frac{1+\slashed{v}}{2}S_\mu^{ij}=S_\mu^{ij}$.
The $s_l=0$ baryons have $J=1/2$ and can be described by a Dirac spinor field $T^{ij}$, which is antisymmetric in $i$ and $j$ and satisfies the constraint $\frac{1+\slashed{v}}{2}T^{ij}=T^{ij}$.
For two quark flavors, one has
\begin{equation}
T=\frac{1}{\sqrt{2}}\left( \begin{array}{cc} 0 & \Lambda_b \\ -\Lambda_b & 0 \end{array} \right).
\end{equation}
In $SU(2)$ chiral perturbation theory, the pions are described by an $SU(2)$-valued field $\Sigma=\exp(2i\Phi/f)$, which transforms under global $SU(2)_L\times SU(2)_R$ transformations as
\begin{equation}
  \Sigma \rightarrow  L\: \Sigma \: R^\dag .
\end{equation}
For the purposes of heavy-hadron chiral perturbation theory, it is convenient to also introduce the field $\xi=\exp(i\Phi/f)$, so that $\Sigma=\xi^2$. The field $\xi$ transforms as
\begin{equation}
 \xi \rightarrow L \: \xi \: U^\dag = U \: \xi \: R^\dag,
\end{equation}
where the transformation matrix $U(x)$ is a function of $L$, $R$ and $\Phi(x)$, implicitly defined through the above equations. Under the vector subgroup $L=R=V$, the field $\xi$
transforms as $\xi \rightarrow V \xi V^\dag$. Therefore, the natural transformation laws for the heavy-hadron fields also involve the matrix $U$:
\begin{eqnarray}
\nonumber H^i &\rightarrow& U^i_{\:\:j} H^j, \\
\nonumber S^{ij}_\mu & \rightarrow & U^i_{\:\:k} U^j_{\:\:l} S^{kl}_\mu, \\
T^{ij} & \rightarrow & U^i_{\:\:k} U^j_{\:\:l} T^{kl}.
\end{eqnarray}
The leading-order heavy-hadron chiral perturbation theory Lagrangian is then given by \cite{Yan:1992gz, Cho:1992gg, Cho:1992cf, Cheng:1993kp}
\begin{eqnarray} 
\nonumber \mathcal{L} &=& \frac{f^2}{8}  (\partial^\mu \Sigma^\dag)_{ij} \partial_\mu \Sigma^{ji} - i\:\mathrm{tr}_{\mathrm{D}}\left[ \overline{H}_i v\cdot\mathscr{D} H^i \right] - i\overline{S}_{ij}^\mu v  \cdot  \mathscr{D} S^{ij}_\mu + i \overline{T}_{ij} v  \cdot  \mathscr{D} T^{ij} \\
\nonumber &&+ \Delta \overline{S}_{ij}^\mu S^{ij}_\mu + g_1 \mathrm{tr}_{\mathrm{D}}\left[ \overline{H}_i (\mathscr{A}^{\mu})^i_{\:\:j} \gamma_\mu \gamma_5 H^j \right]  - i g_2 \epsilon_{\mu\nu\sigma\lambda} \overline{S}^\mu_{ki} v^\nu (\mathscr{A}^\sigma)^i_{\:\:j} (S^\lambda)^{jk} \\
&& + \sqrt{2} g_3 \left[ \overline{S}^\mu_{ki} (\mathscr{A}_\mu)^i_{\:\:j} T^{jk} + \overline{T}_{ki} (\mathscr{A}^\mu)^i_{\:\:j} S_\mu^{jk} \right]  + (m_q\mathrm{\:\:terms}) + (1/m_Q\mathrm{\:\:terms}), \label{eq:HHchPTL}
\end{eqnarray}
where $\mathrm{tr}_{\mathrm{D}}$ denotes the trace in Dirac space, and the covariant derivatives are defined as
\begin{eqnarray}
\nonumber \mathscr{D}^\mu H^i         &=& \partial^\mu H^i + (\mathscr{V}^\mu)^i_{\:\:j} H^j, \\
\nonumber  \mathscr{D}^\mu S^{ij}_\nu &=& \partial^\mu S^{ij}_\nu + (\mathscr{V}^\mu)^i_{\:\:k} S^{kj}_\nu + (\mathscr{V}^\mu)^j_{\:\:k} S^{ik}_\nu, \\
 \mathscr{D}^\mu T^{ij}               &=& \partial^\mu T^{ij} + (\mathscr{V}^\mu)^i_{\:\:k} T^{kj} + (\mathscr{V}^\mu)^j_{\:\:k} T^{ik},
\end{eqnarray}
with the vector and axial-vector fields
\begin{eqnarray}
\nonumber \mathscr{V}^\mu \!\!&=&\!\! \frac12 \left( \xi^\dag \partial^\mu \xi + \xi \partial^\mu \xi^\dag \right), \\
\mathscr{A}^\mu \!\!&=&\!\! \frac i2 \left( \xi^\dag \partial^\mu \xi - \xi \partial^\mu \xi^\dag \right).
\end{eqnarray}
The $H$ and $T$ fields are rescaled such that their masses do not appear in the Lagrangian. The quantity $\Delta$ is the mass difference between the $S$ and $T$ baryons.
This mass difference does not vanish in the chiral limit nor in the heavy-quark limit. From experiment, one has $\Delta\approx 200$ MeV \cite{Aaltonen:2007rw, CDF:2011ac}.

Our definitions of the axial couplings $g_1$, $g_2$, and $g_3$ in Eq.~(\ref{eq:HHchPTL}) are related to the definitions of Yan {\it et~al.} \cite{Yan:1992gz} and Cho \cite{Cho:1992cf} as follows
\cite{Cheng:1997rp}:
\begin{eqnarray}
\nonumber g_1 =& g_1^{\rm(Cho)}  &= g^{\rm(Yan)},\\
\nonumber g_2 =& -g_2^{\rm(Cho)} &= \frac32 g_1^{\rm(Yan)},\\
          g_3 =& g_3^{\rm(Cho)}  &= -\sqrt{3} g_2^{\rm(Yan)}. \label{eq:conventions}
\end{eqnarray}
We introduced the minus sign on $g_2$ relative to the definition by Cho, so that in our conventions all three couplings are positive.

The Lagrangian (\ref{eq:HHchPTL}) has the same form for both $SU(2)$ and $SU(3)$ chiral perturbation theory, the only difference being that the flavor indices run from 1 to 2 and 1 to 3, respectively.
The theory can be generalized to the partially quenched $SU(4|2)$ or $SU(6|3)$ cases, where the valence and sea quarks can have different masses;
for more details see \cite{Detmold:2011rb} and the references therein.

As can be seen by expanding the field $\xi=\exp(i\Phi/f)$ in Eq.~(\ref{eq:HHchPTL}) in terms of the pion field $\Phi$, at lowest order the term with the axial coupling $g_1$ leads to an $H$-$H$-$\Phi$ vertex, the term with the
coupling $g_2$ leads to an $S$-$S$-$\Phi$ vertex, and finally the term with the coupling $g_3$ leads to an $S$-$T$-$\Phi$ vertex.

\subsection{Axial-current matrix elements in heavy-hadron chiral perturbation theory}

\label{sec:axialcurrentmatrixeltshhchPT}

The simplest quantities that depend on the axial couplings are the matrix elements of the axial currents between single-hadron states (an alternative
approach to determine the axial couplings based on static hadron-hadron potentials is discussed in Ref.~\cite{Detmold:2007wk}).
To extract $g_1$, $g_2$, and $g_3$, we will calculate the matrix elements of the axial current in both $\chi$PT and (lattice) QCD, and match the two
with each other:
\begin{equation}
\langle X | \: A_\mu^{(\mathrm{QCD})} \: | Y \rangle_{\mathrm{QCD}} = \langle X | \: A_\mu^{(\chi \mathrm{PT})} \: | Y \rangle_{\chi \mathrm{PT}}.
\end{equation}
In QCD, the axial current is simply given by
\begin{equation}
 A_\mu^{a(\mathrm{QCD})} = \bar q \frac{\tau^a}{2} \gamma_\mu\gamma_5 q.
\end{equation}
To derive the expression for the axial current in heavy-hadron chiral perturbation theory, one can use the Noether procedure.
For an infinitesimal local axial transformation,
\begin{equation}
 R(x)=L^\dag(x)=1+i\alpha_a(x) \tau^a,
\end{equation}
one finds that the change in the leading-order Lagrangian is given by
\begin{equation}
\delta \mathcal{L} = (\partial^\mu \alpha_a)\:A_\mu^{a(\chi \mathrm{PT})}
\end{equation}
with
\begin{eqnarray}
\nonumber A^{a(\chi \mathrm{PT})}_\mu &=& \frac{if^2}{8} \left[ \tau^a \Sigma^\dag \partial_\mu \Sigma + \tau^a (\partial_\mu\Sigma)\Sigma^\dag - \tau^a \Sigma \partial_\mu \Sigma^\dag - \tau^a (\partial_\mu \Sigma^\dag) \Sigma \right]^i_{\:\:i} \\
\nonumber && - \: v_\mu \: \mathrm{tr}_{\mathrm{D}} \left[ \ol{H}_i (\tau^a_{\xi-} )^i_{\:\:j} H^j \right] - v_\mu \left[ \ol{S}^\nu_{ij}  (  \tau^a_{\xi-} )^i_{\:\:k}  S_\nu^{kj} + \ol{S}_{ij}^\nu  (  \tau^a_{\xi-} )^j_{\:\:k}  S_\nu^{ik} \right] + v_\mu \left[ \ol{T}_{ij}  (  \tau^a_{\xi-} )^i_{\:\:k}  T^{kj} + \ol{T}_{ij}  (  \tau^a_{\xi-} )^j_{\:\:k}  T^{ik} \right] \\
\nonumber && + \: {g_1} \: \mathrm{tr}_{\mathrm{D}} \left[ \ol{H}_i (\tau^a_{\xi+} )^i_{\:\:j} \gamma_\mu\gamma_5 H^j \right] - i {g_2} \: \varepsilon_{\mu\nu\sigma\lambda} \ol{S}^\nu_{ki}  v^\sigma ( \tau^a_{\xi+}  )^i_{\:\:j}  (S^\lambda)^{jk}\\
&& + \: \sqrt{2}\:{g_3} \left[ (\ol{S}_\mu)_{ki}  ( \tau^a_{\xi+}  )^i_{\:\:j}  T^{jk} + \ol{T}_{ki}  ( \tau^a_{\xi+}  )^i_{\:\:j}  (S_\mu)^{jk} \right], \label{eq:chPTaxialcurrent}
\end{eqnarray}
where we have introduced the quantities
\begin{eqnarray}
\nonumber  \tau^a_{\xi-} &=& \frac 12 \left( \xi^\dag \tau^a \xi - \xi \tau^a \xi^\dag \right), \\
 \tau^a_{\xi+} &=& \frac 12 \left( \xi^\dag \tau^a \xi + \xi \tau^a \xi^\dag \right).
\end{eqnarray}
Equation (\ref{eq:chPTaxialcurrent}) is the leading-order axial current in the chiral effective theory. In the following, we work with a particular flavor of the axial current,
\begin{equation}
A^{-(\chi \mathrm{PT})}_\mu=A^{1(\chi \mathrm{PT})}_\mu - i\:A^{2(\chi \mathrm{PT})}_\mu, 
\end{equation}
which corresponds to the QCD current $\bar d \gamma_\mu\gamma_5 u$.  To lowest order in the pion fields (zero pion fields), the part of the axial current that will contribute to the matrix elements
we will consider reads
\begin{equation}
A^{\mu-(\chi \mathrm{PT})} = \: g_1 \: \mathrm{tr}_{\mathrm{D}} \left[ \ol{H}_d\: \gamma^\mu\gamma_5 H^u \right]
- i g_2 \: \varepsilon^{\mu\nu\sigma\lambda} v_\sigma \left[ (\ol{S}_\nu)_{dd}   (S_\lambda)^{du} + (\ol{S}_\nu)_{du} (S_\lambda)^{uu} \right]
+ \: \sqrt{2}\:g_3 \left[ -\ol{S}^\mu_{dd}  T^{du} - \ol{T}_{du} (S^\mu)^{uu} \right]. \label{eq:chPTaxialcurrentplus}
\end{equation}
We will now calculate matrix elements of (\ref{eq:chPTaxialcurrentplus}) at leading order. To this end, we canonically quantize heavy-hadron chiral perturbation theory.
The following derivation allows us to determine the correct normalization of the matrix elements (canonical quantization
cannot be performed in the partially quenched theory, but the normalizations can be inferred).
We begin with the heavy mesons. Using Eq.~(\ref{eq:Hfield}), we find that the free part of the heavy-meson kinetic term in Eq.~(\ref{eq:HHchPTL}) is equal to
\begin{eqnarray}
\nonumber \mathcal{L}_H &=& -i v_\mu \: \mathrm{tr}_{\mathrm{D}} \left[ \ol{H}_i \partial^\mu  H^i \right] \\
               &=& -i v_\mu \: \left[ -2 P_i^{\dag} \partial^\mu P^i
+ 2  P^{*\nu\dag}_{i}  \: \partial^\mu \: P^{*i}_\nu \right].
\end{eqnarray}
From this, we deduce the canonical equal-time commutation relations for the field operators (in the following, we use sans-serif font for operators)
\begin{eqnarray}
\nonumber \:[ \mathsf{P}^i(\mathbf{x},t), \: \mathsf{P}_j^{\dag}(\mathbf{x'},t) ] & = & \frac{1}{2 v^0} \delta^i_{\:\:j}\delta^3(\mathbf{x}-\mathbf{x'}), \\
\:[ \mathsf{P}^{*i}_{\mu}(\mathbf{x},t), \: \mathsf{P}_j^{*\nu\dag}(\mathbf{x'},t) ] & = & -\frac{1}{2 v^0} \delta^i_{\:\:j} (g_\mu^{\:\:\nu}-v_\mu v^\nu) \delta^3(\mathbf{x}-\mathbf{x'}). \label{eq:fieldOpCR}
\end{eqnarray}
The field operators of the noninteracting theory can be written as
\begin{eqnarray}
\nonumber \mathsf{P}^i(x) &=& \frac{1}{\sqrt{2v^0}} \int \frac{\mathrm{d}^3 k}{(2\pi)^3} \mathsf{a}_{(P)}^i(\mathbf{k}) e^{-i k \cdot x}, \\
\nonumber  \mathsf{P}^{\dag}_i(x) &=& \frac{1}{\sqrt{2v^0}} \int \frac{\mathrm{d}^3 k}{(2\pi)^3} \mathsf{a}^{(P)\dag}_i (\mathbf{k}) e^{i k \cdot x}, \\
\nonumber  \mathsf{P}^{*i}_{\mu}(x) &=& \frac{1}{\sqrt{2v^0}} \int \frac{\mathrm{d}^3 k}{(2\pi)^3} \sum_{s=1}^3 \mathsf{a}_{(P^*)}^i(\mathbf{k},s) \epsilon_\mu(s) e^{-i k \cdot x}, \\
 \mathsf{P}_i^{*\mu\dag}(x) &=& \frac{1}{\sqrt{2v^0}} \int \frac{\mathrm{d}^3 k}{(2\pi)^3} \sum_{s=1}^3 \mathsf{a}^{(P^*)\dag}_i(\mathbf{k},s) \epsilon^{*\mu}(s) e^{i k \cdot x}, \label{eq:mesonmodedecomp}
\end{eqnarray}
where $k_0 = \mathbf{v}\cdot\mathbf{k}$, and the basis polarization vectors satisfy the spin sum
\begin{equation}
\sum_{s=1}^3 \epsilon_\mu(s) \epsilon^*_\nu(s) = -g_{\mu\nu}+v_\mu v_\nu.  \label{eq:spinsum1}
\end{equation}
Equation (\ref{eq:fieldOpCR}) is satisfied if the commutation relations for the creation and annihilation operators are
\begin{eqnarray}
\nonumber \: [ \mathsf{a}_{(P)}^i(\mathbf{k}), \: \mathsf{a}_j^{(P)\dag}(\mathbf{k'}) ] &=& (2\pi)^3 \delta^i_{\:\:j} \delta^3 (\mathbf{k}-\mathbf{k'}), \\
 \: [ \mathsf{a}_{(P^*)}^i(\mathbf{k},s), \: \mathsf{a}_j^{(P^*)\dag}(\mathbf{k'},s') ] &=& (2\pi)^3 \delta^i_{\:\:j} \delta_{s s'} \delta^3 (\mathbf{k}-\mathbf{k'}).
\end{eqnarray}
We define single-particle states via
\begin{eqnarray}
\nonumber |P^i(\mathbf{k})\rangle = \sqrt{2 v^0} \mathsf{a}_i^{(P)\dag}(\mathbf{k}) |0\rangle, \\
|P^{*i}(\mathbf{k},s)\rangle = \sqrt{2 v^0} \mathsf{a}_i^{(P^*)\dag}(\mathbf{k},s) |0\rangle. \label{eq:mesonstates}
\end{eqnarray}
Note that all the heavy-hadron states and operators also depend on the velocity $v$, which is a parameter in the Lagrangian. However, since $v$ is fixed throughout this paper,
we do not include a label $v$ explicitly.

The states (\ref{eq:mesonstates}) are normalized as
\begin{eqnarray}
\nonumber \langle P^i(\mathbf{k}) | P^j(\mathbf{k}') \rangle &=&  2v^0 (2\pi)^3 \delta^{ij} \delta^3(\mathbf{k}-\mathbf{k}'), \\
 \langle P^{*i}(\mathbf{k},s) | P^{*j}(\mathbf{k}',s') \rangle &=&  2v^0 (2\pi)^3 \delta^{ij}\delta_{s s'} \delta^3(\mathbf{k}-\mathbf{k}').
\end{eqnarray}
We now calculate the matrix element $\langle P^{*d} | \mathsf{A}^{\mu-(\chi \mathrm{PT})} | P^{u}\rangle$. Expressing the mesonic part of the current (\ref{eq:chPTaxialcurrentplus}) using the $P$ and $P^*$ fields, we find
\begin{eqnarray}
\nonumber A^{\mu-(\chi \mathrm{PT})} &=& \: g_1 \: \mathrm{tr}_{\mathrm{D}} \left[ \ol{H}_d \gamma^\mu\gamma_5 H^u \right] + ... \\
\nonumber &=&  g_1 \: \mathrm{tr}_{\mathrm{D}} \left[ P^{*\dag}_{d\nu} \gamma^\nu  \gamma^\mu\gamma_5 \left( - P^u\gamma_5 \right)\frac{1-\slashed{v}}{2} \right] + ... \\
&=& - 2\:g_1 \: P^{*\mu\dag}_d P^{u} + ..., \label{eq:PstarPcurrent}
\end{eqnarray}
where we have only shown the piece that contributes to the matrix element considered here. By inserting the field operators (\ref{eq:mesonmodedecomp})
into Eq.~(\ref{eq:PstarPcurrent}), we immediately obtain, at zero residual momentum,
\begin{equation}
 \langle P^{*d}(0,s) | \mathsf{A}^{-(\chi \mathrm{PT})}_\mu (0) | P^u(0) \rangle|_{\rm LO} = -2\:g_1 \: \varepsilon^*_\mu(s). \label{eq:PPstarmatrixelt}
\end{equation}
Next, we consider the $s_l=1$ baryon field $S^{ij}_\mu$, for which the free part of the kinetic term in Eq.~(\ref{eq:HHchPTL}) is equal to
\begin{eqnarray}
\nonumber \mathcal{L}_S &=&  \ol{S}^\nu_{ij} \left[ -i v_\mu \partial^\mu + \Delta \right]  S^{ij}_\nu \\
&=& \sum_{i\geq j}  (2-\delta_{ij}) \ol{S}^\nu_{ij} \left[ - i v_\mu \partial^\mu + \Delta \right] S^{ij}_\nu.
\end{eqnarray}
In the following we always assume that the flavor indices on the fields $S^{ij}$ and $\overline{S}_{ij}$ are ordered as $i\geq j$. We find the canonical anticommutation relations
\begin{equation}
 \{ \mathsf{S}^{\mu\: ij}(\mathbf{x},t)_\alpha,\:\ol{\mathsf{S}}_{\nu\:kl\:\beta}(\mathbf{x'},t) \} = -\frac{1}{(2-\delta_{ij})v^0} \delta^i_{\:\:k} \delta^j_{\:\:l} \left( \frac{1+\slashed{v}}{2} \right)_{\alpha\beta} (g^\mu_{\:\:\nu}-v^\mu v_\nu) \:\delta^3(\mathbf{x}-\mathbf{x'}).
\end{equation}
The field operators can be written as
\begin{eqnarray}
\nonumber  \mathsf{S}^{\mu\:ij}(x)&=&\frac{1}{\sqrt{(2-\delta_{ij})v^0}} \int \frac{\mathrm{d}^3 k}{(2\pi)^3} \sum_{s=1}^6 \mathsf{a}_{(S)}^{ij}(\mathbf{k},s) U^\mu (s) e^{-i k \cdot x}, \\
  \ol{\mathsf{S}}_{\mu\:ij}(x)&=&\frac{1}{\sqrt{(2-\delta_{ij})v^0}} \int \frac{\mathrm{d}^3 k}{(2\pi)^3} \sum_{s=1}^6 \mathsf{a}^{(S)\dag}_{ij}(\mathbf{k},s) \overline{U}_\mu(s) e^{i k \cdot x}, \label{eq:Smodedecomp}
\end{eqnarray}
where $k_0 = \mathbf{v}\cdot\mathbf{k} + \frac{1}{v^0}\Delta$, and the basis spinors satisfy the spin sum
\begin{equation}
\sum_{s=1}^6 U^\mu_\alpha(s) \overline{U}^\nu_\beta(s) = -(g^{\mu\nu}-v^\mu v^\nu) \left[\frac{1+\slashed{v}}{2} \right]_{\alpha\beta}. \label{eq:spinsum2}
\end{equation}
Note that $U_\mu(s)$ is \emph{not} a Rarita-Schwinger spinor, but rather contains the degrees of freedom of both spin-$1/2$ and spin-$3/2$ baryons.
The creation and annihilation operators satisfy the anticommutation relations
\begin{equation}
 \: \{ \mathsf{a}_{(S)}^{ij}(\mathbf{k},s), \: \mathsf{a}^{(S)\dag}_{kl}(\mathbf{k'},s') \} = (2\pi)^3 \delta^i_{\:\:k}\delta^j_{\:\:l} \delta_{s s'} \delta^3 (\mathbf{k}-\mathbf{k'}).
\end{equation}
Here we define single-particle states via
\begin{equation}
|S^{ij}(\mathbf{k},s)\rangle = \sqrt{v^0} \mathsf{a}^{(S)\dag}_{ij}(\mathbf{k},s) |0\rangle, \label{eq:Sstates}
\end{equation}
which corresponds to the following normalization:
\begin{eqnarray}
\langle S^{ij}(\mathbf{k},s) | S^{kl}(\mathbf{k}',s') \rangle &=&  v^0 (2\pi)^3 \delta^{ik}\delta^{jl} \delta_{s s'} \delta^3(\mathbf{k}-\mathbf{k}').
\end{eqnarray}
Using the expression (\ref{eq:chPTaxialcurrentplus}) for the axial current, and the mode decomposition (\ref{eq:Smodedecomp}), we find the following leading-order matrix element of the
axial current:
\begin{equation}
 \langle S^{dd}(0,s) | \mathsf{A}^{\mu-(\chi \mathrm{PT})}(0) |  S^{du}(0,s') \rangle|_{\rm LO}
= -\frac{i}{\sqrt{2}}\:g_2 \:v_\lambda \: \epsilon^{\lambda \mu \nu\rho}\: \overline{U}_\nu(s) U_\rho(s'). \label{eq:SSmatrixelt}
\end{equation}
Finally, we consider the $s_l=0$ baryon field $T^{ij}$. The free part of the kinetic term in Eq.~(\ref{eq:HHchPTL}) is equal to
\begin{eqnarray}
\nonumber \mathcal{L}_T &=& i v_\mu  \: \ol{T}_{ij} \partial^\mu  T^{ij}  \\
&=& 2\:i v_\mu \sum_{i>j} \ol{T}_{ij} \partial^\mu  T^{ij}. 
\end{eqnarray}
For the $T$ and $\overline{T}$ fields, we also assume in the following that the flavor indices are ordered ($i > j$). We obtain the canonical anticommutation relations
\begin{equation}
 \{ \mathsf{T}^{ij}(\mathbf{x},t)_\alpha,\:\ol{\mathsf{T}}_{kl\:\beta}(\mathbf{x'},t) \} = \frac{1}{2v^0} \delta^i_{\:\:k} \delta^j_{\:\:l} \left( \frac{1+\slashed{v}}{2} \right)_{\alpha\beta} \delta^3(\mathbf{x}-\mathbf{x'}),
\end{equation}
and the field operators can be written as
\begin{eqnarray}
\nonumber  \mathsf{T}^{ij}(x)&=&\frac{1}{\sqrt{2v^0}} \int \frac{\mathrm{d}^3 k}{(2\pi)^3} \sum_{s=1}^2 \mathsf{a}_{(T)}^{ij}(\mathbf{k},s) \mathcal{U}(s) e^{-i k \cdot x}, \\
  \ol{\mathsf{T}}_{ij}(x)&=&\frac{1}{\sqrt{2v^0}} \int \frac{\mathrm{d}^3 k}{(2\pi)^3} \sum_{s=1}^2 \mathsf{a}^{(T)\dag}_{ij}(\mathbf{k},s) \overline{\mathcal{U}}(s) e^{i k \cdot x},
\end{eqnarray}
where $k_0 = \mathbf{v}\cdot\mathbf{k}$, and the basis spinors satisfy the spin sum
\begin{equation}
\sum_{s=1}^2 \mathcal{U}_\alpha(s) \overline{\mathcal{U}}_\beta(s) = \left[\frac{1+\slashed{v}}{2} \right]_{\alpha\beta}. \label{eq:spinsum3}
\end{equation}
The creation and annihilation operators satisfy the anticommutation relations
\begin{equation}
 \: \{ \mathsf{a}_{(T)}^{ij}(\mathbf{k},s), \: \mathsf{a}^{(T)\dag}_{kl}(\mathbf{k'},s') \} = (2\pi)^3 \delta^i_{\:\:k}\delta^j_{\:\:l} \delta_{s s'} \delta^3 (\mathbf{k}-\mathbf{k'}).
\end{equation}
Again, we define single-particle states via
\begin{equation}
|T^{ij}(\mathbf{k},s)\rangle = \sqrt{v^0} \mathsf{a}^{(T)\dag}_{ij}(\mathbf{k},s) |0\rangle, \label{eq:Tstates}
\end{equation}
which are normalized as
\begin{eqnarray}
\langle T^{ij}(\mathbf{k},s) | T^{kl}(\mathbf{k}',s') \rangle &=&  v^0 (2\pi)^3 \delta^{ik}\delta^{jl} \delta_{s s'} \delta^3(\mathbf{k}-\mathbf{k}').
\end{eqnarray}
Now we have all the ingredients to obtain the leading-order $S$-$T$ transition matrix element of the axial current:
\begin{equation}
\langle S^{dd}(0,s) | \mathsf{A}^{\mu-(\chi \mathrm{PT})}(0) |   T^{du}(0,s') \rangle|_{\rm LO} = -g_3 \: \overline{U}^\mu(s) \: \mathcal{U}(s'). \label{eq:STmatrixelt}
\end{equation}
To go beyond leading order, we need to replace $g_1$, $g_2$, $g_3$ in Eqs.~(\ref{eq:PPstarmatrixelt}), (\ref{eq:SSmatrixelt}), (\ref{eq:STmatrixelt}) by ``effective
axial couplings''  $(g_1)_{\rm eff}$, $(g_2)_{\rm eff}$, and $(g_3)_{\rm eff}$:
\begin{eqnarray}
\nonumber \langle P^{*d}(0,s) | \mathsf{A}^{\mu-(\chi \mathrm{PT})} (0) | P^{u}(0) \rangle &=& -2\:(g_1)_{\rm eff} \: \varepsilon^{*\mu}(s), \\
\nonumber \langle S^{dd}(0,s) | \mathsf{A}^{\mu-(\chi \mathrm{PT})}(0) |  S^{du}(0,s') \rangle &=& -\frac{i}{\sqrt{2}}\:(g_2)_{\rm eff} \:v_\lambda \: \epsilon^{\lambda \mu \nu\rho}\: \overline{U}_\nu(s) U_\rho(s'), \\
 \langle S^{dd}(0,s) | \mathsf{A}^{\mu-(\chi \mathrm{PT})}(0) |   T^{du}(0,s') \rangle &=& -(g_3)_{\rm eff} \: \overline{U}^\mu(s) \: \mathcal{U}(s'). \label{eq:matrixelementsCHPT}
\end{eqnarray}
The next-to-leading-order expressions in the partially quenched $SU(4|2)$ theory (for $m_u^{(\rm val)}=m_d^{(\rm val)}$ and $m_u^{(\rm sea)}=m_d^{(\rm sea)}$), calculated via the
perturbative expansion of the path integral, have been derived in Ref.~\cite{Detmold:2011rb} and are given by
\begin{eqnarray}\
\nonumber (g_1)_{\rm eff} &=& g_1-\frac{2\:g_1}{f^2}\:\mathcal{I}(m_\pi^{(\mathrm{vs})}) + \frac{g_1^3}{f^2}\Bigg[ 4 \: \mathcal{H}(m_\pi^{(\mathrm{vs})},\:0)-4 \: \delta^2_{VS} \mathcal{H}_{\eta'}(m_\pi^{(\mathrm{vv})},\:0) \Bigg] + ({\rm analytic\:\:terms}), \\
\nonumber (g_2)_{\rm eff} &=& g_2-\frac{2\:g_2}{f^2}\:\mathcal{I}(m_\pi^{(\mathrm{vs})}) + \frac{g_2^3}{f^2}\Bigg[ \frac32 \: \mathcal{H}(m_\pi^{(\mathrm{vs})},\:0) - \delta^2_{VS} \mathcal{H}_{\eta'}(m_\pi^{(\mathrm{vv})},\:0)  \Bigg] \\
\nonumber                 && + \frac{g_2\:g_3^2}{f^2} \Bigg[ 2\: \mathcal{H}(m_\pi^{(\mathrm{vs})},\:-\Delta) - \mathcal{H}(m_\pi^{(\mathrm{vv})},\:-\Delta) - 2\: \mathcal{K}(m_\pi^{(\mathrm{vs})},\:-\Delta,\:0)   \Bigg] + ({\rm analytic\:\:terms}),  \\
\nonumber (g_3)_{\rm eff} &=& g_3-\frac{2\:g_3}{f^2}\:\mathcal{I}(m_\pi^{(\mathrm{vs})}) + \frac{g_3^3}{f^2} \Bigg[\mathcal{H}(m_\pi^{(\mathrm{vs})},\:-\Delta) -\frac12 \mathcal{H}(m_\pi^{(\mathrm{vv})},\:-\Delta) \\
\nonumber                 &&  + \frac{3}{2} \mathcal{H}(m_\pi^{(\mathrm{vv})},\:\Delta) + 3\: \mathcal{H}(m_\pi^{(\mathrm{vs})},\:\Delta)  - \mathcal{K}(m_\pi^{(\mathrm{vs})},\:\Delta,\:0)           \Bigg] \\
\nonumber                 && + \frac{g_3\:g_2^2}{f^2} \Bigg[ -\mathcal{H}(m_\pi^{(\mathrm{vs})},\:\Delta)-\mathcal{H}(m_\pi^{(\mathrm{vv})},\:\Delta) + \mathcal{H}(m_\pi^{(\mathrm{vs})},\:0) - \delta^2_{VS} \mathcal{H}_{\eta'}(m_\pi^{(\mathrm{vv})},\:0) \Bigg] + ({\rm analytic\:\:terms}). \\
\label{eq:PQchiPTformula}
\end{eqnarray}
Here, $m_\pi^{(\mathrm{vs})}$ denotes the mass of a pion consisting of a valence and a sea quark, $m_\pi^{(\mathrm{vv})}$ denotes the mass of a pion consisting of two valence quarks,
and $\delta^2_{VS}=[m_\pi^{(\mathrm{vv})}]^2-[m_\pi^{(\mathrm{vs})}]^2$. The functions $\mathcal{I}$, $\mathcal{H}$, $\mathcal{H}_{\eta'}$ and $\mathcal{K}$,
which arise from the chiral loops and include the leading effects of the finite spatial volume, are defined in Ref.~\cite{Detmold:2011rb}.
At the order considered here, the analytic terms in Eq.~(\ref{eq:PQchiPTformula}) are linear functions of $[m_\pi^{(\mathrm{vv})}]^2$ and $[m_\pi^{(\mathrm{vs})}]^2$.

\subsection{Axial-current matrix elements in lattice QCD}

\label{sec:axialcurrentmeLQCD}

For the lattice QCD calculation, we construct interpolating fields for the heavy hadrons in terms of the quark fields as follows:
\begin{eqnarray}
\nonumber P^i &=& \overline{Q}_{a\alpha} \: (\gamma_5)_{\alpha\beta} \: \tilde{q}^i_{a\beta}, \\
\nonumber P^{*i}_\mu &=& \overline{Q}_{a\alpha} \: (\gamma_\mu)_{\alpha\beta} \: \tilde{q}^i_{a\beta}, \\
\nonumber S^{ij}_{\mu\:\alpha} &=&  \epsilon_{abc}\:(C\gamma_\mu)_{\beta\gamma}\:\tilde{q}^i_{a\beta}\:\tilde{q}^j_{b\gamma}\: Q_{c\alpha}, \\
T^{ij}_\alpha &=& \epsilon_{abc}\:(C\gamma_5)_{\beta\gamma}\:\tilde{q}^i_{a\beta}\:\tilde{q}^j_{b\gamma}\: Q_{c\alpha}. \label{eq:interpolatingfields}
\end{eqnarray}
Here, $a$, $b$, $c$ are color indices, and $\alpha,\beta,\gamma$ are spinor indices. The light-quark field of flavor $i$ is denoted by $q^i$ (we will also use the notation $u=q^u$, $d=q^d$),
and the heavy-quark (antiquark) field is denoted by $Q$ ($\overline{Q}$). The tilde on the light-quark fields indicates that these are smeared over multiple spatial lattice sites, in order to improve
the overlap of the interpolating fields with the corresponding ground-state hadrons and reduce excited-state contamination. We use gauge-invariant Gaussian smearing
obtained by
\begin{equation}
 \tilde{q} = \left(1 + \frac{\sigma^2}{4\:n_S}\Delta^{(2)}\right)^{n_S} q, \label{eq:qsmearing}
\end{equation}
where $\Delta^{(2)}$ is a three-dimensional gauge-covariant lattice Laplacian which includes gauge links, $\sigma$ is the smearing width and $n_S$ is the number of smearing iterations.

The heavy quark $Q$ is defined in the static limit, and we set $\mathbf{v}=0$. The static
heavy-quark field $Q$ satisfies
\begin{equation}
\frac{1+\gamma^0}{2}Q=Q.
\end{equation}
Note that the interpolating field $S^{ij}_{\mu\:\alpha}$ couples to both the $J=1/2$ and the $J=3/2$ baryons with $s_l=1$, with exactly the same relative amplitudes as (\ref{eq:Sfield}).

We use the local four-dimensional lattice axial current, where the quark and antiquark fields are evaluated at the same lattice point.
This current requires a finite renormalization factor $Z_A$ to match the continuum QCD current:
\begin{equation}
A^{a(\mathrm{LQCD})}_\mu = Z_A\:\overline{q}_{i\:a\alpha} \frac12(\tau^a)^i_{\:\:j} (\gamma_\mu \gamma_5)_{\alpha\beta} q^j_{a\beta}. \label{eq:latticeaxialcurrent}
\end{equation}
The value of $Z_A$ depends on the lattice action used,
as well as the lattice spacing. We will come back to this in Sec.~\ref{sec:latticeparams}. As before, we will work with a specific flavor of the axial current:
\begin{equation}
A^{-(\mathrm{LQCD})}_\mu = Z_A\:\overline{d}_{a\alpha} (\gamma_\mu \gamma_5)_{\alpha\beta} u_{a\beta}. \label{eq:latticeaxialcurrent2}
\end{equation}
In the following, we will omit the superscript ``$-(\mathrm{LQCD})$'' in the axial current (\ref{eq:latticeaxialcurrent2}).
Next, we define the overlap factors, which describe the overlap of the operators corresponding to the interpolating fields (\ref{eq:interpolatingfields}) with the relevant
ground-state hadrons in QCD. Here, we use the same notation and normalization of states as in Sec.~\ref{sec:axialcurrentmatrixeltshhchPT}:
\begin{eqnarray}
\nonumber \langle 0 | \mathsf{P}^i(0) | P^i(k) \rangle &=& Z_{P^i}, \\
\nonumber \langle 0 | \mathsf{P}^{*i}_\mu (0) | P^{*i}(\mathbf{k},s) \rangle &=& Z_{P^{*i}}\:\varepsilon_\mu(s), \\
\nonumber \langle 0 | \mathsf{S}^{ij}_{\mu\:\alpha}(0) | S^{ij}(\mathbf{k},s) \rangle &=& Z_{S^{ij}} \: U_{\mu\:\alpha}(s), \\
\langle 0 | \mathsf{T}^{ij}_\alpha (0) | T^{ij}(\mathbf{k},s) \rangle &=& Z_{T^{ij}} \: \mathcal{U}_\alpha(s).
\end{eqnarray}
We stress that these states are now meant to be hadron states in (lattice) QCD, rather than in the chiral effective theory.

We calculate Euclidean two-point functions of the interpolating fields (\ref{eq:interpolatingfields}), as well as Euclidean three-point functions with an insertion
of the axial current (\ref{eq:latticeaxialcurrent2}). These Euclidean correlation functions are obtained from the lattice path integral,
which is performed numerically using importance sampling. In the following, we assume that the Wick rotation $t \rightarrow -it$ has been performed, so that $t$ denotes the Euclidean time.

We calculate the following three-point functions, where $\langle \: ... \: \rangle$ denotes the path-integral over the gauge- and fermion fields
(for the domain-wall action used in this work, there is also an additional path integral over Pauli-Villars fields \cite{Shamir:1993zy, Antonio:2006px}):
\begin{eqnarray}
\nonumber C[P^{*d}\:A\:P_u^{\dag}]^{\mu\nu}(t,\: t') &=& \sum_\mathbf{x} \sum_\mathbf{x'} \:\langle \: P^{*d\:\mu}(\mathbf{x},t) \:A^\nu(\mathbf{x'},t') \: P_u^{\dag}(0) \: \rangle, \\
\nonumber C[S^{dd}A\:\:\overline{S}_{du}]^{\mu\nu\rho}_{\alpha\beta}(t,\: t') &=& \sum_\mathbf{x} \sum_\mathbf{x'} \:\langle \:S^{dd\:\mu}_\alpha(\mathbf{x},t) \:A^\nu(\mathbf{x'},t') \: \overline{S}_{du\:\beta}^{\rho}(0) \: \rangle,  \\
\nonumber C[S^{dd}A\:\:\overline{T}_{du}]^{\mu\nu}_{\alpha\beta}(t,\: t') &=& \sum_\mathbf{x} \sum_\mathbf{x'} \:\langle \:S^{dd\:\mu}_\alpha(\mathbf{x},t) \:A^{\nu}(\mathbf{x'},t') \: \overline{T}_{du\:\beta}(0) \: \rangle,  \\
 C[T^{du}A^\dag\:\:\overline{S}_{dd}]^{\mu\nu}_{\alpha\beta}(t,\: t') &=& \sum_\mathbf{x} \sum_\mathbf{x'} \:\langle \:T^{du}_\alpha(\mathbf{x},t) \:A^{\mu\dag}(\mathbf{x'},t') \: \overline{S}_{dd\:\beta}^{\nu}(0) \: \rangle. \label{eq:threeptfns}
\end{eqnarray}
In addition, we calculate the two-point functions
\begin{eqnarray}
\nonumber C[P^u\:P_u^{\dag}](t) &=& \sum_\mathbf{x} \:\langle \: P^u(\mathbf{x},t) \: P_u^{\dag}(0) \: \rangle, \\
\nonumber C[P^{*d}\:P_d^{*\dag}]^{\mu\nu}(t) &=& \sum_\mathbf{x} \:\langle \: P^{*d\:\mu}(\mathbf{x},t) \: P_d^{*\:\nu\dag}(0) \: \rangle, \\
\nonumber C[S^{dd}\:\overline{S}_{dd}]^{\mu\nu}_{\alpha\beta}(t) &=& \sum_\mathbf{x} \:\langle \:S^{dd\:\mu}_\alpha(\mathbf{x},t) \: \overline{S}_{dd\:\beta}^{\nu}(0) \: \rangle,  \\
\nonumber C[S^{du}\:\overline{S}_{du}]^{\mu\nu}_{\alpha\beta}(t) &=& \sum_\mathbf{x} \:\langle \:S^{du\:\nu}_\alpha(\mathbf{x},t) \: \overline{S}_{du\:\beta}^{\nu}(0) \: \rangle,  \\
 C[T^{du}\:\overline{T}_{du}]_{\alpha\beta}(t) &=& \sum_\mathbf{x} \:\langle \:T^{du}_\alpha(\mathbf{x},t) \: \overline{T}_{du\:\beta}(0) \: \rangle. \label{eq:twoptfns}
\end{eqnarray}
As an example, we show how the two-point function $C[S^{dd}\:\overline{S}_{dd}]^{\mu\nu}_{\alpha\beta}(t)$ is constructed in terms of quark propagators.
The lattice calculation is performed in the isospin limit of equal up and down quark masses.
Inserting the definitions of the interpolating fields, and performing the Grassmann integrals over the fermion fields explicitly, we have
\begin{eqnarray}
\nonumber C[S^{dd}\:\overline{S}_{dd}]^{\mu\nu}_{\alpha\beta}(t) &=& \epsilon_{abc}\:(C\gamma^\mu)_{\delta\gamma}\:\epsilon_{fgh}\:(\overline{C\gamma^\nu})_{\rho\sigma}\:\sum_\mathbf{x}\left\langle \tilde{d}^a_\delta(x)\:\tilde{d}^b_\gamma(x)\: Q^c_\alpha(x)
\: \overline{\tilde{d}}^g_\rho(0)\:\overline{\tilde{d}}^f_\sigma(0)\: \overline{Q}^h_\beta(0) \right\rangle\\
\nonumber &=&\epsilon_{abc}\:(C\gamma^\mu)_{\delta\gamma}\:\epsilon_{fgh}\:(\overline{C\gamma^\nu})_{\rho\sigma}\:\sum_\mathbf{x}\left\langle\: \tilde{\tilde{G}}_{(q)\delta\sigma}^{af}(x,0)\:\tilde{\tilde{G}}_{(q)\gamma\rho}^{bg}(x,0)\:G_{(Q)\alpha\beta}^{ch}(x,0) \right.\\
\nonumber && \hspace{26ex}-\left.\tilde{\tilde{G}}_{(q)\delta\rho}^{ag}(x,0)\:\tilde{\tilde{G}}_{(q)\gamma\sigma}^{bf}(x,0)\:G_{(Q)\alpha\beta}^{ch}(x,0) \right\rangle_U\\
&=& 2\:\epsilon_{abc}\:(C\gamma^\mu)_{\delta\gamma}\:\epsilon_{fgh}\:(\overline{C\gamma^\nu})_{\rho\sigma}\:\sum_\mathbf{x}\left\langle \tilde{\tilde{G}}_{(q)\delta\sigma}^{af}(x,0)\:\tilde{\tilde{G}}_{(q)\gamma\rho}^{bg}(x,0)\:G_{(Q)\alpha\beta}^{ch}(x,0) \right\rangle_U. \label{eq:twoptcontractions}
\end{eqnarray}
Here, $\tilde{\tilde{G}}_{(q)}$ denotes a smeared-source smeared-sink light-quark propagator and $G_{(Q)}$ a heavy-quark propagator.
The notation $\langle \: ... \: \rangle_U$ indicates the path integral over the gauge fields $U$ only. In the last step in Eq.~(\ref{eq:twoptcontractions}) we have used the symmetry of $(C\gamma_\mu)$ and the antisymmetry of $\epsilon_{abc}$ to equate the two Wick contractions.
Note that the two-point function $C[S^{du}\:\overline{S}_{du}]^{\mu\nu}_{\alpha\beta}(t)$ contains only one Wick contraction, because the two light quarks have different flavors. Therefore, in the isospin limit,
\begin{equation}
 C[S^{du}\:\overline{S}_{du}]^{\mu\nu}_{\alpha\beta}(t) = \frac12 C[S^{dd}\:\overline{S}_{dd}]^{\mu\nu}_{\alpha\beta}(t). \label{Stwoptisospin}
\end{equation}
For the static lattice action we are using, the heavy-quark propagator is equal to
\begin{equation}
G_{(Q)\alpha\beta}(x,0)=\delta_{\mathbf{x},\mathbf{0}} \left[\frac{1+\gamma^0}{2}\right]_{\alpha\beta} \tilde{U}_0^\dag(\mathbf{x}, t-a) \: \tilde{U}_0^\dag(\mathbf{x}, t-2a) \: ... \: \tilde{U}_0^\dag(\mathbf{x}, 0), \label{eq:hqprop}
\end{equation}
where $\tilde{U}_0$ is a spatially smeared gauge link in the temporal direction (more details will be given in Sec.~\ref{sec:latticeparams}).
Because the static heavy-quark propagator (\ref{eq:hqprop}) contains
the Kronecker delta $\delta_{\mathbf{x},\mathbf{0}}$, the sums over $\mathbf{x}$ in all the correlation functions, Eqs. (\ref{eq:threeptfns}) and (\ref{eq:twoptfns}), are eliminated.
To calculate the three-point functions (\ref{eq:threeptfns}) in terms of quark propagators, we use pairs of light-quark propagators
with smeared sources at $(0,0)$ and $(0,t)$. This means that new inversions are needed for each value of $t$.

By using the Hamiltonian and the momentum operator to shift the left interpolating operator from $x$ to $0$ and the axial current from $x'$ to $0$, and inserting complete sets of states, we can show that
\begin{eqnarray}
\nonumber C[P^{*d}\:A\:P_u^{\dag}]^{\mu\nu}(t,\: t') &=& \frac14 Z_{P^{*d}}\:Z^*_{P^u} \sum_{s=1}^3  \:\varepsilon^{\mu}(s)\: \langle P^{*d}(0,s) | \mathsf{A}^\nu (0) | P^u(0) \rangle \: \:e^{-E_{P^{*d}}(t-t')}\:e^{-E_{P^u}t'} + ...\:,\\
\nonumber C[S^{dd}A\:\:\overline{S}_{du}]^{\mu\nu\rho}_{\alpha\beta}(t,\: t') &=& Z_{S^{dd}} Z^*_{S^{du}} \sum_{s=1}^6 \sum_{s'=1}^6 U_\alpha^\mu(s) \: \langle S^{dd}(0,s) | \mathsf{A}^\nu(0) |  S^{du}(0,s') \rangle \: \overline{U}_\beta^\rho(s') \: e^{-E_{S^{dd}}(t-t')}\:e^{-E_{S^{du}}t'} + ...\:, \\
\nonumber C[S^{dd}A\:\:\overline{T}_{du}]^{\mu\nu}_{\alpha\beta}(t,\: t') &=& Z_{S^{dd}} Z^*_{T^{du}} \sum_{s=1}^6 \sum_{s'=1}^2 U_\alpha^\mu(s) \: \langle S^{dd}(0,s) | \mathsf{A}^{\nu}(0) |  T^{du}(0,s') \rangle \: \overline{\mathcal{U}}_\beta(s') \: e^{-E_{S^{dd}}(t-t')}\:e^{-E_{T^{du}}t'} + ...\:, \\
\nonumber C[T^{du}A^\dag\:\:\overline{S}_{dd}]^{\mu\nu}_{\alpha\beta}(t,\: t') &=& Z_{T^{du}} Z^*_{S^{dd}} \sum_{s=1}^2 \sum_{s'=1}^6 \mathcal{U}_\alpha(s) \: \langle T^{du}(0,s) | \mathsf{A}^{\mu\dag}(0) |  S^{dd}(0,s') \rangle \: \overline{U}_\beta^\nu(s') \: e^{-E_{T^{du}}(t-t')}\:e^{-E_{S^{dd}}t'} + ...\: \\
\end{eqnarray}
and
\begin{eqnarray}
\nonumber C[P^u\:P_u^\dag](t) &=& \frac 12 |Z_{P^u}|^2 \:e^{-E_{P_u} t} + ...\:, \\
\nonumber C[P^{*d}\:P_d^{*\dag}]^{\mu\nu}(t) &=& \frac 12 |Z_{P^{*d}}|^2 \sum_{s=1}^3 \varepsilon^\mu(s)\varepsilon^{*\nu}(s)  \:e^{-E_{P^{*d}} t} + ...\:, \\
\nonumber C[S^{dd}\:\overline{S}_{dd}]^{\mu\nu}_{\alpha\beta}(t) &=& \sum_{s=1}^6 U^\mu_\alpha(s) \overline{U}^\nu_\beta(s)\:|Z_{S^{dd}}|^2 \:e^{-E_{S^{dd}} t} + ...\:,  \\
\nonumber C[S^{du}\:\overline{S}_{du}]^{\mu\nu}_{\alpha\beta}(t) &=& \sum_{s=1}^6 U^\mu_\alpha(s) \overline{U}^\nu_\beta(s)\:|Z_{S^{du}}|^2 \:e^{-E_{S^{du}} t} + ...\:,  \\
 C[T^{du}\:\overline{T}_{du}]_{\alpha\beta}(t) &=& \sum_{s=1}^2 \mathcal{U}_\alpha(s) \overline{\mathcal{U}}_\beta(s)\:|Z_{T^{du}}|^2 \:e^{-E_{T^{du}} t} + ...\:,
\end{eqnarray}
where the ellipsis indicates contributions from excited states, whose contribution relative to the ground-state contribution shown here vanishes exponentially for $t\rightarrow\infty$, $t'\rightarrow\infty$, $t-t'\rightarrow\infty$
(here and in the following we assume an infinite temporal extent of the lattice). We will return to the discussion of excited states at the end of this section.
Using the relations (\ref{eq:spinsum1}), (\ref{eq:spinsum2}), and (\ref{eq:spinsum3}) to perform the spin sums (recall that here we have $\mathbf{v}=0$) and Eq.~(\ref{eq:matrixelementsCHPT}) to express the axial-current matrix
elements in terms of the effective axial couplings $(g_1)_{\rm eff}$, $(g_2)_{\rm eff}$, $(g_3)_{\rm eff}$, we obtain
\begin{eqnarray}
\nonumber C[P^{*d}\:A\:P_u^{\dag}]^{\mu\nu}(t,\: t') &=& -\frac12\:(g_1)_{\rm eff}\:(v^\mu v^\nu - g^{\mu \nu})\: Z_{P^{*d}}\:Z^*_{P^u} \:e^{-E_{P^{*d}}(t-t')}\:e^{-E_{P^u}t'} + ...\:,\\
\nonumber C[S^{dd}A\:\:\overline{S}_{du}]^{\mu\nu\rho}_{\alpha\beta}(t,\: t') &=& \frac{i}{\sqrt{2}}\: (g_2)_{\rm eff}\: v_\lambda \epsilon^{\lambda \mu\nu\rho }\:\left[\frac{1 + \gamma^0}{2} \right]_{\alpha\beta} Z_{S^{dd}} Z^*_{S^{du}} \: e^{-E_{S^{dd}}(t-t')}\:e^{-E_{S^{du}}t'} + ...\:, \\
\nonumber C[S^{dd}A\:\:\overline{T}_{du}]^{\mu\nu}_{\alpha\beta}(t,\: t') &=& -(g_3)_{\rm eff} \: (v^\mu v^\nu - g^{\mu \nu})\left[\frac{1+\gamma^0}{2}\right]_{\alpha\beta}\: Z_{S^{dd}} Z^*_{T^{du}} \: e^{-E_{S^{dd}}(t-t')}\:e^{-E_{T^{du}}t'} + ...\:,  \\
C[T^{du}A^\dag\:\:\overline{S}_{dd}]^{\mu\nu}_{\alpha\beta}(t,\: t') &=& -(g_3)_{\rm eff} \: (v^\mu v^\nu - g^{\mu \nu})\left[\frac{1+\gamma^0}{2}\right]_{\alpha\beta}\: Z_{T^{du}} Z^*_{S^{dd}}  \: e^{-E_{T^{du}}(t-t')}\:e^{-E_{S^{dd}}t'} + ... 
\end{eqnarray}
and
\begin{eqnarray}
\nonumber C[P^u\:P_u^\dag](t) &=& \frac 12 |Z_{P^u}|^2 \:e^{-E_{P^u} t} + ...\:, \\
\nonumber C[P^{*d}\:P_d^{*\dag}]^{\mu\nu}(t) &=& \frac12 (v^\mu v^\nu - g^{\mu \nu})\:|Z_{P^{*d}}|^2 \:e^{-E_{P^{*d}} t} + ...\:, \\
\nonumber C[S^{dd}\:\overline{S}_{dd}]^{\mu\nu}_{\alpha\beta}(t) &=& (v^\mu v^\nu - g^{\mu \nu})\:\left[\frac{1+\gamma^0}{2}\right]_{\alpha\beta}\:|Z_{S^{dd}}|^2 \:e^{-E_{S^{dd}} t} + ...\:,  \\
\nonumber C[S^{du}\:\overline{S}_{du}]^{\mu\nu}_{\alpha\beta}(t) &=& (v^\mu v^\nu - g^{\mu \nu})\:\left[\frac{1+\gamma^0}{2}\right]_{\alpha\beta}\:|Z_{S^{du}}|^2 \:e^{-E_{S^{du}} t} + ...\:,  \\
 C[T^{du}\:\overline{T}_{du}]_{\alpha\beta}(t) &=& \left[\frac{1+\gamma^0}{2}\right]_{\alpha\beta}\:|Z_{T^{du}}|^2 \:e^{-E_{T^{du}} t} + ...\:.
\end{eqnarray}
In the following, we remove the trivial spin-structure $\left[\frac{1+\gamma_0}{2}\right]_{\alpha\beta}$, which comes purely from the heavy-quark propagator (\ref{eq:hqprop}),
from all baryon correlation functions.

Because the lattice calculation is performed in the isospin limit (and in the static limit for the heavy quark), we have the relations
\begin{eqnarray}
\nonumber E_{P^u} &=& E_{P^{d*}}, \\
\nonumber Z_{P^u} &=& Z_{P^{d*}}, \\
\nonumber E_{S^{dd}} &=& E_{S^{du}}, \\
 Z_{S^{dd}} &=& \sqrt{2}\:Z_{S^{du}}
\end{eqnarray}
[the factor of $\sqrt{2}$ in the last line comes from Eq.~(\ref{Stwoptisospin})]. As a consequence of the equality of energies, the $t'$-dependence of the ground-state contribution
in the three-point functions $C[P^{*d}\:A\:P_u^{\dag}]^{\mu\nu}(t,\: t')$ and $C[S^{dd}A\:\:\overline{S}_{du}]^{\mu\nu\rho}(t,\: t')$
cancels completely. For these three-point functions, the $t$-dependence as well as the $Z$ factors in the ground-state contribution can be canceled by forming the ratios
\begin{eqnarray}
 R_1(t,\: t') &=& -\frac{\frac13\sum_{\mu=1}^3 C[P^{*d}\:A\:P_u^{\dag}]^{\mu\mu}(t,\: t')}{C[P^u\:P_u^\dag](t)} \label{eq:R1} \\
 \nonumber  &=& (g_1)_{\rm eff} + ...\, ,
\end{eqnarray}
and
\begin{eqnarray}
 R_2(t,\: t') &=& 2 \frac{\frac i6 \sum_{\mu,\nu,\rho=1}^3\epsilon_{0\mu\nu\rho} \: C[S^{dd}A\:\:\overline{S}_{du}]^{\mu\nu\rho}(t,\: t') }{\frac13\sum_{\mu=1}^3 C[S^{dd}\:\overline{S}_{dd}]^{\mu\mu}(t) }  \label{eq:R2} \\
\nonumber &=& (g_2)_{\rm eff} + ...\, ,
\end{eqnarray}
where, as before, the ellipsis indicates contributions from excited states that vanish exponentially when all Euclidean time separations are taken to infinity.
To extract $(g_3)_{\rm eff}$, we use the double ratio
\begin{eqnarray}
R_3(t,\: t') &=& \sqrt{\frac{\left[ \frac13\sum_{\mu=1}^3 C[S^{dd}A\:\:\overline{T}_{du}]^{\mu\mu}(t,\: t') \right] \left[ \frac13\sum_{\mu=1}^3 C[T^{du}A^\dag\:\:\overline{S}_{dd}]^{\mu\mu}(t,\: t') \right] }{ \left[ \frac13\sum_{\mu=1}^3 C[S^{dd}\:\overline{S}_{dd}]^{\mu\mu}(t) \right] \:\left[ C[T^{du}\:\overline{T}_{du}](t) \right]  }} \label{eq:R3} \\
\nonumber &=& (g_3)_{\rm eff} + ... \, .
\end{eqnarray}
The numerical results for (\ref{eq:R1}), (\ref{eq:R2}), and (\ref{eq:R3}) and the subsequent analysis will be described in Sec.~\ref{sec:results}.
In the following, we discuss the contributions from excited states to the ratios. Again, we assume an infinite temporal extent of the lattice; with a finite temporal
extent $T$ this means that the following discussion is only valid for source-sink separations $t$ that are smaller than $T/2$ by a sufficient distance (which is the case in our numerical calculations).
We begin with $R_1$, and define
\begin{eqnarray}
\nonumber \langle 0     | \mathsf{P}^u(0)        | P^u_n \rangle &=& Z_{P,n}, \\
\nonumber \langle 0     | \mathsf{P}^{*d}_\mu(0)  | P^{*d}_n (\varepsilon) \rangle &=& Z_{P,n} \:\varepsilon_\mu, \\
          \langle P^{*d}_n | \mathsf{A}_\mu(0)      | P^{u}_m(\varepsilon) \rangle &=& -2 \: A^{(PP^*)}_{nm} \:\varepsilon^*_\mu,
\end{eqnarray}
where $| P^u_n \rangle$ denotes the $n$-th excited state with a nonzero overlap $\langle 0 | \mathsf{P}^u(0) | P^u_n \rangle$,
and similarly for $| P^{*d}_n (\varepsilon) \rangle$. Because of heavy-quark symmetry
and isospin symmetry, all energies
and $Z$-factors in the $P^u$ sector are equal to those in the $P^{*d}$ sector, and $A^{(PP^*)}_{nm}=A^{*(PP^*)}_{mn}$. Note that
\begin{equation}
 A^{(PP^*)}_{11}=(g_1)_{\rm eff}.
\end{equation}
The complete spectral decomposition of $R_1$ reads
\begin{equation}
R_1(t,\: t') = \frac{ \sum_{n=1}^\infty \sum_{m=1}^\infty Z_{P,n} \: Z_{P,m}^* \: A^{(PP^*)}_{nm}\:e^{-E_{P,n}\, t} \: e^{-E_{P,m} (t-t')} }
{\sum_{n=1}^\infty \big|Z_{P,n}\big|^2 e^{-E_{P,n}\, t}}. \label{eq:R1spectralcomplete}
\end{equation}
Showing only the contributions from the ground states and first excited states, we find that
\begin{equation}
R_1(t,\: t') = A^{(PP^*)}_{11} + \frac{ \big|\frac{Z_{P,2}}{Z_{P,1}}\big|^2 (A^{(PP^*)}_{22}-A^{(PP^*)}_{11})\:e^{-\delta_P\, t} +\frac{Z_{P,1} Z_{P,2}^*}{|Z_{P,1}|^2} \: A^{(PP^*)}_{12} \: e^{-\delta_P\, t'}+ \frac{Z_{P,2} Z_{P,1}^*}{|Z_{P,1}|^2} \: A^{*(PP^*)}_{12} \: e^{-\delta_P (t-t')} }
{1+\big|\frac{Z_{P,2}}{Z_{P,1}}\big|^2 e^{-\delta_P\, t}} + ...\, , \label{eq:R1spectral}
\end{equation}
with the energy gap $\delta_P=E_{P,2}-E_{P,1}$. For a given value of $t$, the smallest contamination from excited states is obtained
at the midpoint $t'=t/2$. Evaluating (\ref{eq:R1spectral}) at $t'=t/2$, we get
\begin{equation}
R_1(t,\: t/2) = A^{(PP^*)}_{11} + \frac{ \big|\frac{Z_{P,2}}{Z_{P,1}}\big|^2 (A^{(PP^*)}_{22}-A^{(PP^*)}_{11})\:e^{-\delta_P\, t} +2\:\Re \left[ \frac{Z_{P,1} Z_{P,2}^*}{|Z_{P,1}|^2} \: A^{(PP^*)}_{12} \right] \: e^{-\frac12 \delta_P\, t} }
{1+\big|\frac{Z_{P,2}}{Z_{P,1}}\big|^2 e^{-\delta_P\, t}} + ...\,,
\end{equation}
where $\Re$ denotes the real part. By using the Taylor expansion $1/(1+\big|\frac{Z_{P,2}}{Z_{P,1}}\big|^2 e^{-\delta_P\, t})=1-\big|\frac{Z_{P,2}}{Z_{P,1}}\big|^2 e^{-\delta_P\, t}+...$,
we obtain
\begin{equation}
R_1(t,\: t/2) = A^{(PP^*)}_{11} + \left|\frac{Z_{P,2}}{Z_{P,1}}\right|^2 (A^{(PP^*)}_{22}-A^{(PP^*)}_{11})\:e^{-\delta_P\, t} +2\:\Re \left[ \frac{Z_{P,1} Z_{P,2}^*}{|Z_{P,1}|^2} \: A^{(PP^*)}_{12} \right] \: e^{-\frac12 \delta_P\, t} + ...\, , \label{eq:R1simplespectral}
\end{equation}
where we have omitted terms that decay like $e^{-\frac32 \delta_P\, t}$ or faster, and are therefore exponentially suppressed relative to the terms shown in (\ref{eq:R1simplespectral}) at large $t$.
The result for $R_2(t,\: t/2)$ has the same form (with suitable definitions of the overlap factors and matrix elements):
\begin{equation}
R_2(t,\: t/2) = A^{(SS)}_{11} + \left|\frac{Z_{S,2}}{Z_{S,1}}\right|^2 (A^{(SS)}_{22}-A^{(SS)}_{11})\:e^{-\delta_S\, t} +2\:\Re \left[ \frac{Z_{S,1} Z_{S,2}^*}{|Z_{S,1}|^2} \: A^{(SS)}_{12} \right] \: e^{-\frac12 \delta_S\, t} + ...\,, \label{eq:R2simplespectral}
\end{equation}
with $A^{(SS)}_{11}=(g_2)_{\rm eff}$ and $\delta_S=E_{S,2}-E_{S,1}$. For the double ratio $R_3(t,\: t/2)$ we obtain, after Taylor-expanding the square root and omitting terms that decay faster than
$e^{-\delta_S\, t}$ or $e^{-\delta_T t}$,
\begin{eqnarray}
\nonumber R_3(t,\: t/2) &=& A^{(ST)}_{11}
-\frac{A^{(ST)}_{11} |Z_{S,2}|^2 e^{-\delta_S\, t}}{2 |Z_{S,1}|^2}
-\frac{A^{(ST)}_{11} |Z_{T,2}|^2 e^{-\delta_T t}}{2 |Z_{T,1}|^2} \\
\nonumber && +  \Re \left[ A^{(ST)}_{21} \: \frac{Z_{S,2}}{Z_{S,1}} \right]  e^{-\frac12 \delta_S\, t}
   +  \Re \left[ A^{(ST)}_{12}   \: \frac{Z_{T,2}}{Z_{T,1}} \right]  e^{-\frac12 \delta_T t} \\
&& +  \Re \left[A^{(ST)}_{22} \: \frac{ Z_{T,2} Z_{S,2}^*}{ Z_{T,1} Z_{S,1}^*} \right] e^{-\frac{1}{2} (\delta_S+\delta_T)t} + ... \,, \label{eq:R3simplespectral}
\end{eqnarray}
with $A^{(ST)}_{11}=(g_3)_{\rm eff}$ and $\delta_S=E_{S,2}-E_{S,1}$, $\delta_T=E_{T,2}-E_{T,1}$. The $t$-dependent terms in the right-hand-sides of Eqs. (\ref{eq:R1simplespectral}), (\ref{eq:R2simplespectral}), and (\ref{eq:R3simplespectral}) are the
leading excited-state contributions to the extraction of $(g_i)_{\rm eff}$ from the ratio method.

\section{Lattice actions and parameters}

\label{sec:latticeparams}

Our calculations are based on gauge field ensembles generated by the RBC/UKQCD collaboration. These ensembles
include $2+1$ dynamical light-quark flavors, implemented with a domain-wall action \cite{Kaplan:1992bt, Furman:1994ky, Shamir:1993zy}. The gluons are implemented
with the Iwasaki action \cite{Iwasaki:1983ck, Iwasaki:1984cj}, which is known to reduce the residual chiral symmetry breaking
of the domain-wall action \cite{Aoki:2002vt}.

The details of the actions used in generating the ensembles can be found in Ref.~\cite{Antonio:2006px}.
Our analysis includes the ensembles of size $24^3\times64$ and $32^3\times 64$, which are described in Ref.~\cite{Aoki:2010dy}.
These ensembles have lattice spacings of approximately 0.112 and 0.085 fm, respectively,
so that the spatial volume is about $(2.7\:{\rm fm})^3$ in both cases.

At the coarse lattice spacing, we used only the ensemble with $am_{u,d}^{(\mathrm{sea})}=0.005$, which
is the lightest available mass. At the fine lattice spacing, we used the ensembles with the lightest two
available values of the sea quark mass, $am_{u,d}^{(\mathrm{sea})}=0.004$ and $am_{u,d}^{(\mathrm{sea})}=0.006$.
The values for the residual quark mass, which is the additive quark-mass renormalization coming from
the residual chiral symmetry breaking at finite extent of the fifth dimension $L_s$, are approximately $am_{\rm res}=0.0032$ at the coarse lattice spacing
and $am_{\rm res}=0.00067$ at the fine lattice spacing. The sea-strange-quark masses are about 10\% above the physical value \cite{Aoki:2010dy},
and we assign a 1.5\% systematic uncertainty to our final results for the axial couplings to account for this,
based on the size of the effect on similar axial-current observables as determined using mass reweighting in Ref.~\cite{Aoki:2010dy}.

\begin{figure}[ht!]
\includegraphics[width=0.35\linewidth]{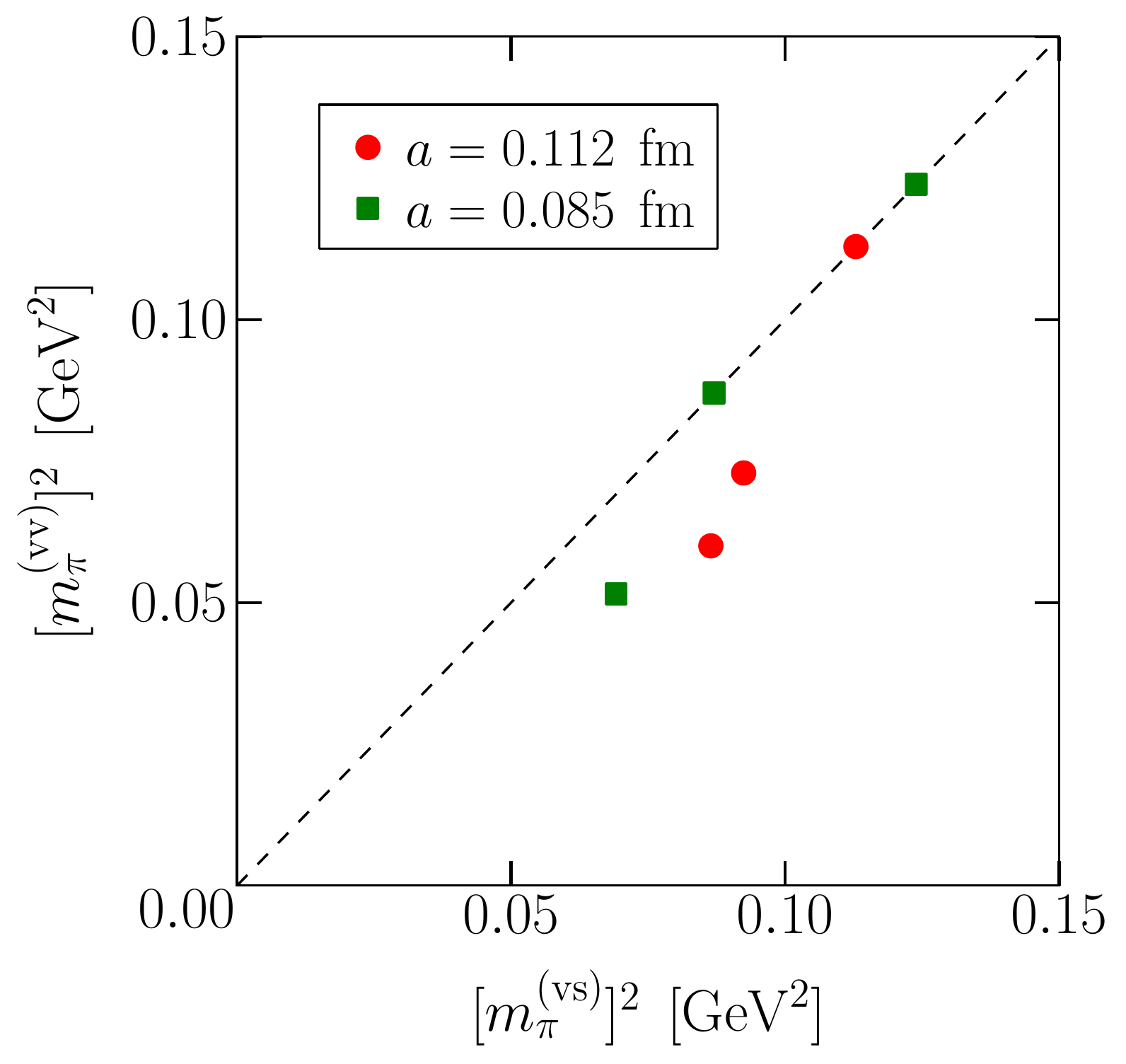}
  \caption{The values of $[m_\pi^{(\mathrm{vv})}]^2$ and $[m_\pi^{(\mathrm{vs})}]^2$ used in our calculation. The dashed line indicates the unquenched case $m_\pi^{(\mathrm{vv})}=m_\pi^{(\mathrm{vs})}$.}
  \label{fig:pionmasses}
\end{figure}

We calculated light-quark propagators using exactly the same domain-wall action that was used by the RBC/UKQCD collaboration for the sea quarks,
with the same domain-wall height of $aM_5=1.8$ and extent of the fifth dimension $L_s=16$. We used propagator sources smeared
according to Eq.~(\ref{eq:qsmearing}), with $\sigma=4.35$ and $n_S=30$.
As summarized in Table \ref{tab:params}, we calculated propagators for valence quark masses $am_{u,d}^{(\mathrm{val})}$ both equal to and lighter than
the sea quark masses. The data with $m_{u,d}^{(\mathrm{val})}<m_{u,d}^{(\mathrm{sea})}$ are referred to as ``partially quenched''.
Also shown in Table \ref{tab:params}, and plotted in Fig.~\ref{fig:pionmasses}, are the corresponding masses of pions composed of the three different possible combinations
of valence and sea quarks. The lightest valence-valence pion mass is $227(3)$ MeV, at the fine lattice spacing.

\begin{table}[ht!]
\begin{tabular}{ccccccccccccccc}
\hline\hline
$L^3\times T$ & & $am_{s}^{(\mathrm{sea})}$  & & $am_{u,d}^{(\mathrm{sea})}$ & & $am_{u,d}^{(\mathrm{val})}$    & & $a$ (fm)   & & $m_\pi^{(\mathrm{ss})}$ (MeV)   & & $m_\pi^{(\mathrm{vs})}$ (MeV)   & & $m_\pi^{(\mathrm{vv})}$ (MeV)  \\
\hline
$24^3\times64$ && $0.04$ && $0.005$ && $0.001$  && $0.1119(17)$  &&  336(5)       &&  294(5)   && 245(4)    \\
$24^3\times64$ && $0.04$ && $0.005$ && $0.002$  && $0.1119(17)$  &&  336(5)       &&  304(5)   && 270(4)    \\
$24^3\times64$ && $0.04$ && $0.005$ && $0.005$  && $0.1119(17)$  &&  336(5)       &&  336(5)   && 336(5)    \\
\\[-2ex]
$32^3\times64$ && $0.03$ && $0.004$ && $0.002$  && $0.0849(12)$  &&  295(4)       &&  263(4)   && 227(3)    \\
$32^3\times64$ && $0.03$ && $0.004$ && $0.004$  && $0.0849(12)$  &&  295(4)       &&  295(4)   && 295(4)    \\
$32^3\times64$ && $0.03$ && $0.006$ && $0.006$  && $0.0848(17)$  &&  352(7)       &&  352(7)   && 352(7)    \\
\hline\hline
\end{tabular}
\caption{\label{tab:params} Parameters of the gauge field ensembles and quark propagators. The lattice spacing values are from Ref.~\protect{\cite{Meinel:2010pv}}.}
\end{table}

For the heavy quark, we use a static action of the form given by Eichten and Hill \cite{Eichten:1989kb}, which corresponds to heavy-quark propagators
of the form (\ref{eq:hqprop}). For the temporal gauge links in this action (or, equivalently, the propagators), we use HYP smeared links \cite{Hasenfratz:2001hp}
with smearing parameters $\alpha_1=\alpha_2=\alpha_3=0.75$. This leads to an exponential improvement
in the signal-to-noise ratio \cite{DellaMorte:2003mn}.
The smearing procedure can be iterated $n_{\rm HYP}$ times, leading to a broader smearing and further improvement of the signal-to-noise ratio.
We generated data for $n_{\rm HYP}=1,2,3,5,10.$ While all of these actions have the same continuum limit, the dependence of the results
on the lattice spacing is expected to be different for different values of $n_{\rm HYP}$. One may naively expect large discretization effects
for large values of $n_{\rm HYP}$, which correspond to a large spatial extent of the heavy-quark-gluon interaction vertex.
We will discuss this in detail when giving the numerical results in Sec.~\ref{sec:results}.
Our final axial coupling results only make use of data generated with $n_{\rm HYP}=1,2,3$.

As mentioned in Sec.~\ref{sec:axialcurrentmeLQCD}, to perform the contractions for the three-point functions (\ref{eq:threeptfns}), we required
pairs of light-quark propagators with sources located at the same spatial point and separated by $t/a$ steps in the time direction. The numbers
of measurements (propagator pairs) for each value of $t/a$ are given in Table \ref{tab:nmeas}. At the coarse lattice
spacing, our data come from typically 120 statistically independent gauge field configurations; at the fine lattice spacing we used
about 240 statistically independent gauge configurations of the $am_{u,d}^{(\mathrm{sea})}=0.004$ ensemble and 180 for the $am_{u,d}^{(\mathrm{sea})}=0.006$ ensemble.
In most cases, we have more measurements than configurations, because we generated propagators from multiple source points on the lattice.
In those cases, we have averaged over the source locations before the further analysis to remove possible autocorrelations.

\begin{table}[ht!]
\begin{tabular}{llllllc}
\hline\hline
$L^3\times T$ & & $am_{u,d}^{(\mathrm{val})}$ & & $t/a$ & & $N_{\rm meas}$ (approx.) \\
\hline
$24^3\times64$ && $0.001$  && 10             &&  550      \\
$24^3\times64$ && $0.001$  && 9, 8, 7, 6     &&  240      \\
$24^3\times64$ && $0.001$  && 5              &&  460      \\
$24^3\times64$ && $0.001$  && 4              &&  120      \\
\\[-2ex]
$24^3\times64$ && $0.002$  && 10             &&  880      \\
$24^3\times64$ && $0.002$  && 9, 8, 7, 6, 4  &&  240      \\
$24^3\times64$ && $0.002$  && 5              &&  480      \\
\\[-2ex]
$24^3\times64$ && $0.005$  && 10             &&  960      \\
$24^3\times64$ && $0.005$  && 9, 8, 7, 6, 4  &&  240      \\
$24^3\times64$ && $0.005$  && 5              &&  480      \\
\\[-2ex]
$32^3\times64$ && $0.002$  && 12             &&  1200     \\
$32^3\times64$ && $0.002$  && 9, 6           &&  480      \\
\\[-2ex]
$32^3\times64$ && $0.004$  && 12             &&  1200     \\
$32^3\times64$ && $0.004$  && 9, 6           &&  480      \\
\\[-2ex]
$32^3\times64$ && $0.006$  && 13             &&  700     \\
\hline\hline
\end{tabular}
\caption{\label{tab:nmeas} Number of propagator pairs used for the three-point functions for various values of the source-sink separation $t/a$.}
\end{table}

Within each of the three gauge field ensembles that we used, the data from different source-sink separations, different valence quark masses, and different values of $n_{\rm HYP}$
are correlated with each other. In our analysis, we properly took into account these correlations using the statistical bootstrap procedure. The initial averaging over source locations
mentioned above was also required to reduce all data from the same ensemble to matching ordered sets of measurements, as necessary to calculate the covariance matrices.
It turned out that the correlations between the $am_{u,d}^{(\mathrm{val})}=0.001$ and $am_{u,d}^{(\mathrm{val})}=0.002/0.005$ data at the coarse lattice spacing
were very weak even though the data came from the same ensemble of gauge field configurations. The reason was that all source locations used for the $am_{u,d}^{(\mathrm{val})}=0.001$ propagators were distinct from those used at $am_{u,d}^{(\mathrm{val})}=0.002/0.005$.
In contrast, the data from $am_{u,d}^{(\mathrm{val})}=0.002$ and $am_{u,d}^{(\mathrm{val})}=0.005$ came from almost identical source locations, resulting in very strong correlations (these correlations were advantageous
in constraining the quark-mass dependence in our chiral fits). Similarly, at the fine lattice spacing, the data from $am_{u,d}^{(\mathrm{val})}=0.002$ and $am_{u,d}^{(\mathrm{val})}=0.004$ came
from identical source locations, leading to strong correlations.

For the axial-current renormalization parameter, we use the values obtained nonperturbatively by the RBC/UKQCD collaboration, which are \cite{Aoki:2010dy}
\begin{equation}
Z_A=\left\{ \begin{array}{l} 0.7019(26) \:\:\:\:\mathrm{for}\:\: a=0.112\:\:\mathrm{fm}, \\ 0.7396(17)\:\:\:\:\mathrm{for}\:\:a=0.085\:\:\mathrm{fm}.  \end{array} \right.
\end{equation}

\section{Data analysis}
\label{sec:results}

\subsection{Ratios}
\label{sec:resultsratios}

\begin{figure*}[ht!]
 \includegraphics[width=0.48\linewidth]{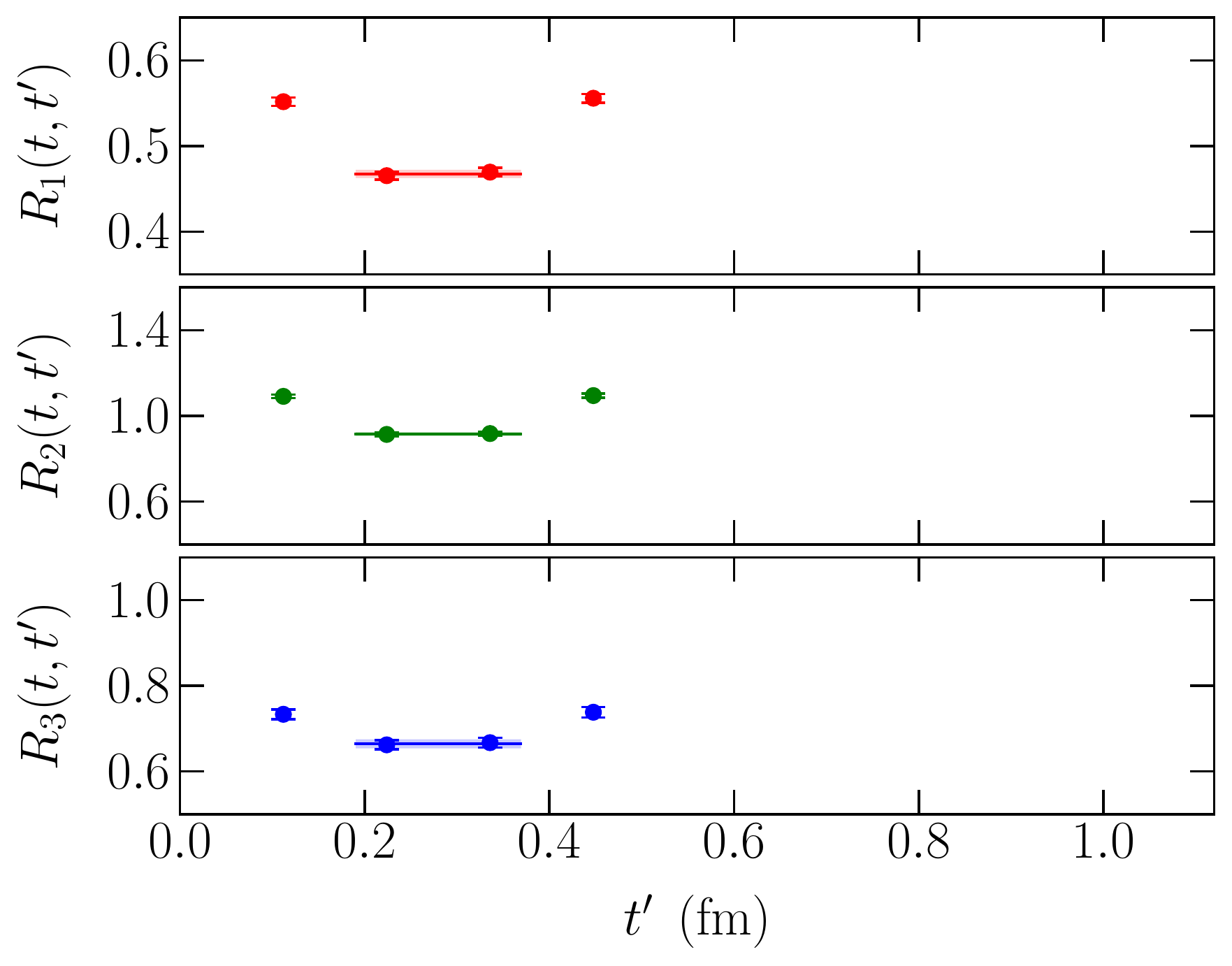}  \hfill \includegraphics[width=0.48\linewidth]{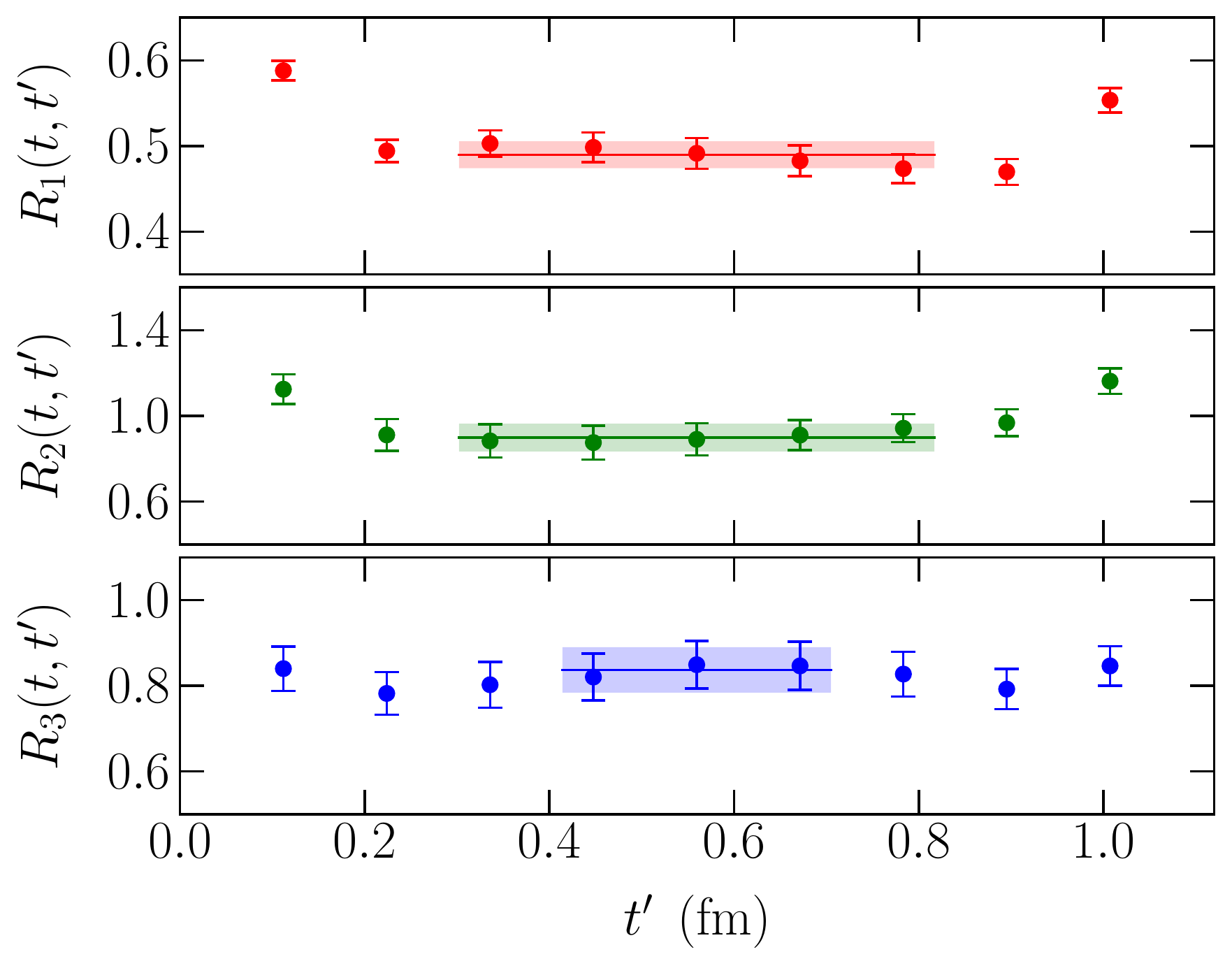}
\caption{\label{fig:ratiosL24}Ratios $R_1$, $R_2$, $R_3$ as a function of the current insertion time slice $t'$, for the source-sink separations $t/a=5$ (left) and $t/a=10$ (right).
The data shown are for $a=0.112$ fm, $n_{\rm HYP}=3$, and $am_{u,d}^{(\mathrm{val})}=0.002$.}
\end{figure*}

\begin{figure*}[ht!]
 \includegraphics[width=0.48\linewidth]{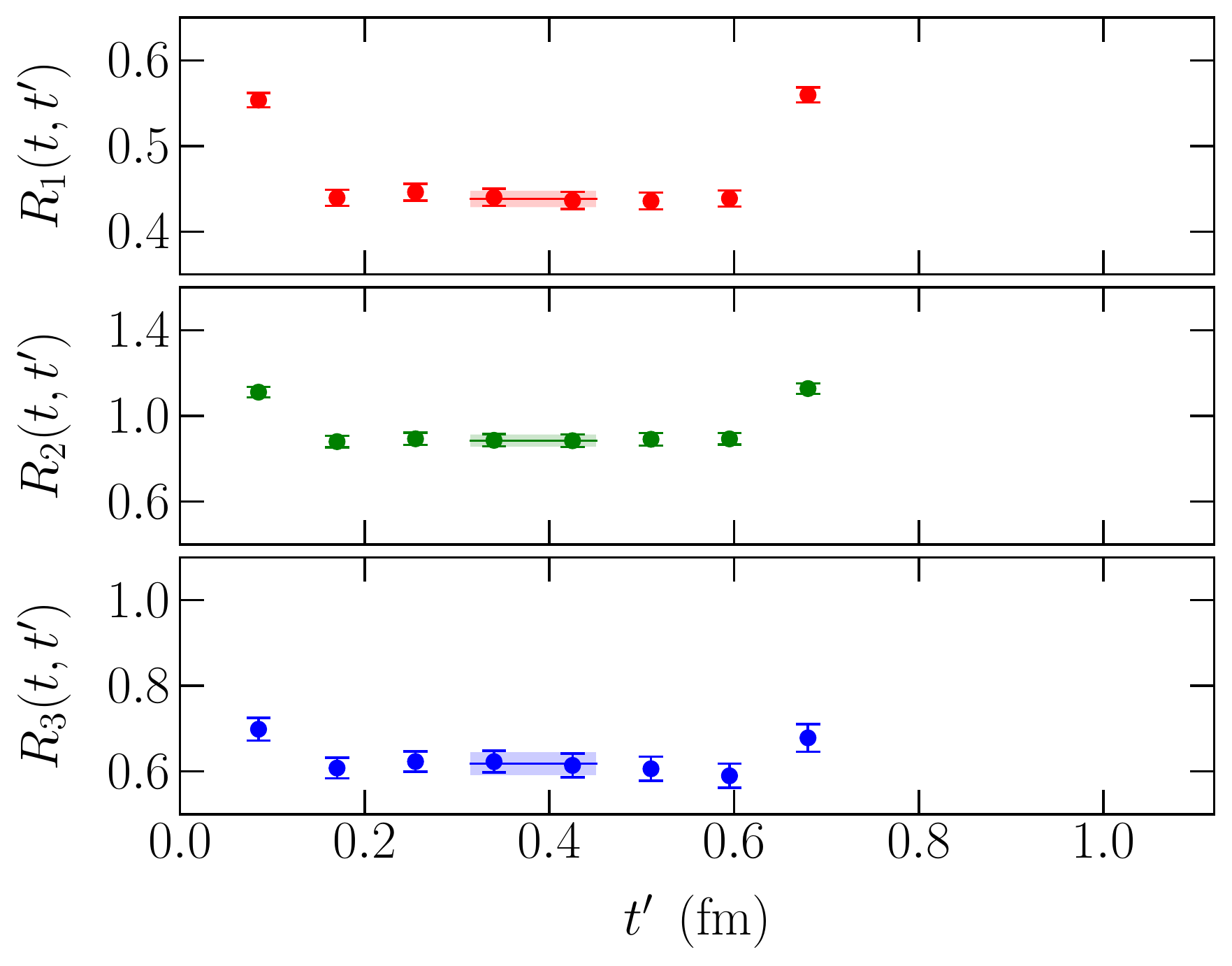}  \hfill \includegraphics[width=0.48\linewidth]{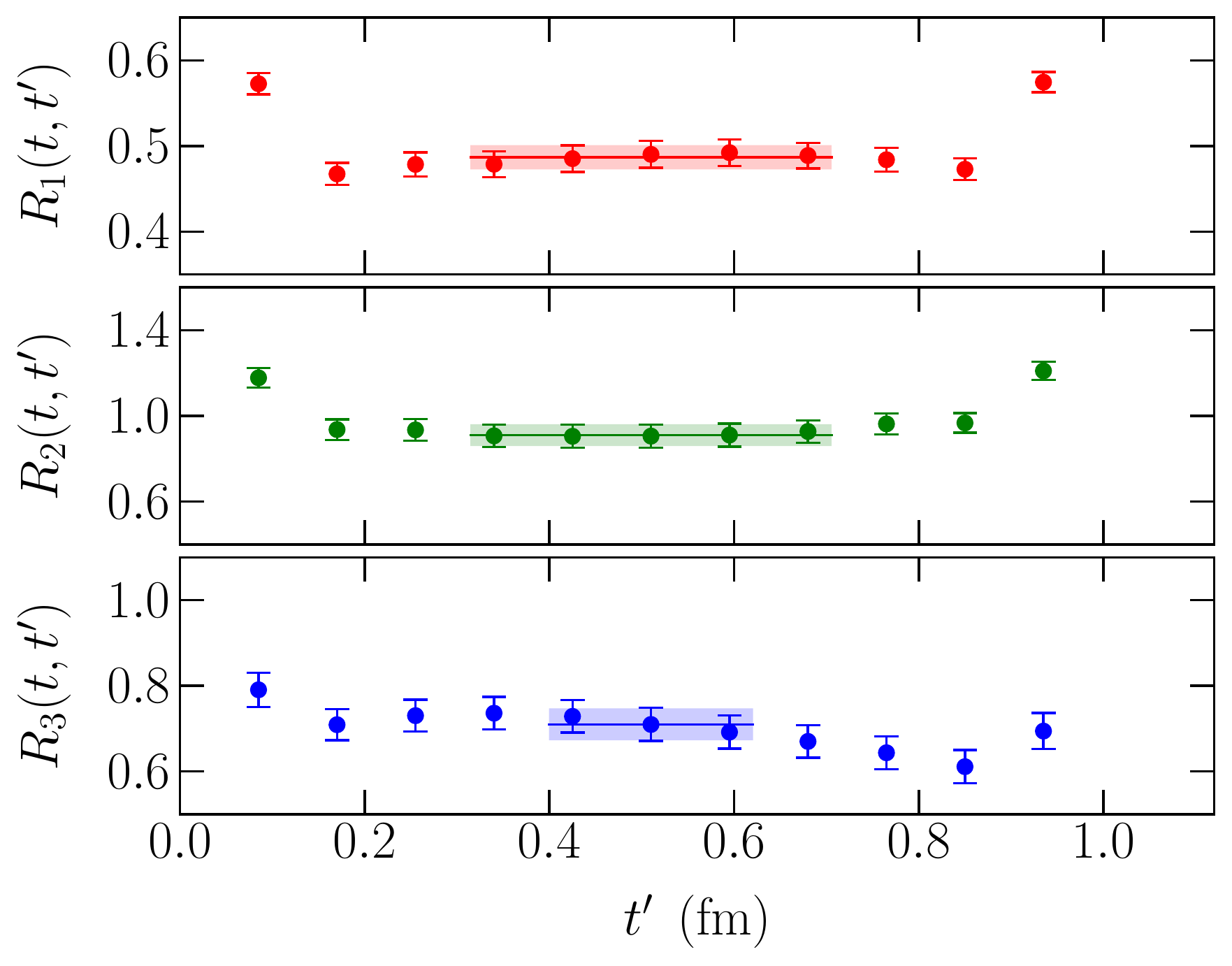}
\caption{\label{fig:ratiosL32}Ratios $R_1$, $R_2$, $R_3$ as a function of the current insertion time slice $t'$, for the source-sink separations $t/a=9$ (left) and $t/a=12$ (right).
The data shown here are for $a=0.085$ fm and $n_{\rm HYP}=3$, and $am_{u,d}^{(\mathrm{val})}=0.002$.}
\end{figure*}

Examples of numerical results for the ratios (\ref{eq:R1}), (\ref{eq:R2}), and (\ref{eq:R3}) are shown in Fig.~\ref{fig:ratiosL24} (for $a=0.112$ fm) and Fig.~\ref{fig:ratiosL32} (for $a=0.085$ fm).
These ratios were calculated using statistical bootstrap to take into account the strong correlations between the three-point and two-point functions in numerator and denominator.
Because of these correlations, the statistical uncertainties in the ratios are found to be smaller than those in the three-point functions themselves. To maximize correlations, it
is essential to use the two-point functions from the source locations matching those of the three-point functions.
The figures show results at two different values of the source-sink separation $t$ in each case, for $am_{u,d}^{(\mathrm{val})}=0.002$ and $n_{\rm HYP}=3$.

We observe clear plateaus of $R_i(t,\: t')$ as a function of $t'$, whose extent appears to be slightly larger for the simple ratios $R_1$ and $R_2$ compared to the double ratio $R_3$.
The plateaus in $t'$ indicate that in these regions contributions from off-diagonal matrix elements of the axial current between a ground-state hadron and an excited state are negligible
in comparison to the statistical uncertainties, because such a transition matrix element would introduce $t'$-dependence [see Eqs. (\ref{eq:R1spectralcomplete}) and (\ref{eq:R1spectral})].
For $R_1$, the flatness with respect to $t'$ has previously been observed and discussed in Refs.~\cite{Abada:2003un, Becirevic:2009yb}.

We averaged $R_i(t,\: t')$ over a symmetric range of $t'$ values around $t/2$
in a region where there was no statistically significant $t'$ dependence (requiring that the $\chi^2/{\rm d.o.f.}$ of correlated constant fits be of order 1).
These regions and the extracted values, which we denote as $R_i(t)$, are indicated in Figs.~\ref{fig:ratiosL24} and \ref{fig:ratiosL32} for representative data sets.
The averaging in the plateau region is essentially equivalent to using
\begin{equation}
 R_i(t, \:t/2). \label{eq:standardratio}
\end{equation}
Indeed, because of the strong corrrelation between neighboring $t'$ points, we found that the averaging over $t'$ in the plateau region (which is again performed using bootstrap)
gives almost the same result and uncertainty as $R_i(t,\: t/2)$. An alternative method for defining $R_i(t)$ is discussed in Appendix \ref{sec:summationmethod}.

Summary plots of all the extracted values for $R_1(t)$, $R_2(t)$, and $R_3(t)$ at the different quark masses, lattice spacings,
and numbers of heavy-quark smearing iterations $n_{\rm HYP}$ are given in the appendix in Fig.~\ref{fig:rawdatag2g3}. Note that the numbers of measurements vary (see Table \ref{tab:nmeas}).
The statistical uncertainties are found to grow quickly when $t$ is increased or $am_{u,d}^{(\mathrm{val})}$ is decreased, as expected \cite{Lepage:1991ui}.
Furthermore, the statistical uncertainties are reduced with every iteration of HYP smearing in the static heavy-quark action, which is also expected \cite{DellaMorte:2003mn}.
While the results for $(g_i)_{\rm eff}=\lim_{t\rightarrow\infty} R_i(t)$ from all fixed values of $n_{\rm HYP}$ will become equal when the lattice spacing is taken to zero,
at nonzero lattice spacing different values of $n_{\rm HYP}$ will have different discretization errors. Indeed, $R_1$ and $R_2$ are seen to increase with $n_{\rm HYP}$.
Remarkably, the results for $R_3$ are almost independent of $n_{\rm HYP}$ within the statistical uncertainties even at nonzero lattice spacing. See also Fig.~\ref{fig:tinfdata}
in Sec.~\ref{sec:sourcesinkextrap} for the $n_{\rm HYP}$-dependence of $(g_i)_{\rm eff}=\lim_{t\rightarrow\infty} R_i(t)$.

\subsection{Extrapolation to infinite source-sink separation}
\label{sec:sourcesinkextrap}

The effective axial couplings $(g_i)_{\rm eff}$ at given values of the lattice spacing, the quark masses, and $n_{\rm HYP}$, are defined as the infinite-time limits of $R_i(t)$:
\begin{equation}
(g_i)_{\rm eff}=\lim_{t\rightarrow \infty} R_i(t).
\end{equation}
The $t$-dependence of $R_i(t)$ is caused by excited states, and at sufficiently large $t$, the contributions from the first relevant excitation dominates.
The expected functional form of $R_1(t)$, $R_2(t)$, and $R_3(t)$ in this regime is shown in Eqs.~(\ref{eq:R1simplespectral}), (\ref{eq:R2simplespectral}), and (\ref{eq:R3simplespectral}),
respectively. As can be seen in these equations, the ``diagonal'' contributions proportional to the matrix elements $A_{11}$ and $A_{22}$ decay exponentially
like
\begin{equation}
e^{-\delta\:t},
\end{equation}
where $\delta$ is the energy gap to the first excited state that has a nonzero overlap with the corresponding interpolating field as defined in Eq.~(\ref{eq:interpolatingfields});
$\delta=\delta_P$ for $R_1$, $\delta=\delta_S$ for $R_2$, and $\delta=\min(\delta_S,\:\delta_T)$ for $R_3$. Additionally, there are ``off-diagonal'' contributions
proportional to the matrix elements $A_{12}$ and $A_{21}$, which decay only like
\begin{equation}
e^{-\frac12\delta\:t}.
\end{equation}
However, as discussed in Sec.~\ref{sec:resultsratios} and Ref.~\cite{Becirevic:2009yb}, these off-diagonal contributions appear to be small, because the numerical results
for $R(t,\:t')$ show plateaus as a function of $t'$.

We performed correlated fits to the lattice data for $R_i(t,a,m,n_{\rm HYP})$, which depends on the source-sink separation $t$,
the lattice spacing $a$, the quark masses (here denoted generically as $m$), and $n_{\rm HYP}$, using the following functions:
\begin{eqnarray}
\nonumber R_1(t,a,m,n_{\rm HYP}) &=& (g_1)_{\rm eff}(a,m,n_{\rm HYP}) - A_1(a,m,n_{\rm HYP}) \:e^{-\delta_1(a,m,n_{\rm HYP})\:t}, \\
\nonumber R_2(t,a,m,n_{\rm HYP}) &=& (g_2)_{\rm eff}(a,m,n_{\rm HYP}) - A_2(a,m,n_{\rm HYP}) \:e^{-\delta_2(a,m,n_{\rm HYP})\:t}, \\
          R_3(t,a,m,n_{\rm HYP}) &=& (g_3)_{\rm eff}(a,m,n_{\rm HYP}) - A_3(a,m,n_{\rm HYP}) \:e^{-\delta_3(a,m,n_{\rm HYP})\:t}. \label{eq:tsepextrap}
\end{eqnarray}
Here $\delta_i$ is the energy gap to the dominant excitation in $R_i$. Because $\delta_i$ is not constrained, it does not
matter whether or not we include a factor of $1/2$ in the exponent. Since the energy gap $\delta_i$ is positive by definition, we choose
to parametrize it as
\begin{equation}
 a \:\delta_i(a,m,n_{\rm HYP})=e^{l_i(a,m,n_{\rm HYP})},
\end{equation}
using the logarithm $l_i(a,m,n_{\rm HYP})$ as the fit parameter.

\begin{figure*}[ht!]
 \includegraphics[width=0.48\linewidth]{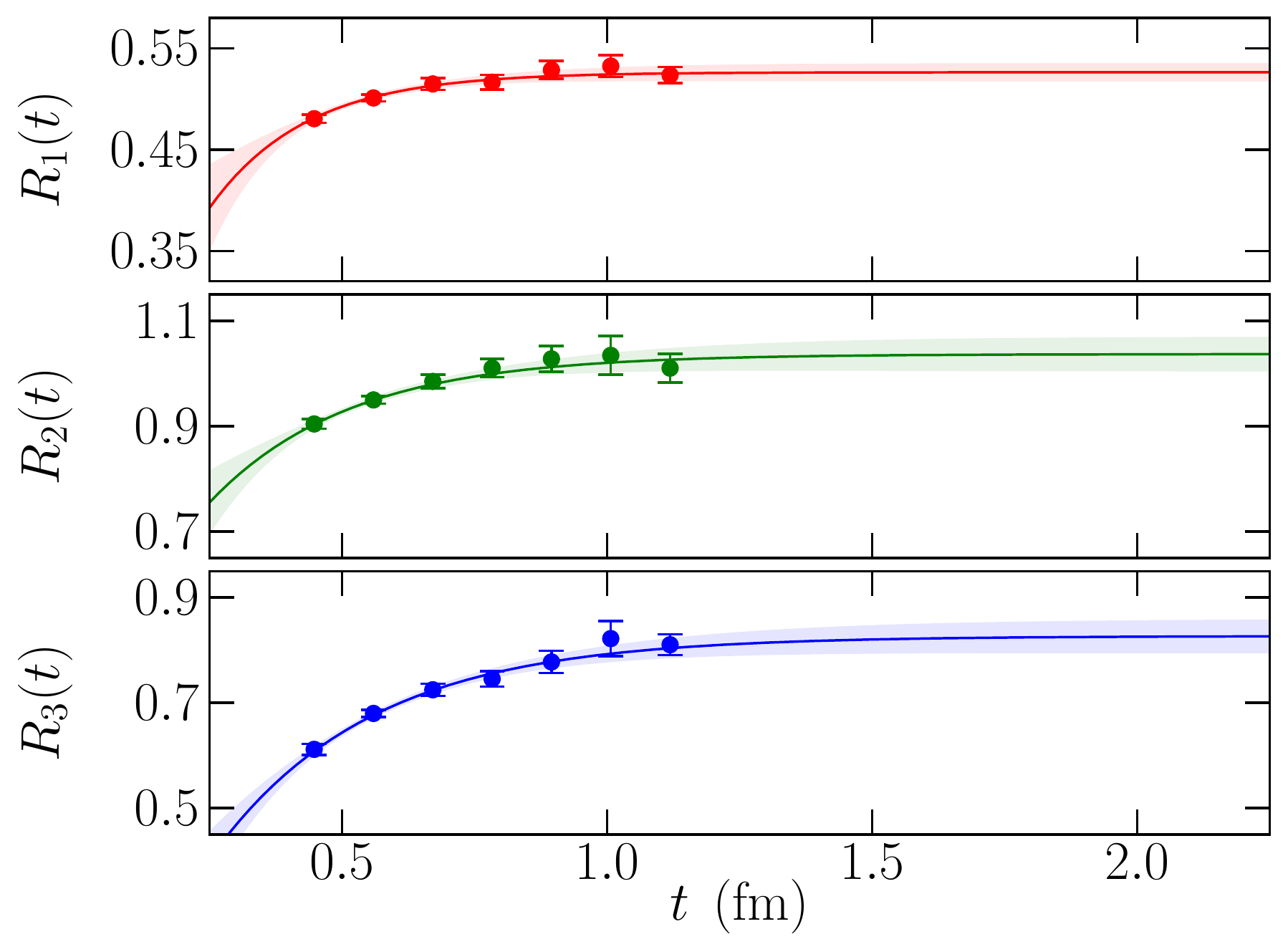}  \hfill \includegraphics[width=0.48\linewidth]{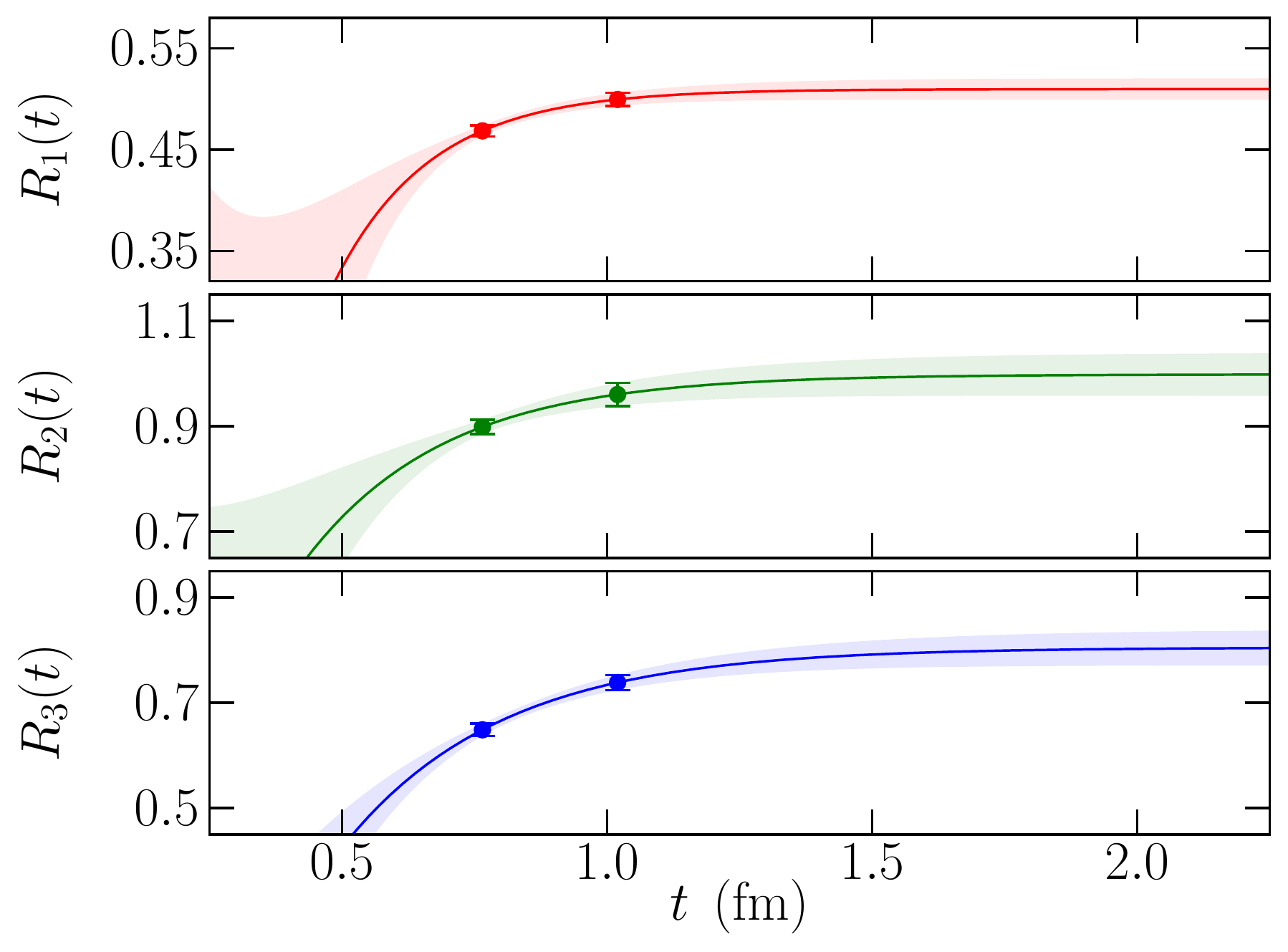}
\caption{\label{fig:tsepExtrapolationHyp10} Fits of the $t$-dependence of $R_1(t)$, $R_2(t)$, $R_3(t)$, for $n_{\rm HYP}=10$, using the functions defined in Eq.~(\ref{eq:tsepextrap}).
The left-hand side shows unconstrained fits of the $a=0.112$ fm, $am_{u,d}^{(\mathrm{val})}=0.005$ data; the right-hand side
shows fits of the $a=0.085$ fm, $am_{u,d}^{(\mathrm{val})}=0.004$ data, where the energy gap was constrained using information
from (\ref{eq:coarsenhyp10tsepextrap}).}
\end{figure*}

Because the statistical uncertainties in $R_i$ grow exponentially as $t$ is increased, we were only able to perform the lattice QCD calculations in the range $t\lesssim$ 1.1 fm.
As can be seen in Table \ref{tab:nmeas}, at the coarse lattice spacing ($a=0.112$ fm) we have data for $t/a=4,5,6,7,8,9,10$. We found
that the functions (\ref{eq:tsepextrap}) described the data from the coarse lattice spacing well for all these values of $t/a$.
The smallest statistical uncertainties are obtained for $am_{u,d}^{(\mathrm{val})}=0.005$ and $n_{\rm HYP}=10$, and therefore
we first performed unconstrained fits to this data set, obtaining the following fit results ($l_i$ converted to $\delta_i=e^{l_i}/a$):
\begin{eqnarray}
\nonumber && (g_1)_{\rm eff} = 0.5264   \pm 0.0090,   \:\:A_1 = 0.53 \pm 0.39, \:\:\delta_1 = ( 1.08 \pm 0.38 )\:\:{\rm GeV},\:\:\chi^2/{\rm d.o.f.}=1.07, \\
\nonumber && (g_2)_{\rm eff} = 1.037\nb \pm 0.033,\nb \:\:A_2 = 0.73 \pm 0.35, \:\:\delta_2 = ( 0.75 \pm 0.29 )\:\:{\rm GeV},\:\:\chi^2/{\rm d.o.f.}=0.31, \\
          && (g_3)_{\rm eff} = 0.827\nb \pm 0.032,\nb \:\:A_3 = 0.98 \pm 0.25, \:\:\delta_3 = ( 0.66 \pm 0.16 )\:\:{\rm GeV},\:\:\chi^2/{\rm d.o.f.}=0.41. 
\label{eq:coarsenhyp10tsepextrap}
\end{eqnarray}
The corresponding fits and the data points are shown in the left-hand side of Fig.~\ref{fig:tsepExtrapolationHyp10}. Notice that the fit to $R_1$ gives
an energy gap consistent with recent lattice QCD results for the $2S$ radial excitation energy of about 0.75 GeV \cite{Burch:2008qx, Michael:2010iv}.
This is expected, because our interpolating fields (\ref{eq:interpolatingfields}) for the heavy-light mesons have $S$-wave-type smearing and
therefore do not couple to the lower-lying $1P$ state. The fit result (and the flatness of the plateau as a function of $t'$) indicates that the data for $R_1$
are dominated by
\begin{equation}
 R_1(t) \approx (g_1)_{\rm eff} + \left|\frac{ Z_{P(2S)} }{ Z_{P(1S)} }\right|^2 (A^{(PP^*)}_{2S,2S}-A^{(PP^*)}_{1S,1S})\:e^{-\delta_P\, t},
\end{equation}
with a negligible off-diagonal matrix element $A^{(PP^*)}_{2S,1S}\approx 0$.

At the fine lattice spacing, we only have data for $t/a=6,9,12$ (for $am_{u,d}^{(\mathrm{val})}=0.002,\:0.004$) or $t/a=13$ (for $am_{u,d}^{(\mathrm{val})}=0.006$).
Because the energy gap in physical units is not expected to have an $a$-dependence that is larger than the statistical uncertainties in (\ref{eq:coarsenhyp10tsepextrap}),
it is possible to use the fit results for $l_i$ from the coarse lattice spacing to constrain the parameters $l_i$ at the fine lattice spacing (at similar pion masses).
As a first step, we performed fits to the data from the fine lattice spacing with $am_{u,d}^{(\mathrm{val})}=0.004$ and $n_{\rm HYP}=10$, where only
the parameters $l_i$ were constrained using an augmented $\chi^2$ with a Gaussian prior for $l_i$,
\begin{equation}
 \chi^2 \rightarrow \chi^2 + \frac{ (l_i - \tilde{l}_i )^2 }{ \tilde{\sigma}^2_{l_i} }.
\end{equation}
Here, $\tilde{l}_i$ and $\tilde{\sigma}_{l_i}$ are the central values and uncertainties of the (scaled) energy gap parameters
from the fit to the coarse $am_{u,d}^{(\mathrm{val})}=0.005$, $n_{\rm HYP}=10$ data, Eq.~(\ref{eq:coarsenhyp10tsepextrap}).
The fits to the data from the fine lattice spacing with $am_{u,d}^{(\mathrm{val})}=0.004$ and $n_{\rm HYP}=10$ then gave
\begin{eqnarray}
\nonumber && (g_1)_{\rm eff} = 0.510 \pm 0.011 ,\:\:A_1 = 2.7 \pm 3.6,\:\:\delta_1 = ( 1.08 \pm 0.38 )\:\:{\rm GeV}, \\
\nonumber && (g_2)_{\rm eff} = 0.998 \pm 0.041 ,\:\:A_2 = 1.8 \pm 1.7,\:\:\delta_2 = ( 0.75 \pm 0.29 )\:\:{\rm GeV}, \\
          && (g_3)_{\rm eff} = 0.805 \pm 0.034 ,\:\:A_3 = 2.0 \pm 1.0,\:\:\delta_3 = ( 0.66 \pm 0.16 )\:\:{\rm GeV}.
\label{eq:finehyp10tsepextrap}
\end{eqnarray}
At the fine lattice spacing, it was necessary to remove the data points with the shortest separation $t/a=6$ to obtain acceptable single-exponential fits.
Therefore, the resulting gap parameter matches exactly the prior, and $\chi^2/{\rm d.o.f.}$ is undefined. The central values of the
overlap parameters $A_i$ in (\ref{eq:finehyp10tsepextrap})
are larger than in (\ref{eq:coarsenhyp10tsepextrap}), indicating a stronger overlap of the interpolating fields with excited states at the fine lattice
spacing. Different overlap factors were expected here, because the smearing width of the light-quark fields in physical units was different
(we used the same smearing width in lattice units for both lattice spacings). The fit curves are shown in the right-hand side of Fig.~\ref{fig:tsepExtrapolationHyp10}.

\begin{figure*}[ht!]
 \includegraphics[width=0.48\linewidth]{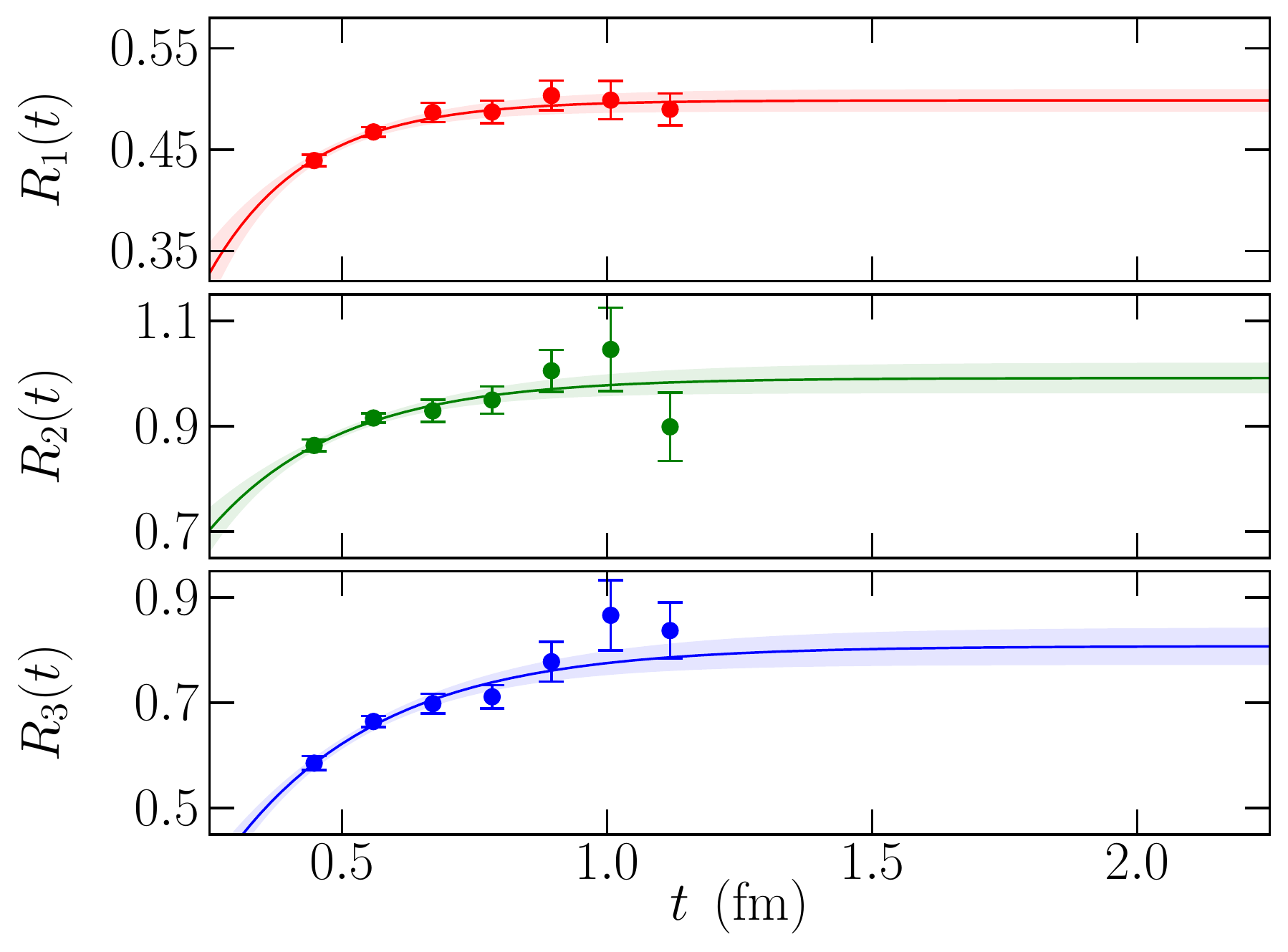}  \hfill \includegraphics[width=0.48\linewidth]{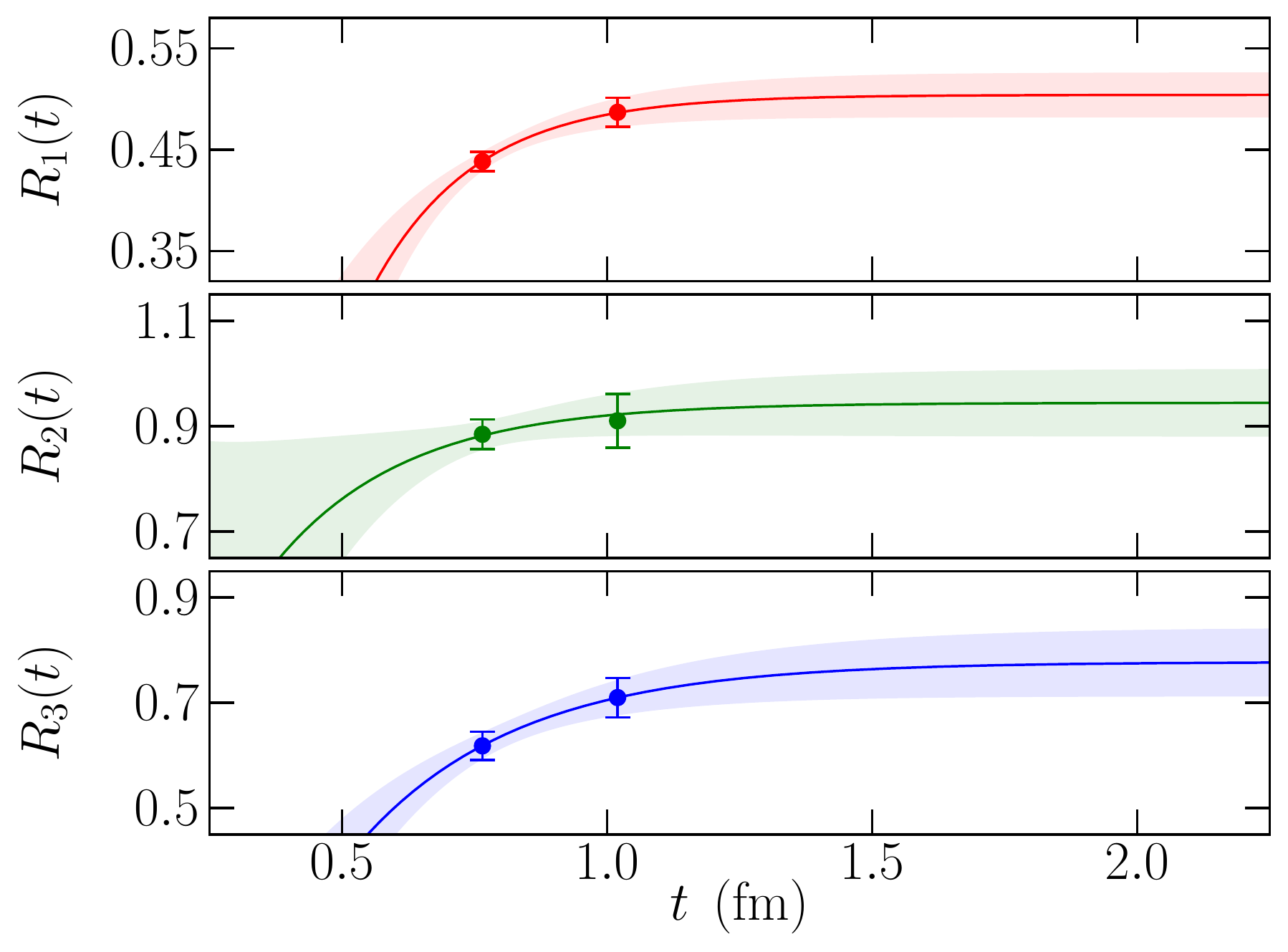}
\caption{\label{fig:tsepExtrapolationHyp3}  Fits of the $t$-dependence of $R_1(t)$, $R_2(t)$, $R_3(t)$, for $n_{\rm HYP}=3$, using the functions defined in Eq.~(\ref{eq:tsepextrap}).
The left-hand side shows the $a=0.112$ fm, $am_{u,d}^{(\mathrm{val})}=0.002$  data, and the right-hand side shows the $a=0.085$ fm, $am_{u,d}^{(\mathrm{val})}=0.002$ data.
In both cases, the overlap parameters $A_i$ and energy gaps $\delta_i$ were constrained using information from (\ref{eq:coarsenhyp10tsepextrap}) and (\ref{eq:finehyp10tsepextrap}).}
\end{figure*}

\begin{figure}[ht!]
 \includegraphics[width=0.5\linewidth]{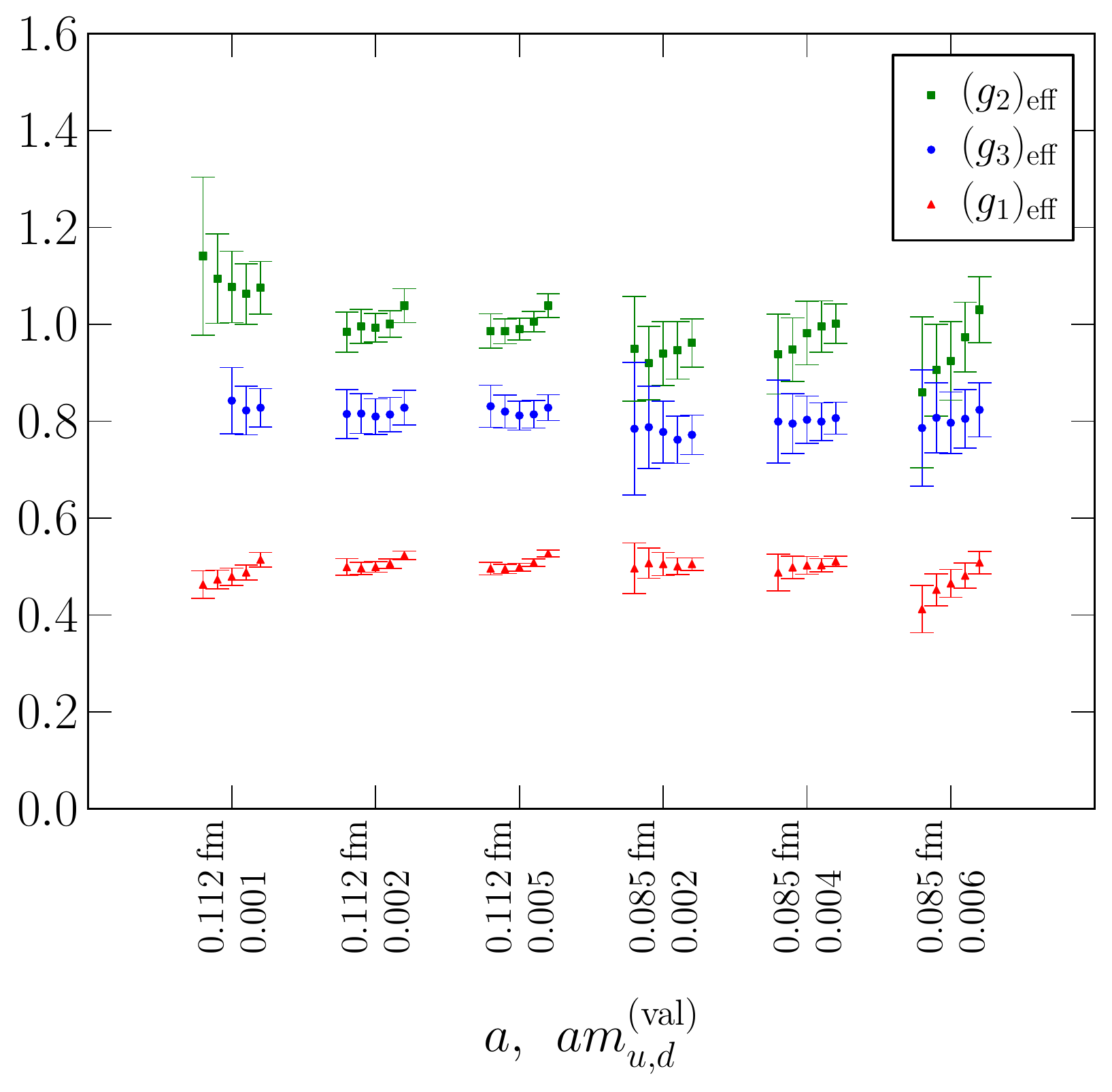}
\caption{\label{fig:tinfdata}Plot of the results for $(g_i)_{\rm eff}$ from Tables \ref{tab:L24tinfdata} and \ref{tab:L32tinfdata}.
For each combination of $a$ and $am_{u,d}^{(\mathrm{val})}$, results from up to five different values of $n_{\rm HYP}$ are shown
(from left to right: $n_{\rm HYP}=1,2,3,5,10$; points offset horizontally for legibility).}
\end{figure}

We then performed new fits to the data for all values of the quark masses and $n_{\rm HYP}$. For those fits, the parameters $(g_i)_{\rm eff}(a,m,n_{\rm HYP})$ were left
unconstrained, but Gaussian priors were used for both $l_i$ and $A_i$, with central values and widths as taken from the initial fits (\ref{eq:coarsenhyp10tsepextrap})
for the coarse lattice spacing and (\ref{eq:finehyp10tsepextrap}) for the fine lattice spacing. Examples of these fits are shown in Fig.~\ref{fig:tsepExtrapolationHyp3}.
The only assumption made by using the priors is that the dependence of $l_i$ and $A_i$ on $n_{\rm HYP}$ and on the quark masses is smaller than the width of the priors as determined by the statistical uncertainties in (\ref{eq:coarsenhyp10tsepextrap})
and (\ref{eq:finehyp10tsepextrap}). Given that these widths were 25\% or larger in all cases, this appears to be a reasonable assumption.
As a test, we also performed unconstrained fits where possible, which gave consistent results but were less stable.
The results for $(g_i)_{\rm eff}(a,m,n_{\rm HYP})$ are given in Tables \ref{tab:L24tinfdata} and \ref{tab:L32tinfdata}, and plotted in Fig.~\ref{fig:tinfdata}.
The central values and uncertainties shown there are bootstrap averages and 68\% widths, respectively, from repeated correlated fits of the $t$-dependence for a bootstrap ensemble of data.

\begin{table}[ht!]
\begin{tabular}{cccccccccccc}
\hline\hline
$am_{u,d}^{(\mathrm{sea})}$ & &  $am_{u,d}^{(\mathrm{val})}$  & &  $n_{\rm HYP}$  & &  $(g_1)_{\rm eff}$ & & $(g_2)_{\rm eff}$ & & $(g_3)_{\rm eff}$   \\
\hline
0.005  && 0.001  &&  1    && 0.463(28)\nb && 1.14(16)\nb &&  $\hdots$   \\
0.005  && 0.001  &&  2    && 0.473(20)\nb && 1.094(92)   &&  $\hdots$   \\
0.005  && 0.001  &&  3    && 0.479(18)\nb && 1.077(74)   &&  0.843(68)  \\
0.005  && 0.001  &&  5    && 0.488(15)\nb && 1.063(62)   &&  0.822(50)  \\
0.005  && 0.001  &&  10   && 0.514(15)\nb && 1.075(54)   &&  0.828(40)  \\
\\[-2ex]
0.005  && 0.002  &&  1    && 0.499(17)\nb && 0.984(42)   &&  0.815(51)  \\
0.005  && 0.002  &&  2    && 0.496(13)\nb && 0.996(35)   &&  0.816(41)  \\
0.005  && 0.002  &&  3    && 0.499(11)\nb && 0.993(29)   &&  0.810(37)  \\
0.005  && 0.002  &&  5    && 0.5059(96)   && 1.001(27)   &&  0.814(35)  \\
0.005  && 0.002  &&  10   && 0.5230(89)   && 1.039(35)   &&  0.828(36)  \\
\\[-2ex]
0.005  && 0.005  &&  1    && 0.496(13)\nb && 0.986(36)   &&  0.831(43)  \\
0.005  && 0.005  &&  2    && 0.4950(94)   && 0.986(25)   &&  0.820(34)  \\
0.005  && 0.005  &&  3    && 0.4987(80)   && 0.990(23)   &&  0.812(30)  \\
0.005  && 0.005  &&  5    && 0.5080(74)   && 1.006(21)   &&  0.814(28)  \\
0.005  && 0.005  &&  10   && 0.5270(71)   && 1.039(25)   &&  0.828(27)  \\
\hline\hline
\end{tabular}
\caption{\label{tab:L24tinfdata} Effective axial couplings $(g_i)_{\rm eff}$ at $a=0.112$ fm, obtained by extrapolating $R_i(t)$ to $t=\infty$.
At $am_{u,d}^{(\mathrm{val})}=0.001$ we do not have results for $n_{\rm HYP}=1,2$, because the statistical fluctuations were too large to calculate
the square root of the double ratio, Eq.~(\ref{eq:R3}).}
\end{table}

\begin{table}[ht!]
\begin{tabular}{cccccccccccc}
\hline\hline
$am_{u,d}^{(\mathrm{sea})}$ & &  $am_{u,d}^{(\mathrm{val})}$  & &  $n_{\rm HYP}$  & &  $(g_1)_{\rm eff}$ & & $(g_2)_{\rm eff}$ & & $(g_3)_{\rm eff}$   \\
\hline
0.004  && 0.002  &&  1    && 0.496(52) && 0.95(11)\nb &&  0.78(14)\nb \\
0.004  && 0.002  &&  2    && 0.507(31) && 0.920(75)   &&  0.788(85)   \\
0.004  && 0.002  &&  3    && 0.505(24) && 0.940(66)   &&  0.778(64)   \\
0.004  && 0.002  &&  5    && 0.501(17) && 0.946(59)   &&  0.762(49)   \\
0.004  && 0.002  &&  10   && 0.505(13) && 0.962(50)   &&  0.772(41)   \\
\\[-2ex]
0.004  && 0.004  &&  1    && 0.488(38) && 0.939(82)   &&  0.799(85)   \\
0.004  && 0.004  &&  2    && 0.498(23) && 0.948(65)   &&  0.795(62)   \\
0.004  && 0.004  &&  3    && 0.502(18) && 0.982(66)   &&  0.803(49)   \\
0.004  && 0.004  &&  5    && 0.503(14) && 0.995(53)   &&  0.799(39)   \\
0.004  && 0.004  &&  10   && 0.511(10) && 1.001(41)   &&  0.807(33)   \\
\\[-2ex]
0.006  && 0.006  &&  1    && 0.412(49) && 0.86(16)\nb &&  0.79(12)\nb \\
0.006  && 0.006  &&  2    && 0.452(33) && 0.905(95)   &&  0.807(73)   \\
0.006  && 0.006  &&  3    && 0.465(29) && 0.925(81)   &&  0.797(64)   \\
0.006  && 0.006  &&  5    && 0.481(26) && 0.974(72)   &&  0.805(60)   \\
0.006  && 0.006  &&  10   && 0.508(23) && 1.030(68)   &&  0.824(56)   \\
\hline\hline
\end{tabular}
\caption{\label{tab:L32tinfdata} Effective axial couplings $(g_i)_{\rm eff}$ at $a=0.085$ fm, obtained by extrapolating $R_i(t)$ to $t=\infty$.}
\end{table}

To estimate the systematic uncertainties caused by higher excited states, we calculated
the shifts in $(g_i)_{\rm eff}$ for a representative data set at the coarse lattice spacing
($am_{u,d}^{(\mathrm{val})}=0.002$, $n_{\rm HYP}=3$) when removing one or two data points with the smallest
$t/a$ ($=4,5$) from the fits, or adding a second exponential to the fit function,
\begin{equation}
 R_i(t) = (g_i)_{\rm eff} - A_i \:e^{-\delta_i\:t} - B_i\: A_i \:e^{-(\delta_i+\delta^{(2)}_i)\:t}. \label{eq:secondexp}
\end{equation}
Because the available data were not sufficient to determine the new parameters $B_i$ and $\delta^{(2)}_i$, we used Gaussian
priors to constrain these parameters to physically reasonable values.
The parameters $B_i$ describe the amplitudes of the second-excited-state
contribution, relative to the first-excited-state contribution, and we set
$\tilde{B}_i=0$, $\tilde{\sigma}_{B_i}=2$. For the energy gaps $a\:\delta^{(2)}_i=e^{l_i^{(2)}}$ we used priors
with central values equal to $2/3$ times the fit results for $\delta_i$ in (\ref{eq:coarsenhyp10tsepextrap}),
and widths of 100\%. The fitted parameters $(g_i)_{\rm eff}$ and the corresponding shifts $\delta(g_i)_{\rm eff}$
for the different cases are shown in Table \ref{tab:excitedstatessyst}.
Since the shifts $\delta(g_i)_{\rm eff}$ themselves have
statistical uncertainties $\sigma\delta(g_i)_{\rm eff}$, we choose to quote the maximum
value of
\begin{equation}
\sqrt{ [\delta(g_i)_{\rm eff}]^2 +  [\sigma\delta(g_i)_{\rm eff}]^2} \label{eq:excitedstatesystematicerror}
\end{equation}
from the three different methods (removing $t/a=4$, removing $t/a=4,5$, or adding a second exponential)
as our estimate of the systematic uncertainties in $(g_i)_{\rm eff}$ caused by higher excited states.
The final estimates are 1.7\%, 2.8\%, and 4.9\% for $g_1$, $g_2$, and $g_3$, respectively.

Because the shifts $\delta(g_i)_{\rm eff}$ in Table \ref{tab:excitedstatessyst}
are consistent with zero in most cases, an alternative way of estimating the systematics is to consider only the increase (calculated with quadrature)
in the \emph{uncertainties} of the fitted parameters when the fits are modified by removing data points or including higher-order terms.
We will use that method for the chiral fits at the end of Sec.~\ref{sec:chiralfits}, see Eq.~(\ref{eq:sigmaincrease}).
The two different methods (\ref{eq:excitedstatesystematicerror}) and (\ref{eq:sigmaincrease}) for calculating the size of the systematic uncertainties give consistent values.

\begin{table}[ht!]
 \begin{tabular}{lcccccccccccc}
\hline\hline
Fit                            & \hspace{2ex} & $(g_1)_{\rm eff}$  & \hspace{2ex} & $\delta(g_1)_{\rm eff}$ & \hspace{2ex} & $(g_2)_{\rm eff}$ & \hspace{2ex} & $\delta(g_2)_{\rm eff}$ & \hspace{2ex} & $(g_3)_{\rm eff}$ & \hspace{2ex} & $\delta(g_3)_{\rm eff}$ \\
\hline
Original                       &&  0.499(11) &&    0         &&  0.993(29) &&     0      &&  0.810(36) &&     0      \\
$t/a=4$ removed                &&  0.496(13) &&  0.0030(76)  &&  0.975(35) &&  0.016(19) &&  0.783(43) &&  0.026(15) \\
$t/a=4,5$ removed              &&  0.494(12) &&  0.0041(76)  &&  0.984(41) &&  0.009(26) &&  0.807(54) &&  0.003(30) \\
Second exponential added       &&  0.498(11) &&  0.0009(77)  &&  0.988(30) &&  0.005(21) &&  0.796(40) &&  0.014(36) \\
\hline\hline 
\end{tabular}
\caption{\label{tab:excitedstatessyst} Fits used to estimate systematic uncertainties from higher excited states. Data for $a=0.112$ fm, $am_{u,d}^{(\mathrm{val})}=0.002$, $n_{\rm HYP}=3$.
Shown are the fit results for $(g_i)_{\rm eff}$, as well as the differences $\delta(g_i)_{\rm eff}$ to the original fit result, calculated using bootstrap.}
\end{table}

\clearpage

\subsection{Extraction of the axial couplings $g_1$, $g_2$, and $g_3$ using HH$\chi$PT fits of the data}

\label{sec:chiralfits}

In the previous section, we obtained results for the effective axial couplings $(g_i)_{\rm eff}(a,m,n_{\rm HYP})$ at two different lattice
spacings $a$, multiple values for the quark masses $am_{u,d}^{(\mathrm{sea})}$ and $am_{u,d}^{(\mathrm{val})}$, and multiple values for the
heavy-quark gauge-link smearing parameter $n_{\rm HYP}$ (corresponding to multiple heavy-quark lattice discretizations). All data are for a finite spatial
volume of about $(2.7\:\:{\rm fm})^3$.
In the following, we discuss how we extracted the axial couplings $g_1$, $g_2$, $g_3$, which are the parameters of the continuum
heavy-hadron chiral perturbation theory Lagrangian (\ref{eq:HHchPTL}), from the data for $(g_i)_{\rm eff}(a,m,n_{\rm HYP})$.

To fit the quark-mass and volume dependence of $(g_i)_{\rm eff}$ we use the next-to-leading-order predictions from HH$\chi$PT \cite{Detmold:2011rb}, which
were already shown in Eq.~(\ref{eq:PQchiPTformula}). Here, we extend these formulas to include the leading effects of the nonzero lattice spacing $a$. These
leading effects are quadratic in $a$, with coefficients $d_{i,\:n_{\rm HYP}}$ that depend on $n_{\rm HYP}$. We do not expect $O(a)$ errors because of the chiral symmetry of
the domain-wall action used for the light quarks (neglecting the small effects caused by the residual chiral symmetry breaking at finite $L_s$ \cite{Aoki:2010dy}) and
the automatic $O(a)$ improvement of the static heavy-quark action. Higher-order effects in the $a$- and $m_\pi$-dependence
of $(g_i)_{\rm eff}$ are discussed at the end of this section.

For $(g_1)_{\rm eff}$, the fit function is given by
\begin{eqnarray}
\nonumber (g_1)_{\rm eff}(a, m, n_{\rm HYP}) &=& g_1\Bigg[1 -\frac{2}{f^2}\:\mathcal{I}(m_\pi^{(\mathrm{vs})}) + \frac{g_1^2}{f^2}\Big\{ 4 \: \mathcal{H}(m_\pi^{(\mathrm{vs})},\:0)-4 \: \delta^2_{VS} \mathcal{H}_{\eta'}(m_\pi^{(\mathrm{vv})},\:0) \Big\} \\
                 && \phantom{g_1\Bigg[1} + c_1^{(\mathrm{vv})}\: [m_\pi^{(\mathrm{vv})}]^2 + c_1^{(\mathrm{vs})}\: [m_\pi^{(\mathrm{vs})}]^2 + d_{1,\:n_{\rm HYP}}\:a^2 \Bigg].
\label{eq:g1chiralcontinuumfit}
\end{eqnarray}
Similarly, for $(g_2)_{\rm eff}$ and $(g_3)_{\rm eff}$, we use
\begin{eqnarray}
\nonumber (g_2)_{\rm eff}(a, m, n_{\rm HYP}) &=& g_2\Bigg[1-\frac{2}{f^2}\:\mathcal{I}(m_\pi^{(\mathrm{vs})}) + \frac{g_2^2}{f^2}\Big\{ \frac32 \: \mathcal{H}(m_\pi^{(\mathrm{vs})},\:0) - \delta^2_{VS} \mathcal{H}_{\eta'}(m_\pi^{(\mathrm{vv})},\:0)  \Big\} \\
\nonumber        && \phantom{g_2\Bigg[1} + \frac{g_3^2}{f^2} \Big\{ 2\: \mathcal{H}(m_\pi^{(\mathrm{vs})},\:-\Delta) - \mathcal{H}(m_\pi^{(\mathrm{vv})},\:-\Delta) - 2\: \mathcal{K}(m_\pi^{(\mathrm{vs})},\:-\Delta,\:0)   \Big\}  \\
                 && \phantom{g_2\Bigg[1} + c_2^{(\mathrm{vv})}\: [m_\pi^{(\mathrm{vv})}]^2 + c_2^{(\mathrm{vs})}\: [m_\pi^{(\mathrm{vs})}]^2 + d_{2,\:n_{\rm HYP}}\:a^2 \Bigg], \label{eq:g2chiralcontinuumfit}
\end{eqnarray}
\begin{eqnarray}
\nonumber (g_3)_{\rm eff}(a, m, n_{\rm HYP}) &=& g_3\Bigg[1-\frac{2}{f^2}\:\mathcal{I}(m_\pi^{(\mathrm{vs})}) + \frac{g_3^2}{f^2} \Big\{\mathcal{H}(m_\pi^{(\mathrm{vs})},\:-\Delta) -\frac12 \mathcal{H}(m_\pi^{(\mathrm{vv})},\:-\Delta) \\
\nonumber        && \phantom{g_3\Bigg[1} + \frac{3}{2} \mathcal{H}(m_\pi^{(\mathrm{vv})},\:\Delta) + 3\: \mathcal{H}(m_\pi^{(\mathrm{vs})},\:\Delta)  - \mathcal{K}(m_\pi^{(\mathrm{vs})},\:\Delta,\:0)           \Big\} \\
\nonumber        && \phantom{g_3\Bigg[1} + \frac{g_2^2}{f^2} \Big\{ -\mathcal{H}(m_\pi^{(\mathrm{vs})},\:\Delta)-\mathcal{H}(m_\pi^{(\mathrm{vv})},\:\Delta) + \mathcal{H}(m_\pi^{(\mathrm{vs})},\:0) - \delta^2_{VS} \mathcal{H}_{\eta'}(m_\pi^{(\mathrm{vv})},\:0) \Big\} \\
                 && \phantom{g_3\Bigg[1} + c_3^{(\mathrm{vv})}\: [m_\pi^{(\mathrm{vv})}]^2 + c_3^{(\mathrm{vs})}\: [m_\pi^{(\mathrm{vs})}]^2 + d_{3,\:n_{\rm HYP}}\:a^2 \Bigg].
\label{eq:g3chiralcontinuumfit}
\end{eqnarray}
The functions $\mathcal{I}$, $\mathcal{H}$, $\mathcal{H}_{\eta'}$ and $\mathcal{K}$ are the chiral loop corrections \cite{Detmold:2011rb}. They include finite-volume effects  and therefore
they also depend on the lattice size.
Furthermore, these functions depend on the renormalization scale $\mu$, but this scale-dependence is absorbed by the fit parameters $c_i^{(\mathrm{vv})}$
and $c_i^{(\mathrm{vs})}$, as we checked explicitly by varying $\mu$ in the fits. We set the pion decay constant to $f=132$ MeV and the $S-T$ mass splitting
in Eqs.~(\ref{eq:g2chiralcontinuumfit}) and (\ref{eq:g3chiralcontinuumfit}) to $\Delta=200$ MeV. This value of $\Delta$ is consistent with experiments
\cite{Aaltonen:2007rw, CDF:2011ac} and with our lattice data (we also checked that varying $\Delta$ within a few percent does not significantly
affect the results for the axial couplings).
We calculated the covariances of all correlated data points in Tables \ref{tab:L24tinfdata} and \ref{tab:L32tinfdata} using bootstrap, and performed
fully correlated fits using the inverse of the covariance matrix in the definition of $\chi^2$. This method propagates the uncertainties and correlations
of $(g_i)_{\rm eff}$, as obtained from the fits to the ratios $R_i$, into the extracted parameters $g_i$ of the HH$\chi$PT Lagrangian.

\begin{figure}[ht!]
 \includegraphics[width=0.5\linewidth]{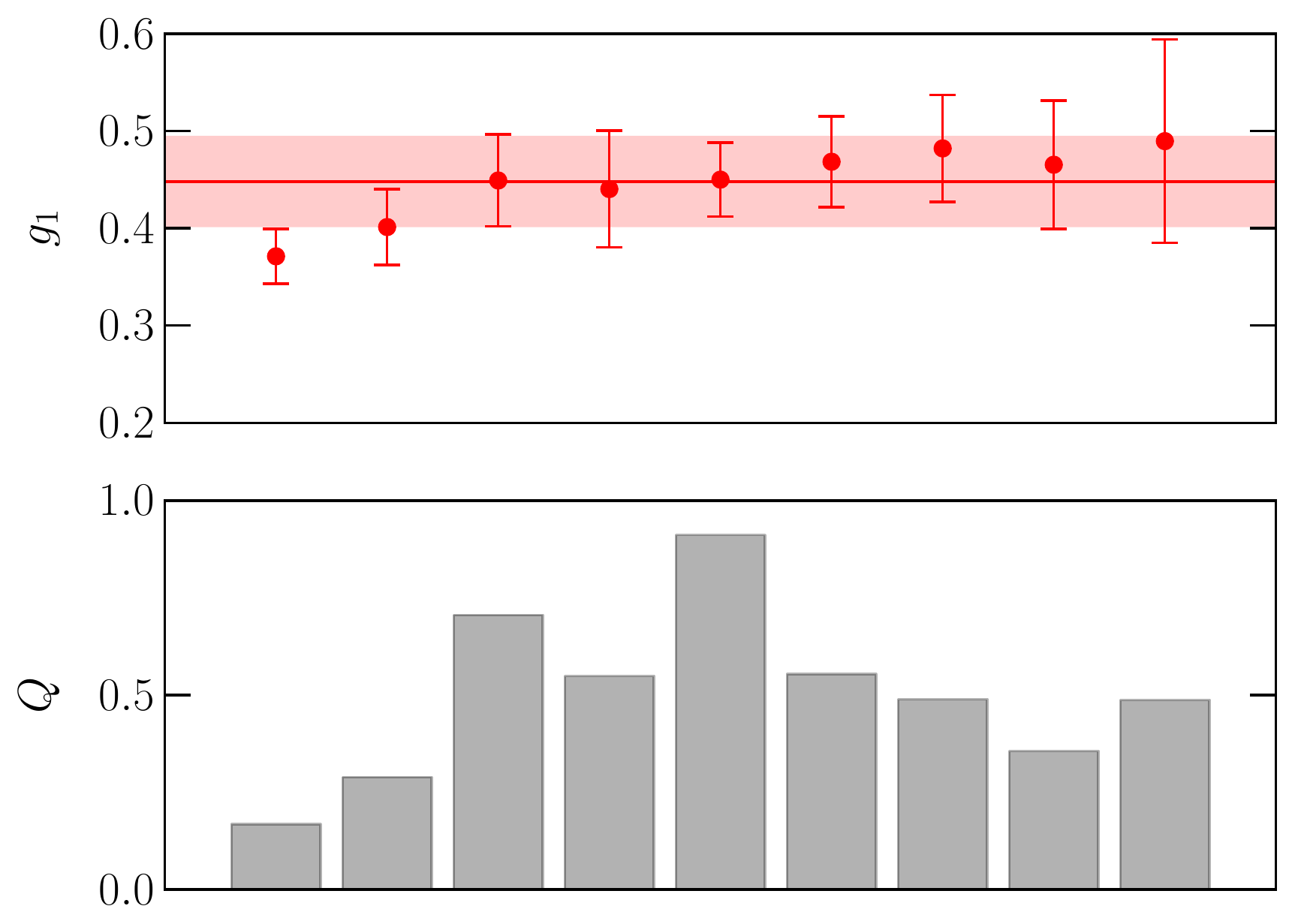}
\caption{\label{fig:g1fitresults}Graphical representation of the fit results for $g_1$ from Table \ref{tab:g1fitresults}. The horizontal axis corresponds to the different fits, ordered
(from left to right) in the same way as the rows in the table (from top to bottom). The line and shaded region in the upper plot indicate the selected result and its uncertainty, which is taken from the third fit
(the fit that includes data with $n_{\rm HYP}=1, 2, 3$).}
\end{figure}

Results from fits of the $(g_1)_{\rm eff}$ data
using the function (\ref{eq:g1chiralcontinuumfit}) are given in Table \ref{tab:g1fitresults} and Fig.~\ref{fig:g1fitresults}.
The fit parameters are $g_1$, $c_1^{(\mathrm{vv})}$, $c_1^{(\mathrm{vs})}$, and $\{ d_{1,\:n_{\rm HYP}} \}$
(the latter for all values of $n_{\rm HYP}$ that were included in the fit). We performed fits that included data with multiple values of $n_{\rm HYP}$,
as well as individual fits including only data with one value of $n_{\rm HYP}$. The fits that included multiple values of $n_{\rm HYP}$
enforced a common continuum limit of the data with different $n_{\rm HYP}$, but with separate $a^2$-coefficients $d_{1,\:n_{\rm HYP}}$
for each $n_{\rm HYP}$. While we know that the actual continuum limit for all values of $n_{\rm HYP}$ has to be the same (if we
took $a$ to zero in the numerical calculations), we only have data for two different values of $a$, and
one may question whether the approach of the continuum limit is described by a simple $a^2$-dependence as assumed in Eq.~(\ref{eq:g1chiralcontinuumfit}).
In particular, one may be worried that large values of $n_{\rm HYP}$, which correspond to more spatially extended heavy-quark actions,
could lead to non-negligible contributions from higher powers of $a$ \cite{DellaMorte:2003mn}. To investigate this, we started from a fit that included all values of $n_{\rm HYP}$ (1, 2, 3, 5, 10),
and then successively removed the data with the largest values of $n_{\rm HYP}$. As can be seen in Table \ref{tab:g1fitresults} and Fig.~\ref{fig:g1fitresults},
the fit including the data from all values of $n_{\rm HYP}$ had a poor quality, $Q=0.17$, and gave a somewhat low value for $g_1$. After excluding $n_{\rm HYP}=10$ and $n_{\rm HYP}=5$,
the fits had a good quality and the results for $g_1$ were stable under further exclusions of the largest $n_{\rm HYP}$-values.
The fit including $n_{\rm HYP}=1, 2, 3$, which has $Q=0.70$, gave the result
\begin{equation}
 g_1 = 0.449\pm0.047_{\:\rm stat}. \label{eq:g1finalresult}
\end{equation}
Estimates of the systematic uncertainties in (\ref{eq:g1finalresult}) will be given at the end of this section. The results from the fits including only one value of $n_{\rm HYP}$ were all consistent with (\ref{eq:g1finalresult}), even for $n_{\rm HYP}=10$.
This suggests that higher powers of $a^2$ are actually negligible for the values of the lattice spacings considered here ($a=0.085$ fm and $a=0.112$ fm).
The deviating value of $g_1$ as well as the poor $Q$ for the fit that included all values of $n_{\rm HYP}$ simultaneously are likely caused
by technical issues with the covariance-fitting of highly correlated data, associated with small eigenvalues of the data correlation matrix
\cite{Yoon:2011fp}. We will return to the discussion of higher-order discretization effects at the end of this section.

\begin{table}[ht!]
\begin{tabular}{lcccccccc}
\hline\hline
$n_{\rm HYP}$   & \hspace{2ex} &  $g_1$  & \hspace{2ex} & ${\rm d.o.f.}$  & \hspace{2ex} & $\chi^2/{\rm d.o.f.}$ & \hspace{2ex} & $Q$  \\
\hline
1, 2, 3, 5, 10     & &  0.371(28)    & &  $30-8$             & & 1.3\nb                & & 0.17     \\
1, 2, 3, 5         & &  0.401(39)    & &  $24-7$             & & 1.2\nb                & & 0.29     \\ 
1, 2, 3            & &  0.449(47)    & &  $18-6$             & & 0.75                  & & 0.70     \\
1, 2               & &  0.440(60)    & &  $12-5$             & & 0.85                  & & 0.54     \\
10                 & &  0.450(38)    & &  $6-4$              & & 0.09                  & & 0.91     \\
5                  & &  0.468(47)    & &  $6-4$              & & 0.61                  & & 0.55     \\
3                  & &  0.482(55)    & &  $6-4$              & & 0.73                  & & 0.49     \\
2                  & &  0.465(66)    & &  $6-4$              & & 1.0\nb                & & 0.36     \\
1                  & &  0.49(10)\nb  & &  $6-4$              & & 0.72                  & & 0.49     \\
\hline\hline
\end{tabular}
\caption{\label{tab:g1fitresults}Results for the mesonic axial coupling $g_1$, obtained by fitting
the data for $(g_1)_{\rm eff}$ using the function (\ref{eq:g1chiralcontinuumfit}). The first four
rows show the results from fits which include data with multiple values of the heavy-quark smearing parameter $n_{\rm HYP}$.
The remaining rows show the results from fits with only one value of $n_{\rm HYP}$. The number of degrees of freedom (${\rm d.o.f.}$) is given
in the form (number of data points)$-$(number of fit parameters). The last column of the table gives the quality of the fit $Q=\Gamma\left({\rm d.o.f.}/2,\:\chi^2/2\right)$.}
\end{table}

\begin{figure}[ht!]
 \includegraphics[width=0.5\linewidth]{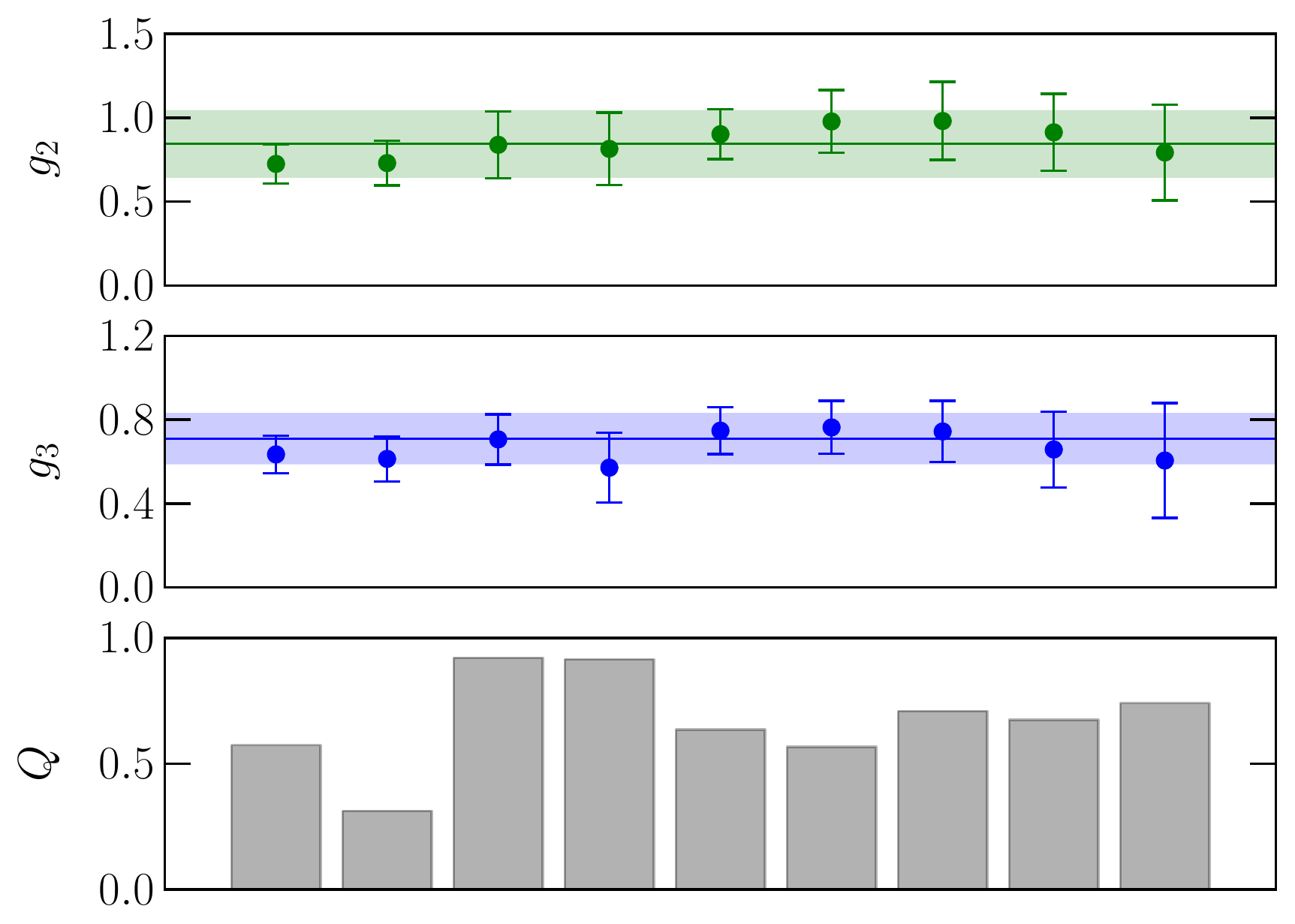}
\caption{\label{fig:g2g3fitresults}Graphical representation of the fit results for $g_2$ and $g_3$ from Table \ref{tab:g2g3fitresults}. The horizontal axis corresponds to the different fits, ordered
(from left to right) in the same way as the rows in the table (from top to bottom). The lines and shaded regions in the upper two plots indicate the selected results and their uncertainties, which are taken from the third fit
(the fit that includes data with $n_{\rm HYP}=1, 2, 3$).}
\end{figure}

\begin{figure}[ht!]
 \includegraphics[width=0.5\linewidth]{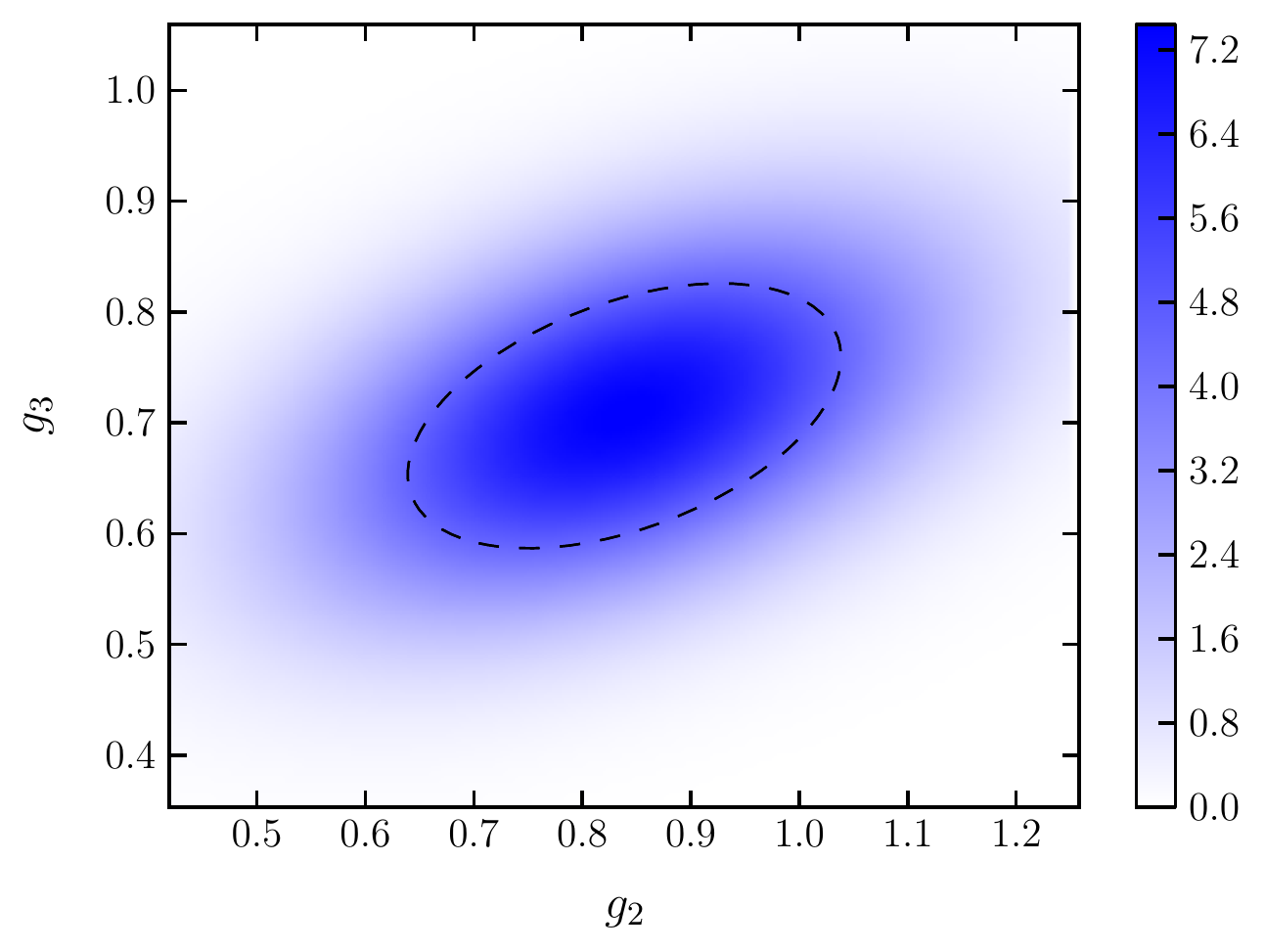}
\caption{\label{fig:g2g3likelihood}Likelihood function for $g_2$ and $g_3$, equal to $L(g_2, g_3)=(2\pi)^{-1} \: \mathrm{det}(\mathrm{Cov})^{-1/2} \exp\left\{-\frac12(g_i-g_i^{(0)})[{\rm Cov}^{-1}]_{ij}(g_j-g_j^{(0)})\right\}$
where $g_i^{(0)}$ are the central values of our fit results (\ref{eq:g2g3finalresult}) and ${\rm Cov}$ is the covariance matrix (\ref{eq:g2g3cov}).
The dashed curve indicates the standard error ellipse.}
\end{figure}

For the baryonic axial couplings, we performed simultaneous, fully correlated fits to the data for $(g_2)_{\rm eff}$ and $(g_3)_{\rm eff}$
using the functions (\ref{eq:g2chiralcontinuumfit}) and (\ref{eq:g3chiralcontinuumfit}), with the fit parameters
$g_2$, $g_3$, $c_2^{(\mathrm{vv})}$, $c_3^{(\mathrm{vv})}$, $c_2^{(\mathrm{vs})}$, $c_3^{(\mathrm{vs})}$ and $\{ d_{2,\:n_{\rm HYP}},\: d_{3,\:n_{\rm HYP}} \}$
(the latter for all values of $n_{\rm HYP}$ that were included in the fit). As already discussed in the fits for $(g_1)_{\rm eff}$,
we performed fits that included data with multiple values of $n_{\rm HYP}$, as well as individual fits including only data with one value of $n_{\rm HYP}$.
The results are shown in Table \ref{tab:g2g3fitresults} and Fig.~\ref{fig:g2g3fitresults}. Again, we select the fit that includes $n_{\rm HYP}=1, 2, 3$,
which gives
\begin{eqnarray}
 \nonumber g_2 &=& 0.84 \pm 0.20_{\:\rm stat}, \\
           g_3 &=& 0.71 \pm 0.12_{\:\rm stat}. \label{eq:g2g3finalresult}
\end{eqnarray}
This fit also had the highest value of the quality of fit, $Q=0.92$. Estimates of the systematic uncertainties in (\ref{eq:g2g3finalresult})
will be given at the end of this section. As can be seen in Fig.~\ref{fig:g2g3fitresults}, the results (\ref{eq:g2g3finalresult}) are in fact consistent with the results
from all other fits within the statistical uncertainties, demonstrating that heavy-quark discretization errors are under good control. The covariance matrix
for $g_2$ and $g_3$ is
\begin{equation}
{\rm Cov} = \left(\begin{array}{cc}  0.040 & 0.011 \\ 0.011 & 0.014 \end{array} \right). \label{eq:g2g3cov}
\end{equation}
The corresponding likelihood function is plotted in Fig.~\ref{fig:g2g3likelihood}.

\begin{table}[ht!]
\begin{tabular}{lcccccccccc}
\hline\hline
$n_{\rm HYP}$  & \hspace{2ex} &  $g_2$  & \hspace{2ex} &  $g_3$  & \hspace{2ex} & ${\rm d.o.f.}$  & \hspace{2ex} & $\chi^2/{\rm d.o.f.}$ & \hspace{2ex} & $Q$  \\
\hline
1, 2, 3, 5, 10     & &  0.72(12)  & &  0.635(90)     & &   $58 - 16$     & &   0.94     & &   0.57      \\
1, 2, 3, 5         & &  0.73(13)  & &  0.61(11)\nb   & &   $46 - 14$     & &   1.1\nb   & &   0.31      \\
1, 2, 3            & &  0.84(20)  & &  0.71(12)\nb   & &   $34 - 12$     & &   0.61     & &   0.92      \\
1, 2               & &  0.81(22)  & &  0.57(17)\nb   & &   $22 - 10$     & &   0.50     & &   0.91      \\
10                 & &  0.90(15)  & &  0.75(11)\nb   & &   $12 - 8$      & &   0.64     & &   0.64      \\
5                  & &  0.98(19)  & &  0.76(13)\nb   & &   $12 - 8$      & &   0.74     & &   0.57      \\
3                  & &  0.98(23)  & &  0.74(15)\nb   & &   $12 - 8$      & &   0.54     & &   0.71      \\
2                  & &  0.91(23)  & &  0.66(18)\nb   & &   $12 - 8$      & &   0.51     & &   0.67      \\
1                  & &  0.79(29)  & &  0.61(27)\nb   & &   $12 - 8$      & &   0.42     & &   0.74      \\
\hline\hline
\end{tabular}
\caption{\label{tab:g2g3fitresults}Results for the baryonic axial couplings $g_2$ and $g_3$, obtained by simultaneously fitting
the data for $(g_2)_{\rm eff}$ and $(g_3)_{\rm eff}$ using the functions (\ref{eq:g2chiralcontinuumfit}) and (\ref{eq:g3chiralcontinuumfit}). The first four
rows show the results from fits which include data with multiple values of the heavy-quark smearing parameter $n_{\rm HYP}$.
The remaining rows show the results from fits with only one value of $n_{\rm HYP}$. The number of degrees of freedom (${\rm d.o.f.}$) is given
in the form (number of data points)$-$(number of fit parameters). The last column of the table gives the quality of the fit $Q=\Gamma\left({\rm d.o.f.}/2,\:\chi^2/2\right)$.}
\end{table}

As another check, we performed fits of $(g_i)_{\rm eff}$ where we excluded all the partially quenched data
(i.e keeping only the unitary data with $m_\pi^{(\mathrm{vv})}=m_\pi^{(\mathrm{vs})}$).
In that case, only one analytic counterterm is needed for each coupling, and we removed the terms $c_i^{(\mathrm{vs})}\: [m_\pi^{(\mathrm{vs})}]^2$
from Eqs.~(\ref{eq:g1chiralcontinuumfit}), (\ref{eq:g2chiralcontinuumfit}), and (\ref{eq:g3chiralcontinuumfit}). These fits, again using $n_{\rm HYP}=1, 2, 3$,
then gave $g_1=0.467\pm0.056$, $g_2=0.92\pm0.22$, and $g_3=0.72\pm0.14$, in full agreement with (\ref{eq:g1finalresult}) and (\ref{eq:g2g3finalresult})
and with slightly larger uncertainties.

\begin{figure*}[ht!]
 \hspace{-3ex} \includegraphics[width=0.57\linewidth]{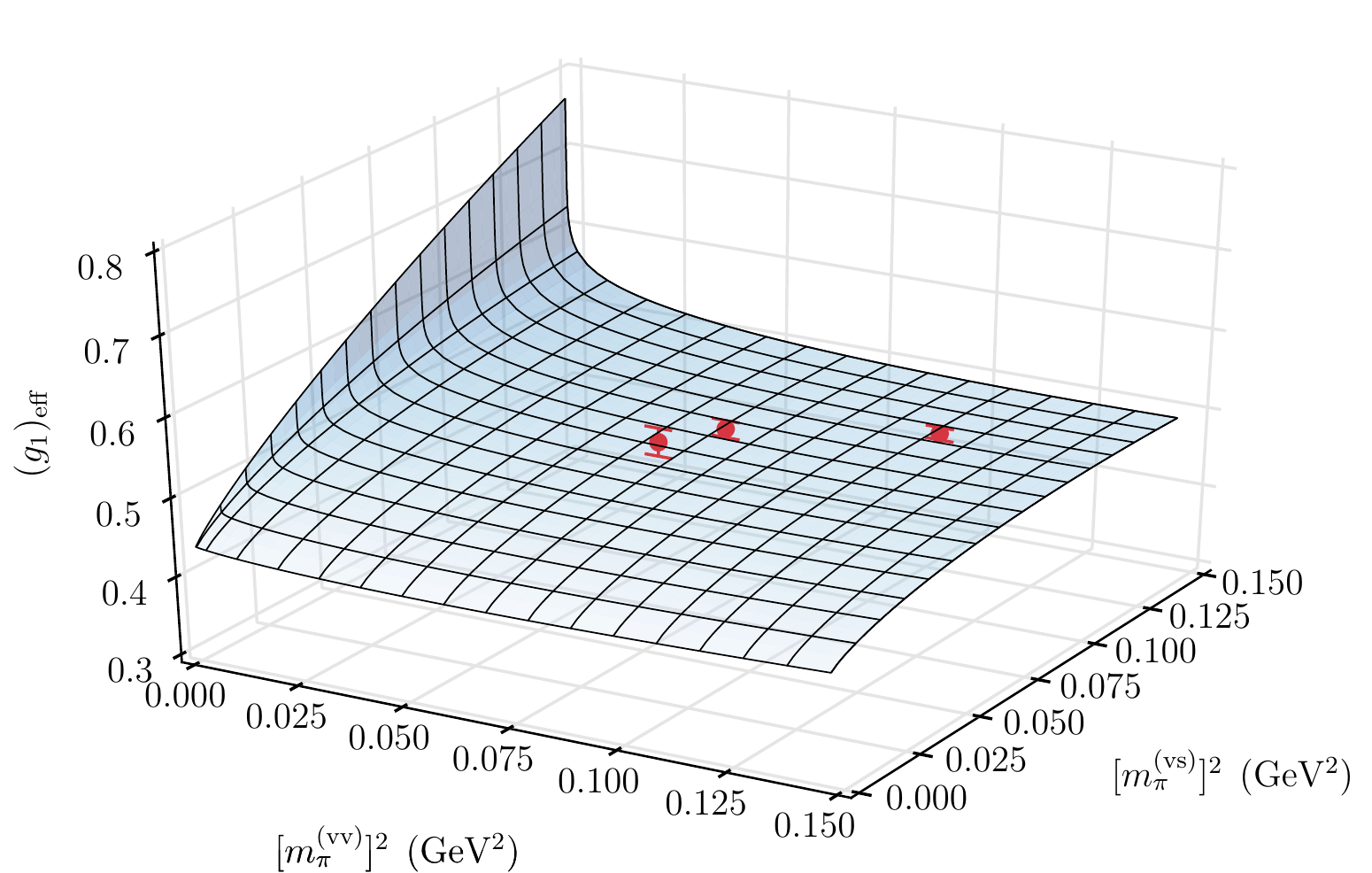}  \hfill \includegraphics[width=0.42\linewidth]{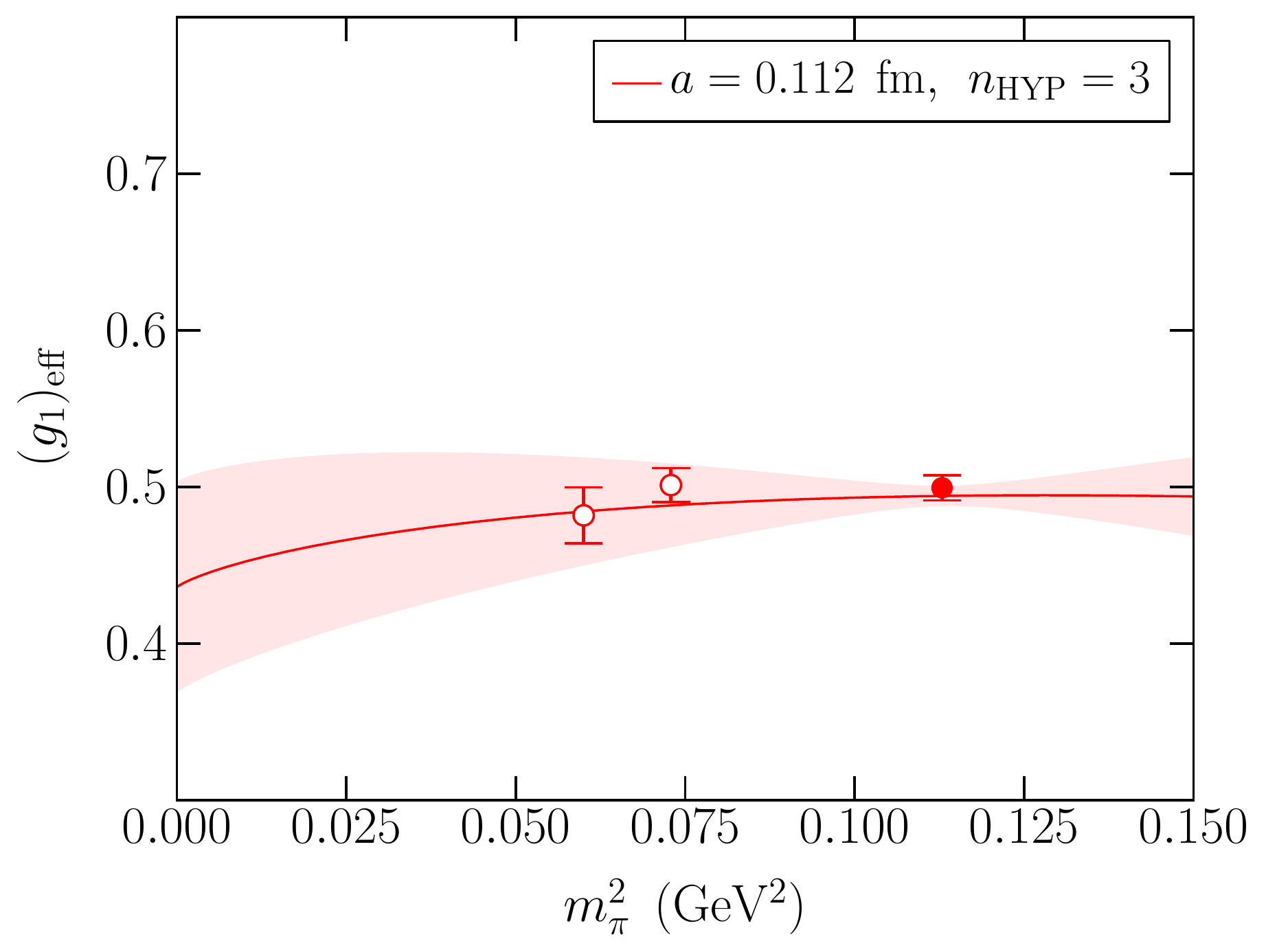}
 \hspace{-3ex} \includegraphics[width=0.57\linewidth]{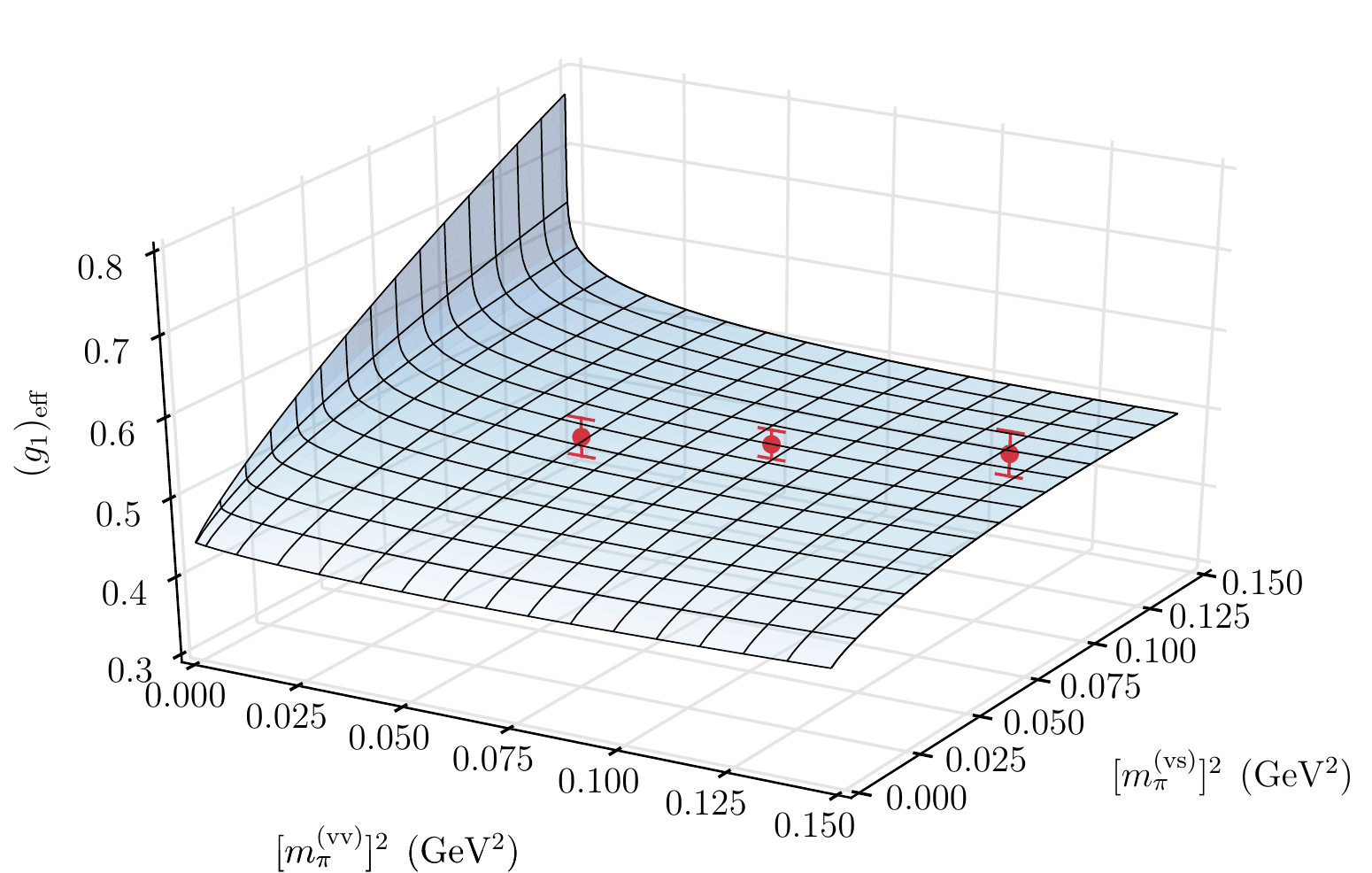}  \hfill \includegraphics[width=0.42\linewidth]{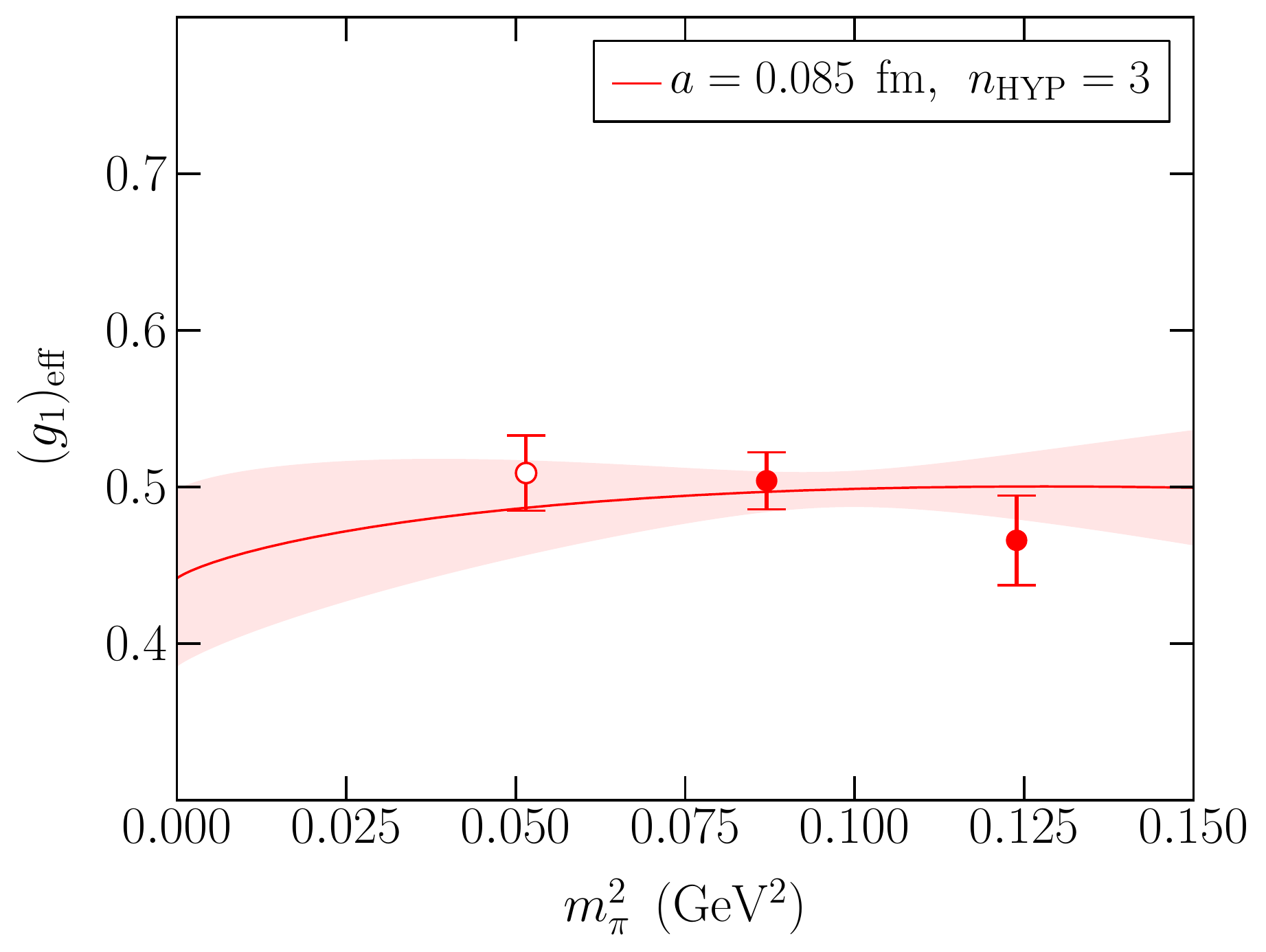}
 \hspace{-3ex} \includegraphics[width=0.57\linewidth]{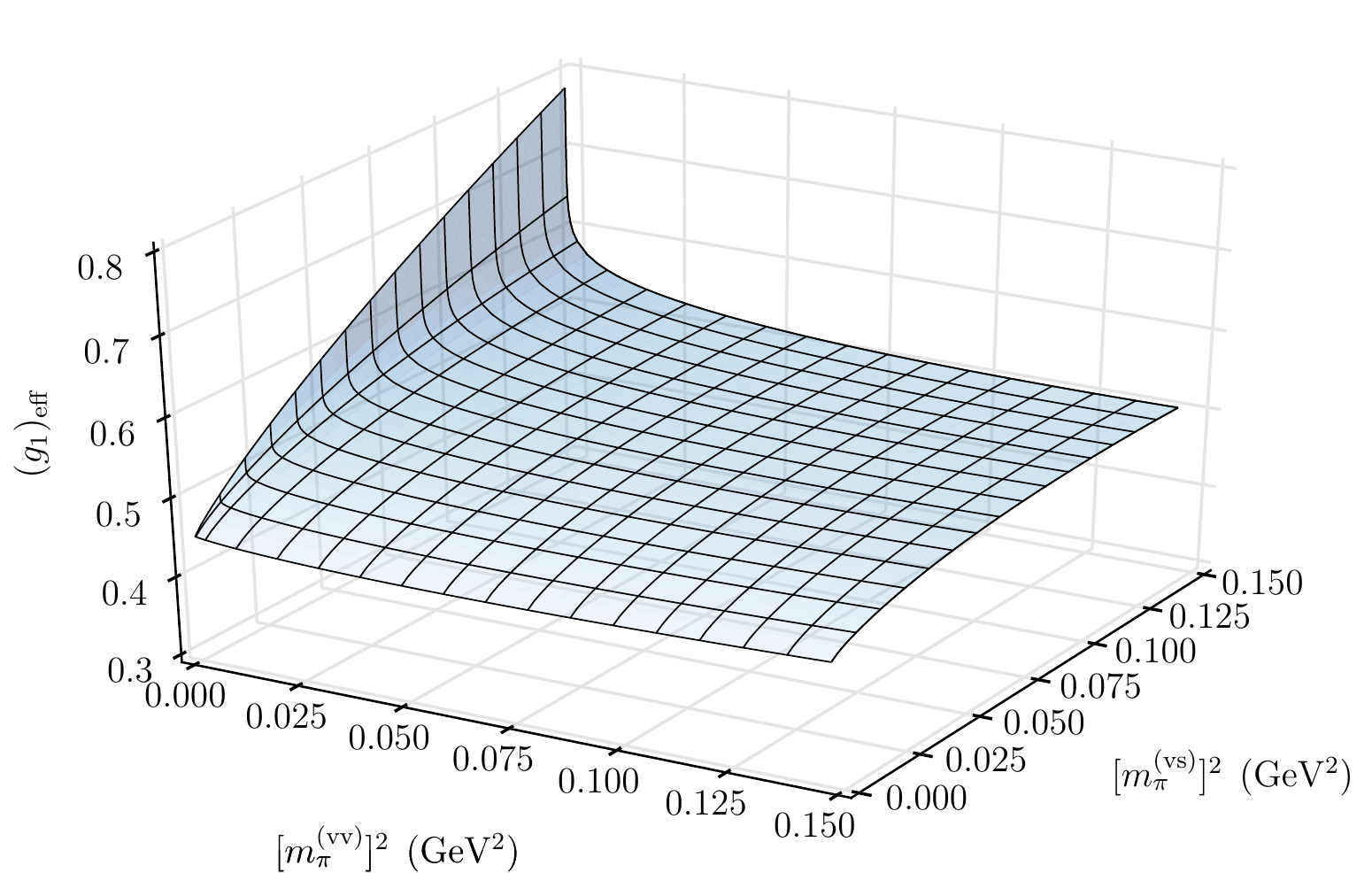}  \hfill \includegraphics[width=0.42\linewidth]{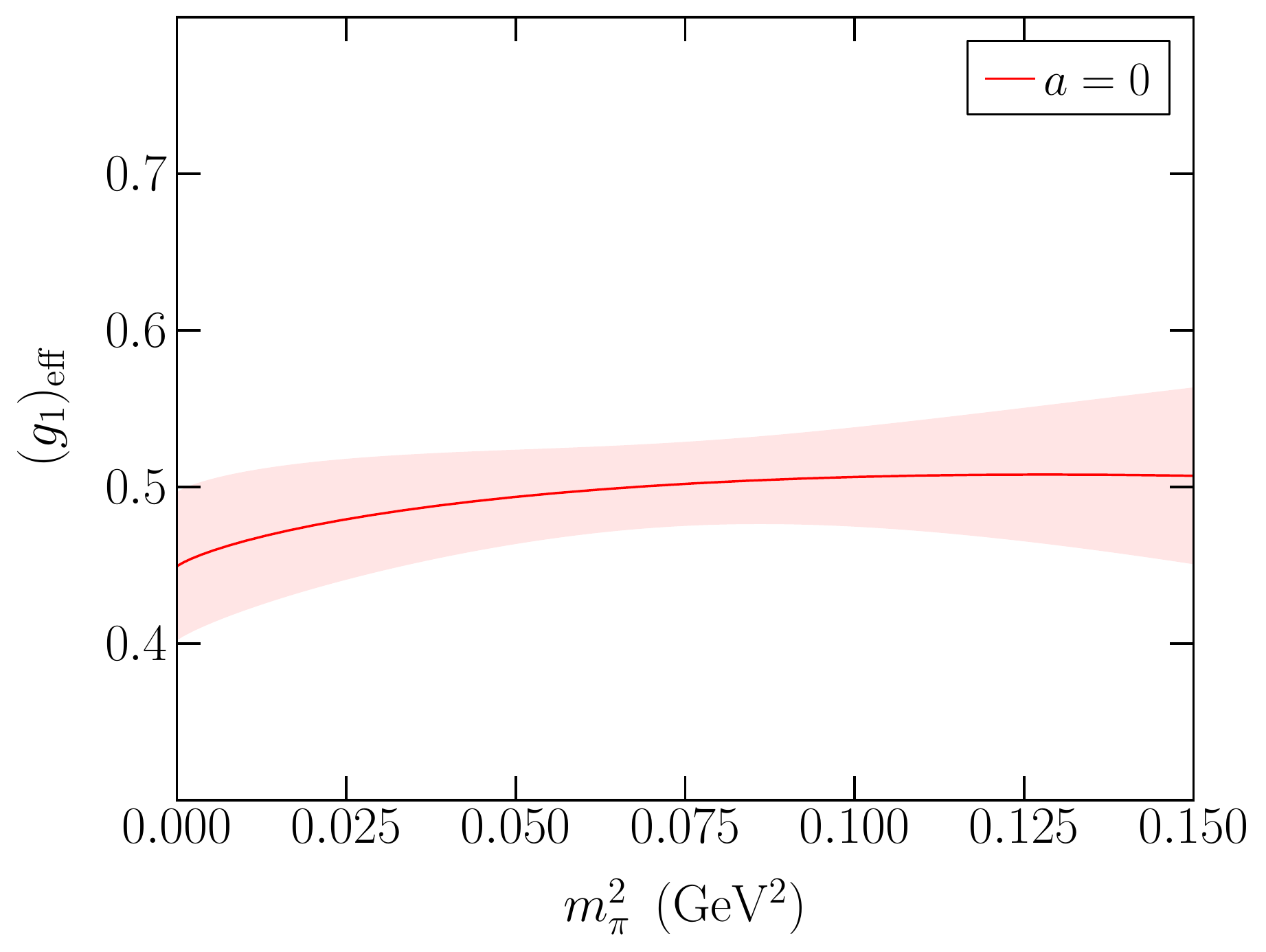}
\caption{\label{fig:g1} The fitted function $(g_1)_{\rm eff}$, evaluated in infinite volume, for $n_{\rm HYP}=3$, at different lattice spacings (from top to bottom: $a=0.112$ fm, $a=0.085$ fm, $a=0$), along with
the data points (shifted to infinite volume). The left-hand side shows the dependence on both $m_\pi^{(\mathrm{vv})}$ and $m_\pi^{(\mathrm{vs})}$.
The right-hand side shows the function (and its statistical uncertainty) evaluated at $m_\pi^{(\mathrm{vv})}=m_\pi^{(\mathrm{vs})}$.
In the plots on the right-hand side, the partially quenched data points, which have $m_\pi^{(\mathrm{vv})}<m_\pi^{(\mathrm{vs})}$, are indicated with open symbols. They are shown
at $m_\pi=m_\pi^{(\mathrm{vv})}$, even though the fitted function $(g_1)_{\rm eff}$ actually has slightly different values for these points. }
\end{figure*}

\begin{figure*}[ht!]
 \hspace{-3ex} \includegraphics[width=0.57\linewidth]{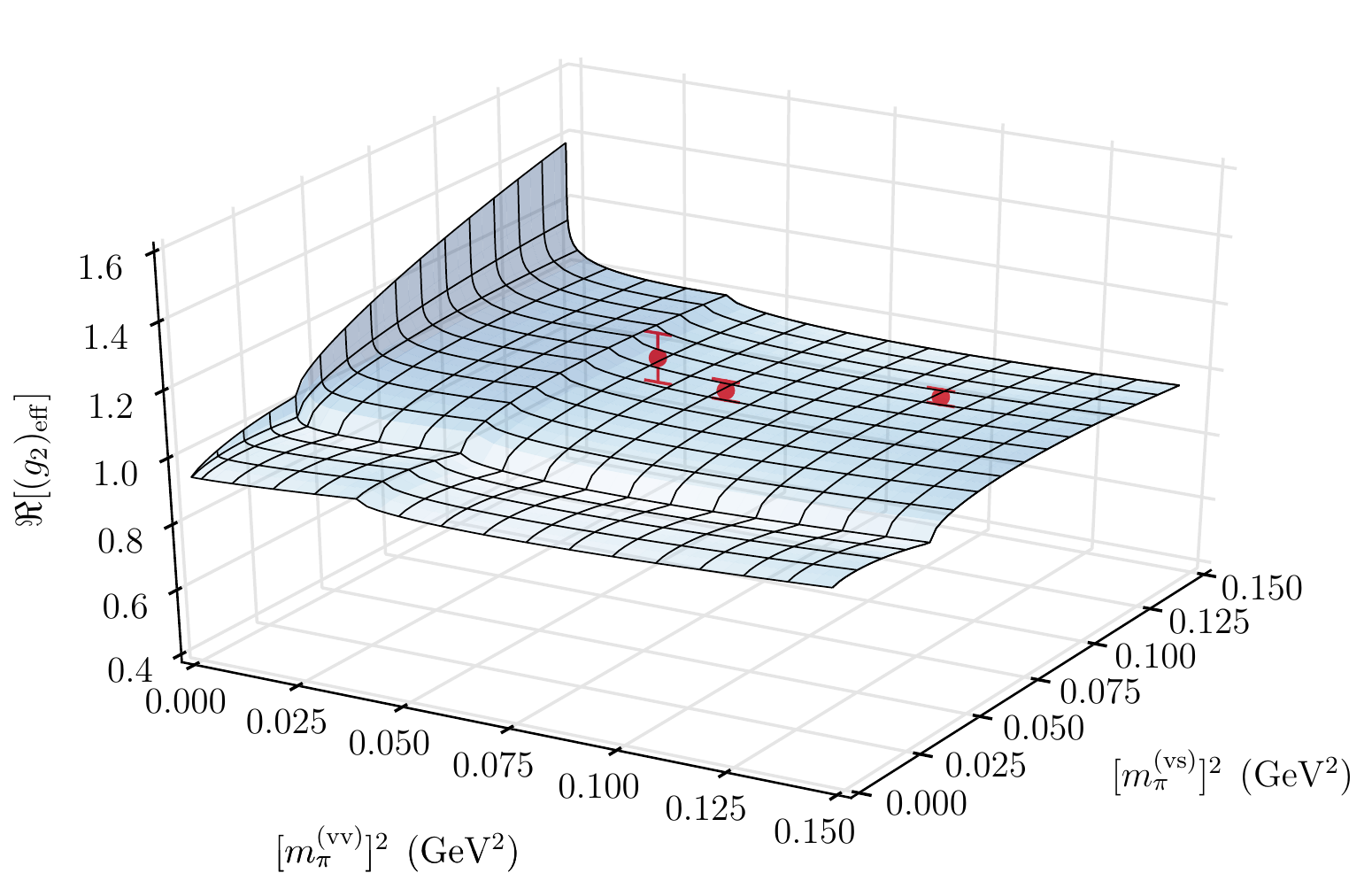}  \hfill \includegraphics[width=0.42\linewidth]{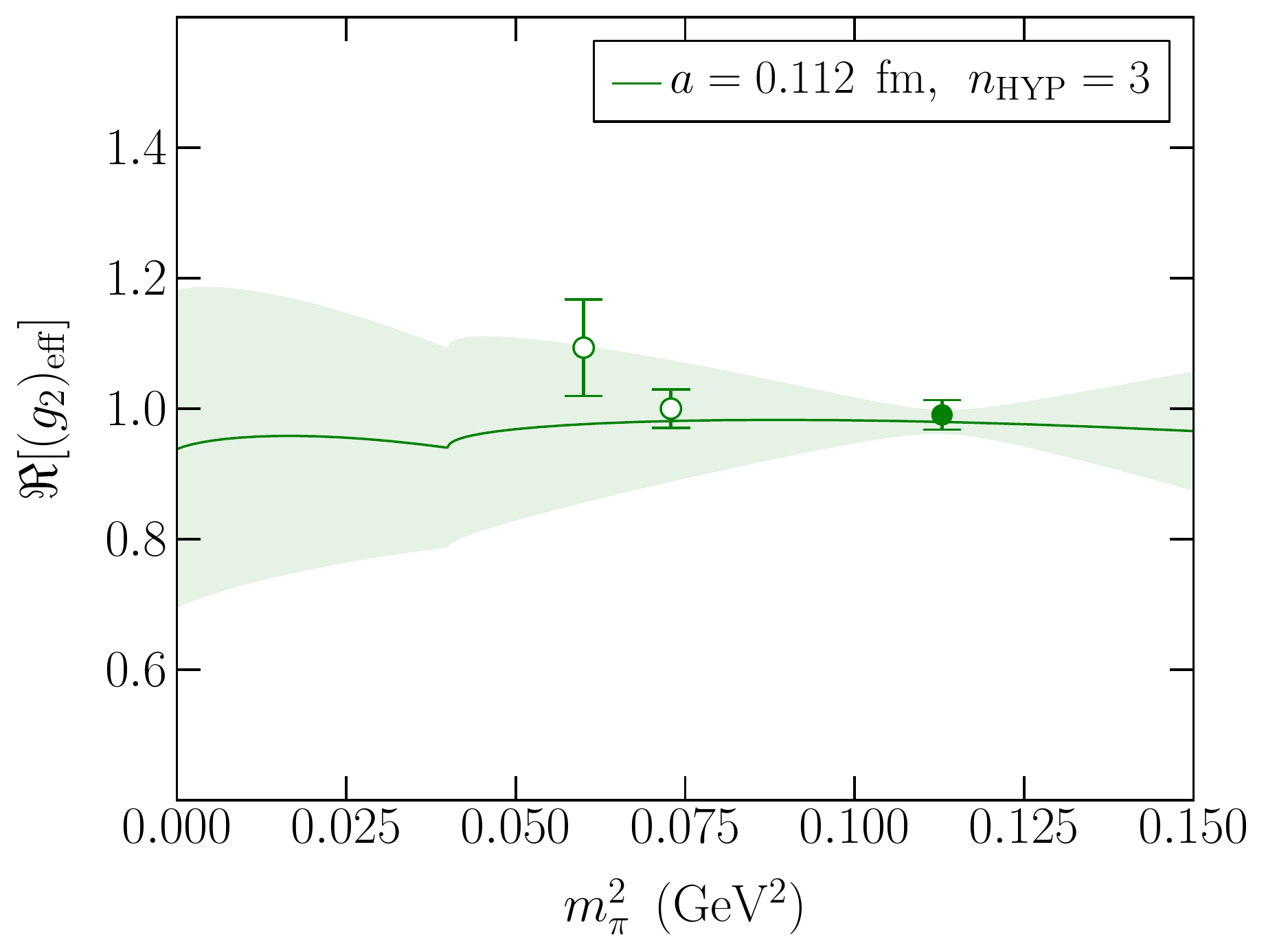}
 \hspace{-3ex} \includegraphics[width=0.57\linewidth]{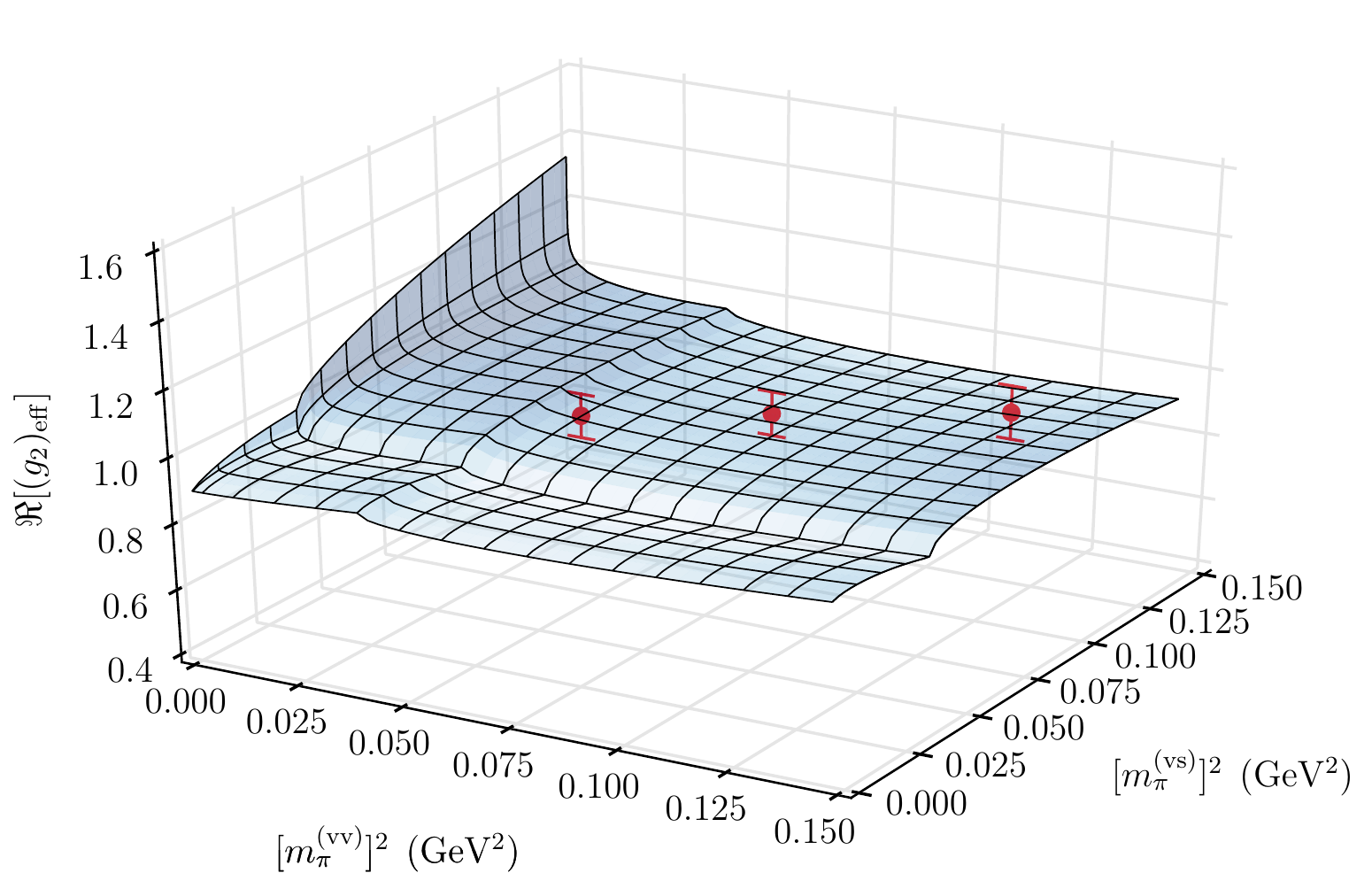}  \hfill \includegraphics[width=0.42\linewidth]{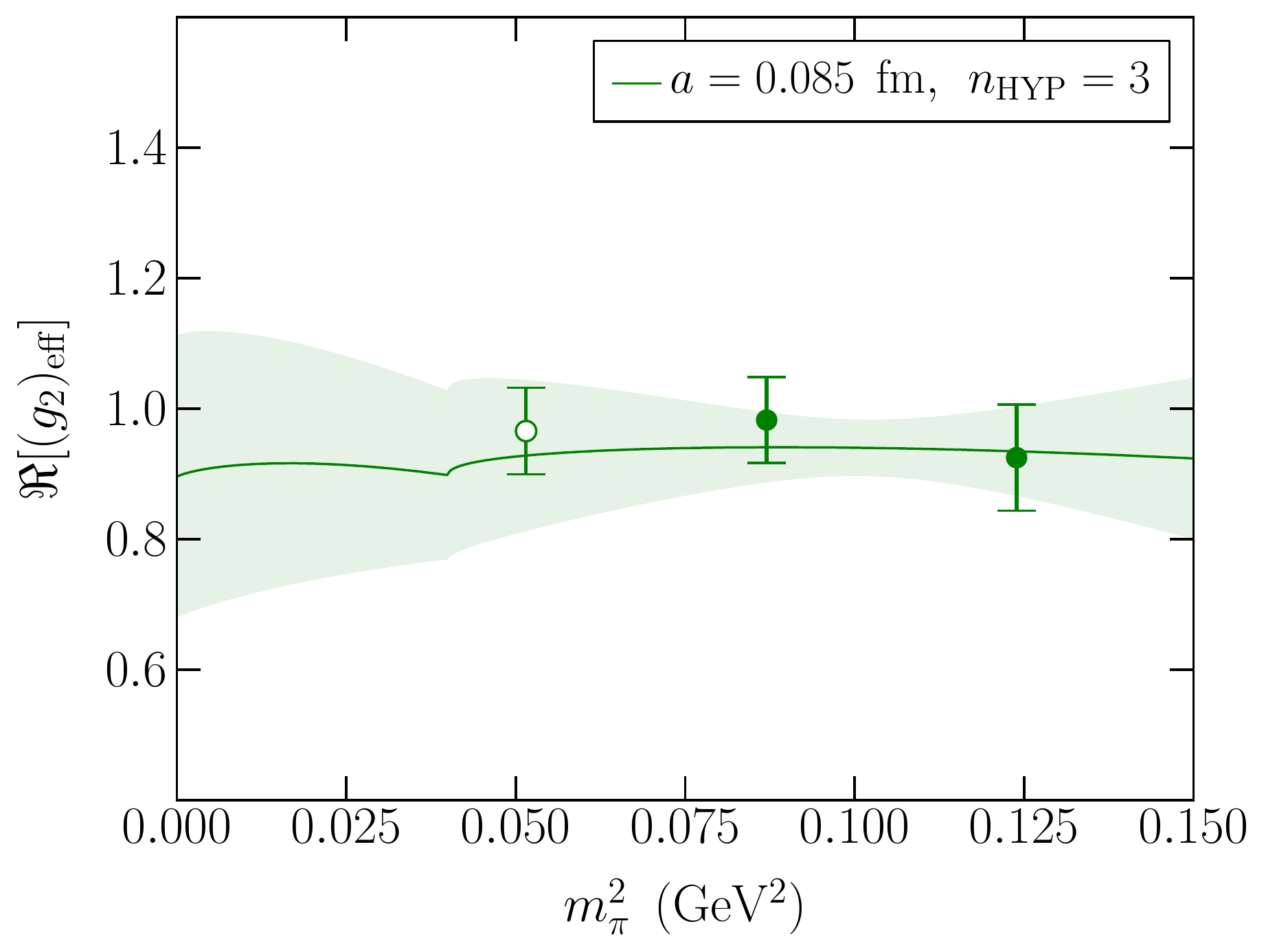}
 \hspace{-3ex} \includegraphics[width=0.57\linewidth]{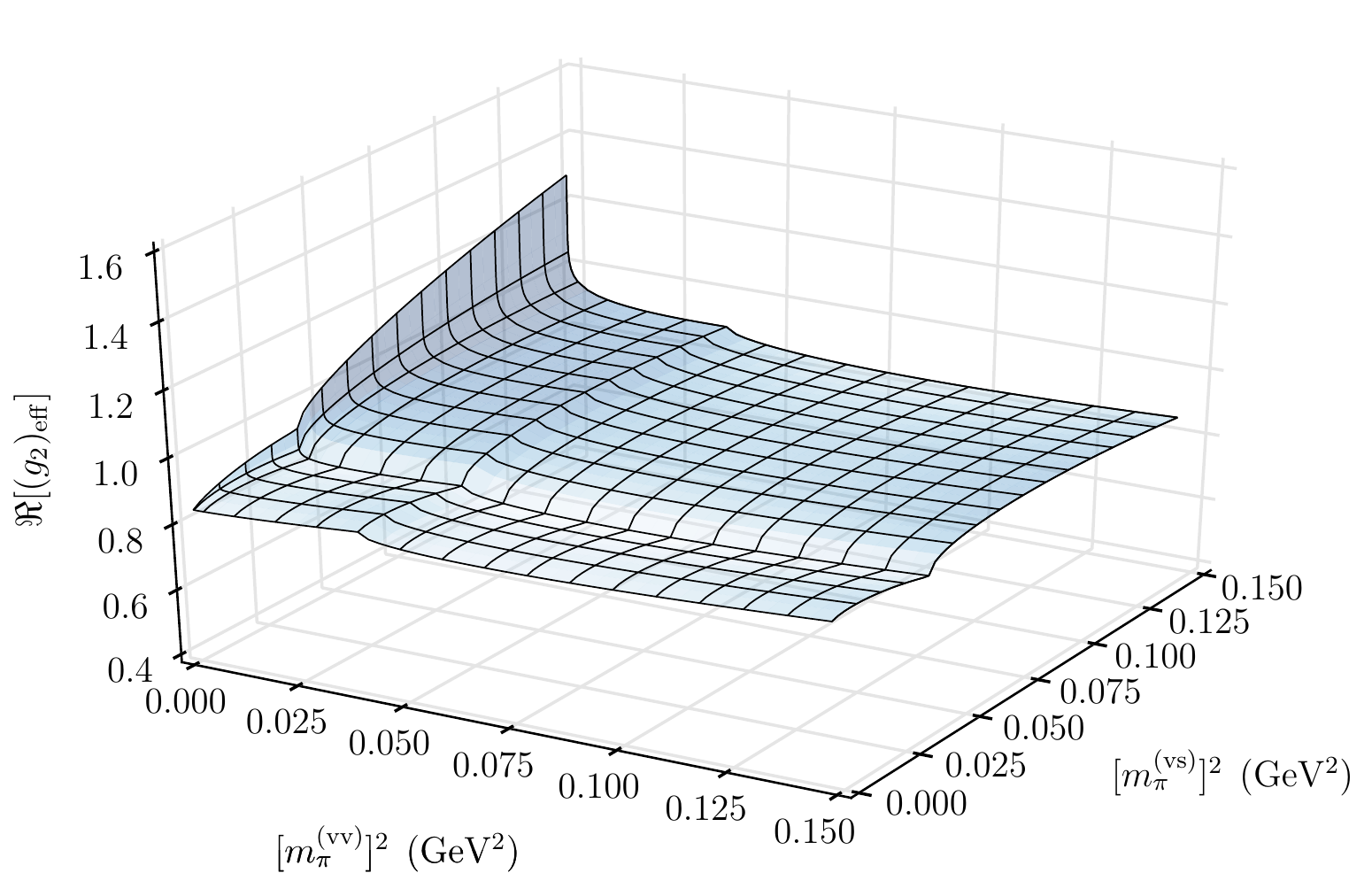}  \hfill \includegraphics[width=0.42\linewidth]{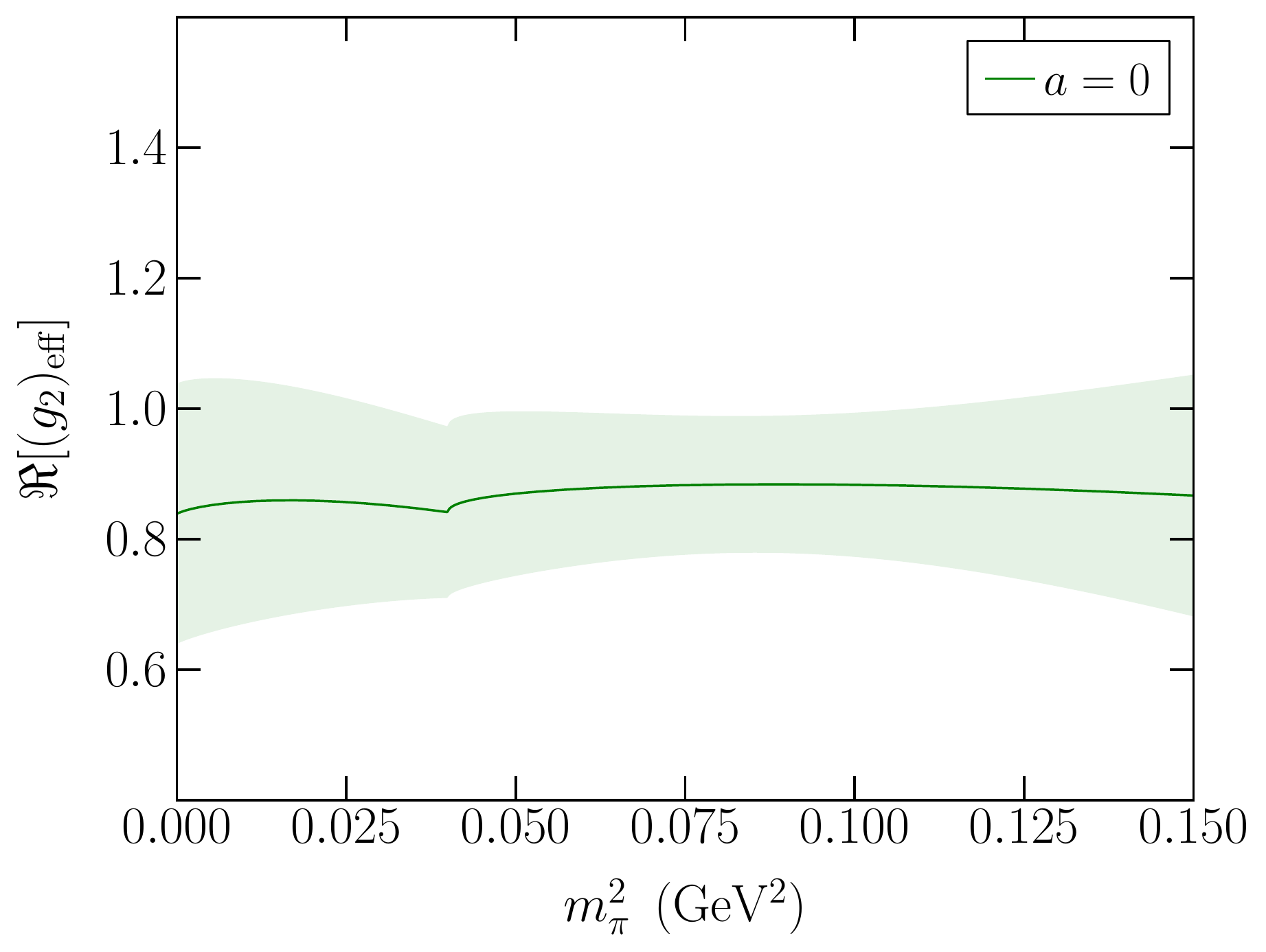}
\caption{\label{fig:g2} Like Fig.~\ref{fig:g1}, but for the real part of $(g_2)_{\rm eff}$.}
\end{figure*}

\begin{figure*}[ht!]
 \hspace{-3ex} \includegraphics[width=0.57\linewidth]{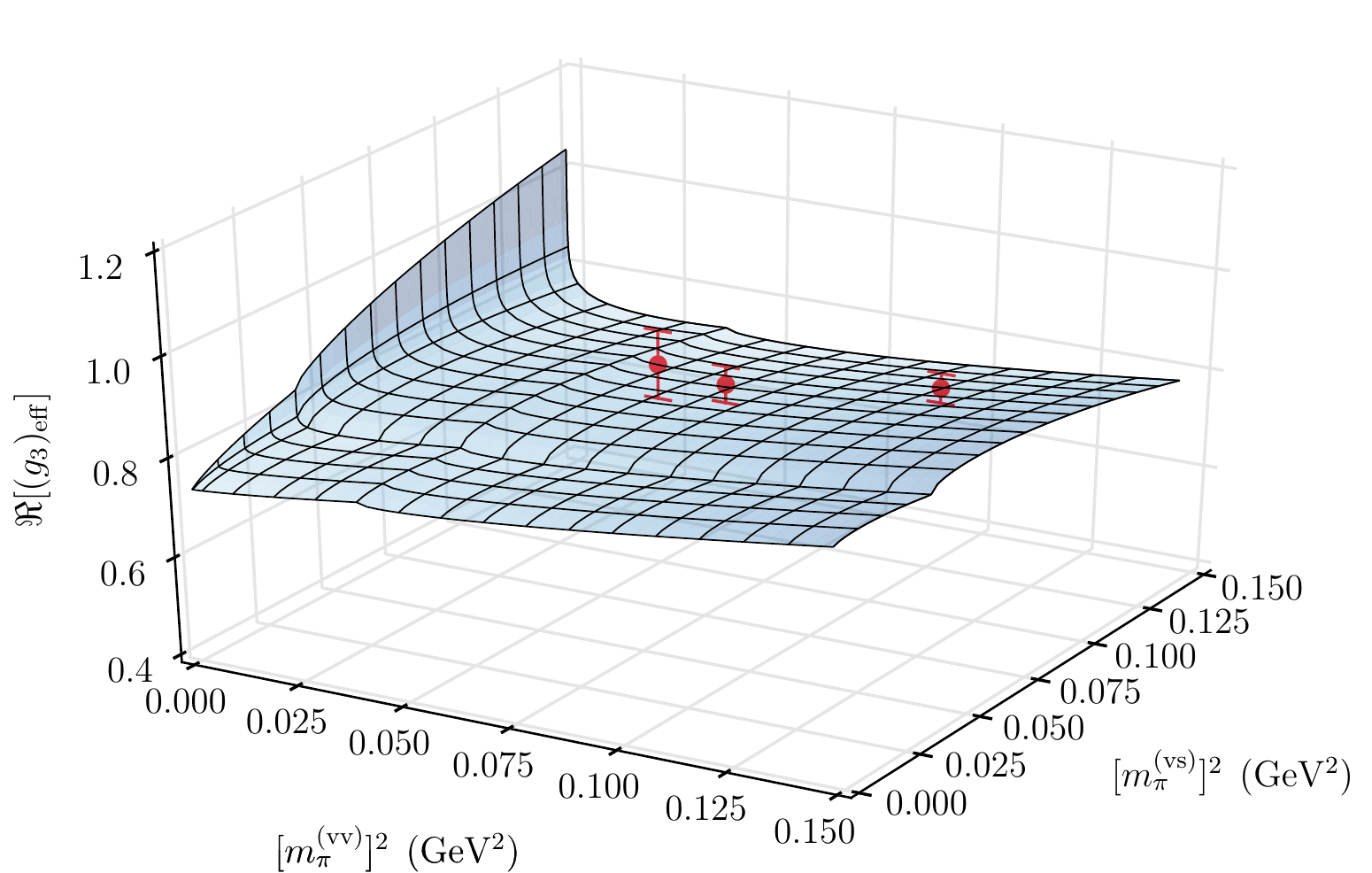}  \hfill \includegraphics[width=0.42\linewidth]{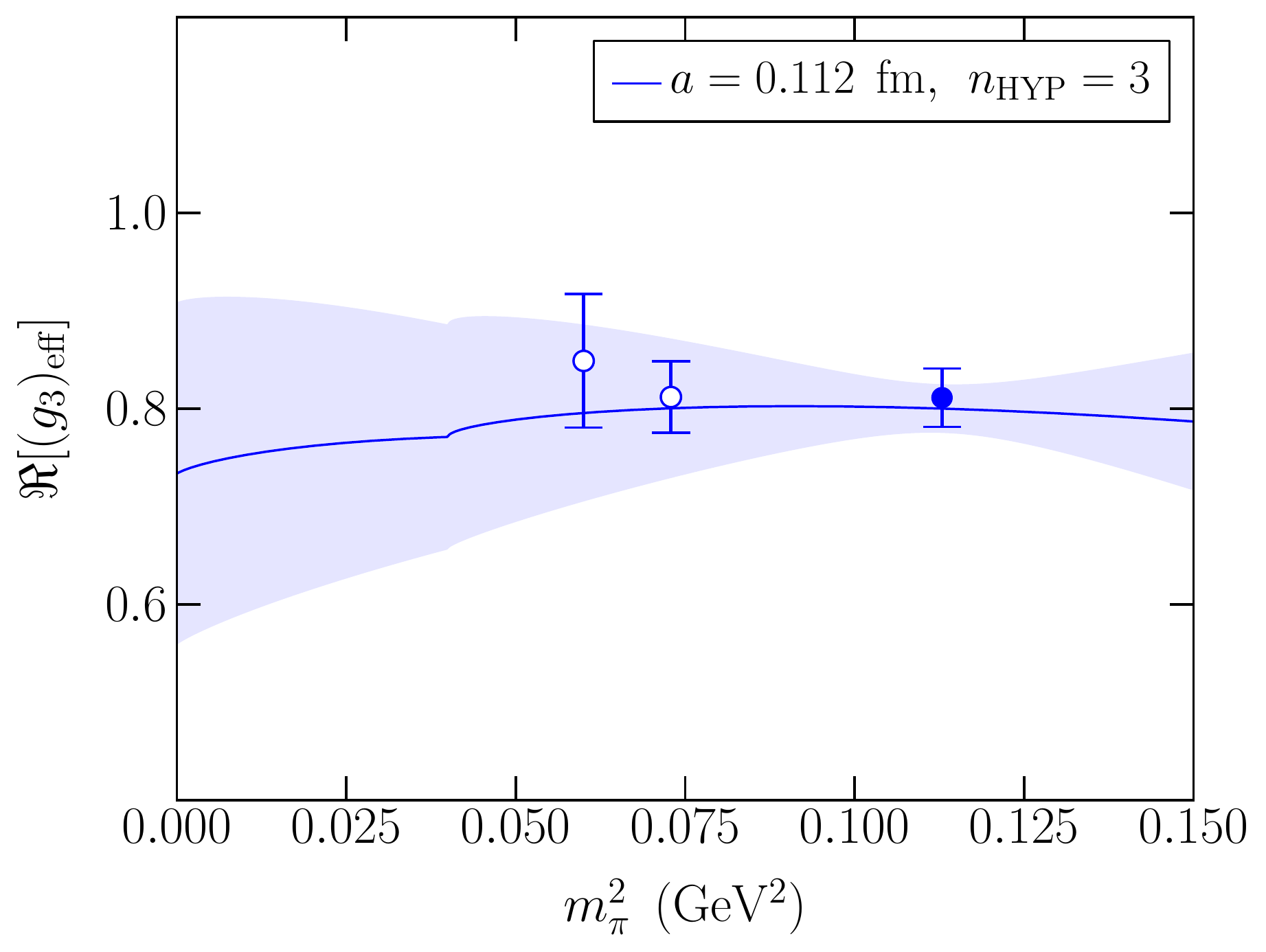}
 \hspace{-3ex} \includegraphics[width=0.57\linewidth]{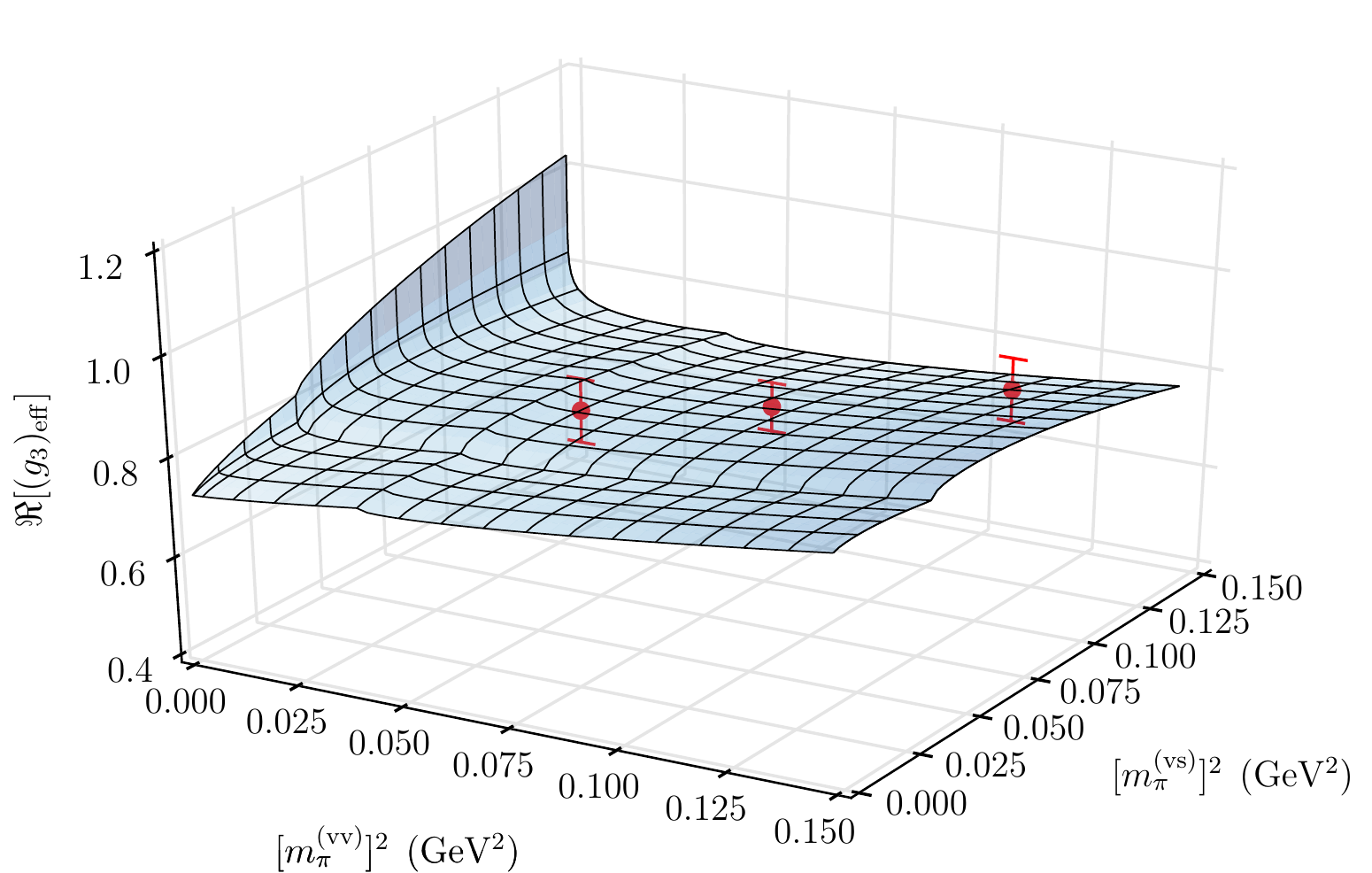}  \hfill \includegraphics[width=0.42\linewidth]{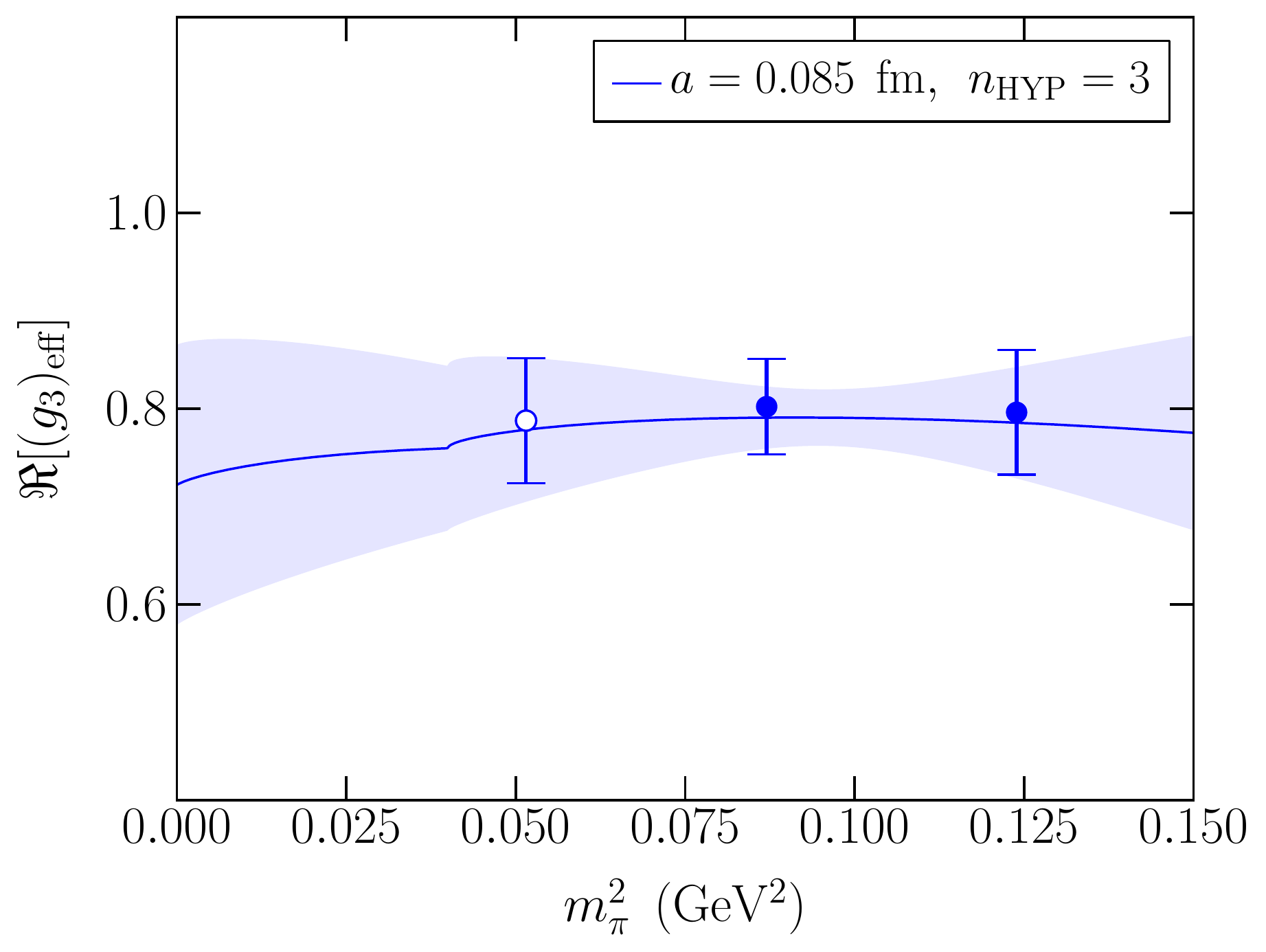}
 \hspace{-3ex} \includegraphics[width=0.57\linewidth]{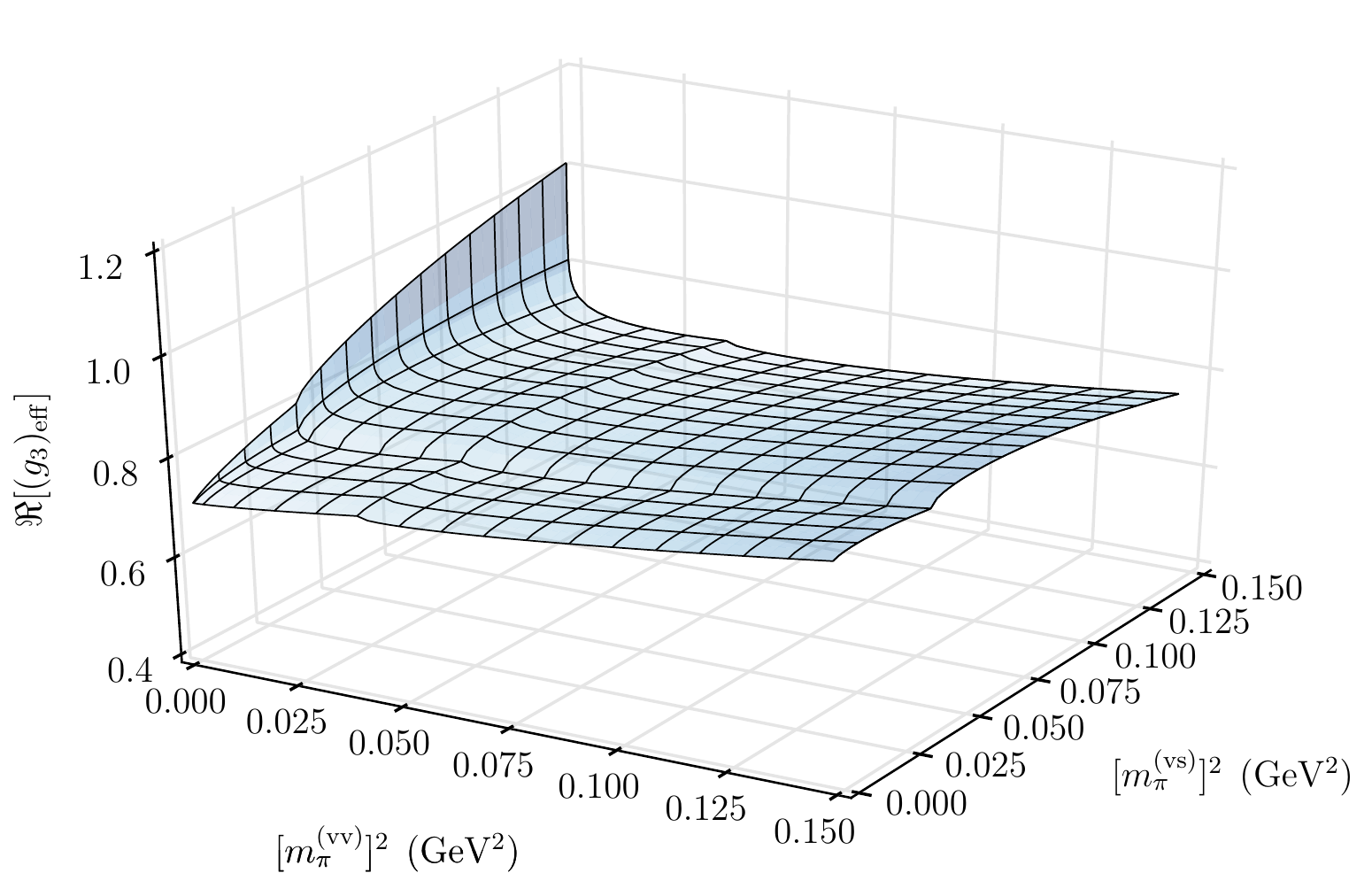}  \hfill \includegraphics[width=0.42\linewidth]{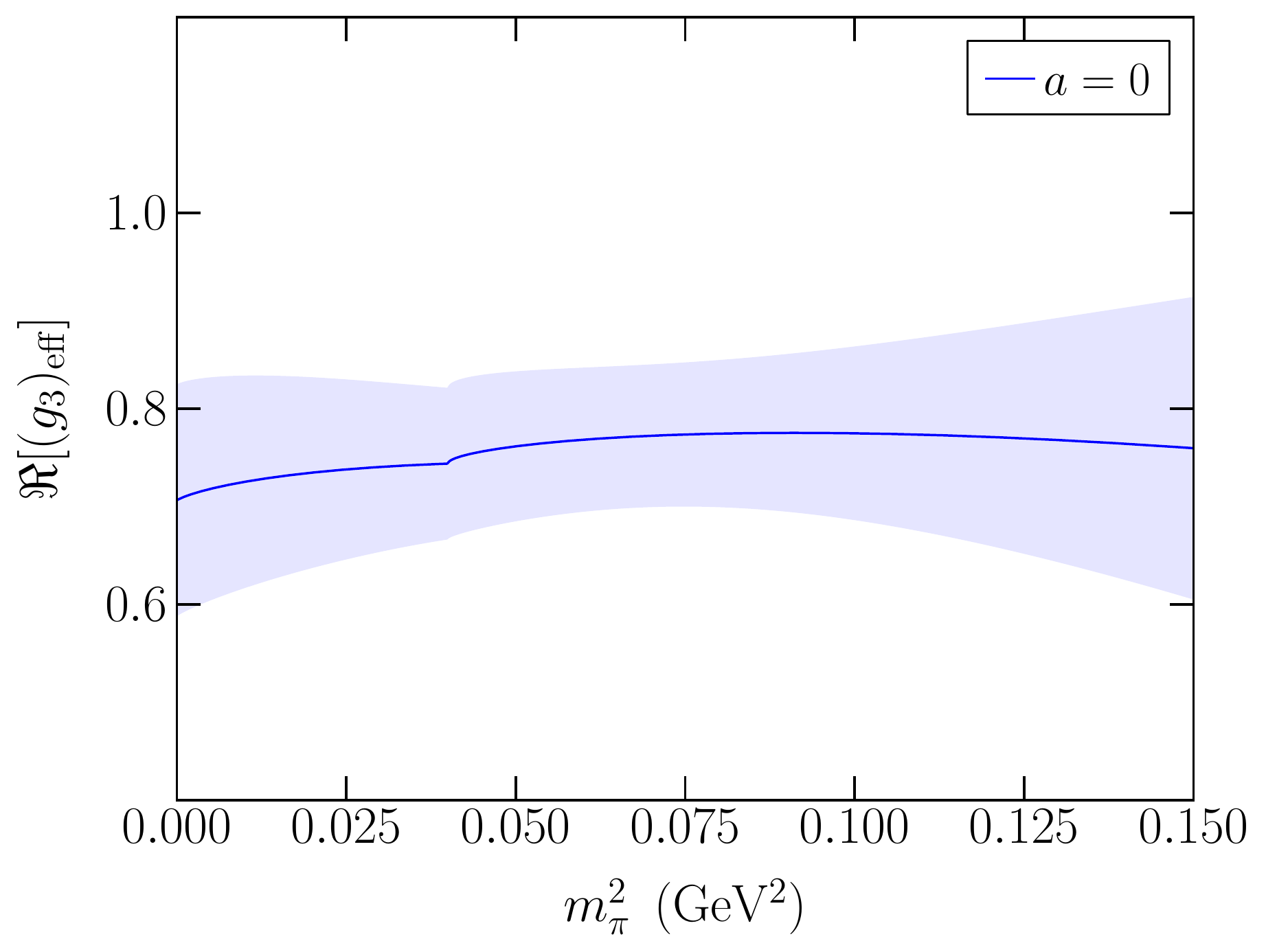}
\caption{\label{fig:g3} Like Fig.~\ref{fig:g1}, but for the real part of $(g_3)_{\rm eff}$. }
\end{figure*}

Plots of the functions $(g_1)_{\rm eff}$, $(g_2)_{\rm eff}$, and $(g_3)_{\rm eff}$, with the parameters from the fits
including the complete data with $n_{\rm HYP}=1, 2, 3$ [i.e. the fit that gives the results (\ref{eq:g1finalresult}) and (\ref{eq:g2g3finalresult})] are shown in Figs.~\ref{fig:g1}, \ref{fig:g2},
and \ref{fig:g3}. For the figures, the functions were evaluated in infinite volume, for the lattice spacings $a=0.112$ fm, $a=0.085$ fm, and $a=0$.
The right-hand sides of the figures show the values and uncertainties of the fitted functions for the unitary case $m_\pi^{(\mathrm{vv})}=m_\pi^{(\mathrm{vs})}$,
while the left-hand sides show the dependence on both $m_\pi^{(\mathrm{vv})}$ and $m_\pi^{(\mathrm{vs})}$. At the two nonzero values of $a$, the functions were
evaluated for $n_{\rm HYP}=3$ and the corresponding data points are also shown (in the continuum limit, the functions for $n_{\rm HYP}=1,2,3$
are all equal). To allow the inclusion in these plots, the data points were shifted to infinite volume using
\begin{equation}
 (g_i)_{\rm eff, data}(m, L=\infty)= (g_i)_{\rm eff, data}(m, L=2.7\:{\rm fm}) + \left[ (g_i)_{\rm eff, fit}(m, L=\infty) - (g_i)_{\rm eff, fit}(m, L=2.7\:{\rm fm})\right],
\end{equation}
where we use the notation $m=(m_\pi^{(\mathrm{vv})}, m_\pi^{(\mathrm{vs})})$. The numerical values of the volume shifts are given in
Table \ref{tab:volumeshifts}. The largest volume shift (2.8\%) occurred for $(g_2)_{\rm eff}$ at $m_\pi^{(\mathrm{vv})}=227$ MeV.

\begin{table}[ht!]
\begin{tabular}{cccccccccc}
\hline\hline
  $m_\pi^{(\mathrm{vs})}$ (MeV)   & & $m_\pi^{(\mathrm{vv})}$ (MeV) & \hspace{1ex} & $\frac{(g_1)_{\rm eff}^{(\infty)}\!-(g_1)_{\rm eff}^{(L)}}{(g_1)_{\rm eff}^{(\infty)}}$ & \hspace{1ex} & $\frac{(g_2)_{\rm eff}^{(\infty)}\!-(g_2)_{\rm eff}^{(L)}}{(g_2)_{\rm eff}^{(\infty)}}$ & \hspace{1ex} & $\frac{(g_3)_{\rm eff}^{(\infty)}\!-(g_3)_{\rm eff}^{(L)}}{(g_3)_{\rm eff}^{(\infty)}}$ \\
\hline
  294   && 245  && 0.0057  && 0.015\nb\nb &&  $\phantom{-}$0.0074\nb   \\
  304   && 270  && 0.0040  && 0.0070\nb   &&  $\phantom{-}$0.0027\nb   \\
  336   && 336  && 0.0016  && 0.00037     &&  $-$0.00079               \\
\\[-2ex]
  263   && 227  && 0.0072  && 0.028\nb\nb &&  $\phantom{-}$0.013\nb\nb \\
  295   && 295  && 0.0031  && 0.00027     &&  $-$0.0012\nb             \\
  352   && 352  && 0.0013  && 0.00033     &&  $-$0.00071               \\
\hline\hline
\end{tabular}
\caption{\label{tab:volumeshifts}Size of the finite-volume corrections for the pion masses where we have data.}
\end{table}

The functions $(g_2)_{\rm eff}$ and $(g_3)_{\rm eff}$ develop small imaginary parts for pion masses below the $S\to T \pi$ threshold at $m_\pi = \Delta$ \cite{Detmold:2011rb}
(the lattice data are all above the threshold). The extracted parameters $g_{1,2,3}$ are real.
Figures \ref{fig:g2}, and \ref{fig:g3} show the real parts of $(g_2)_{\rm eff}$ and $(g_3)_{\rm eff}$ only, which have kinks at the thresholds.

The fit results for the parameters $c_i^{(\mathrm{vv})}$, $c_i^{(\mathrm{vs})}$, which describe the analytic contributions, were natural-sized, i.e.~of order $1/\Lambda_\chi^2$
with $\Lambda_\chi\approx 4\pi f_\pi$, for the renormalization scale $\mu=4\pi f_\pi$. The fit results for the parameters $d_{i,\:n_{\rm HYP}}$, which describe the lattices-spacing dependence, were
also of natural size and consistent with zero within the statistical uncertainties.
The absence of significant $a$-dependence can also be seen in Figs.~\ref{fig:g1}, \ref{fig:g2}, and \ref{fig:g3}.

\begin{figure*}[ht!]
 \centerline{\includegraphics[width=0.49\linewidth]{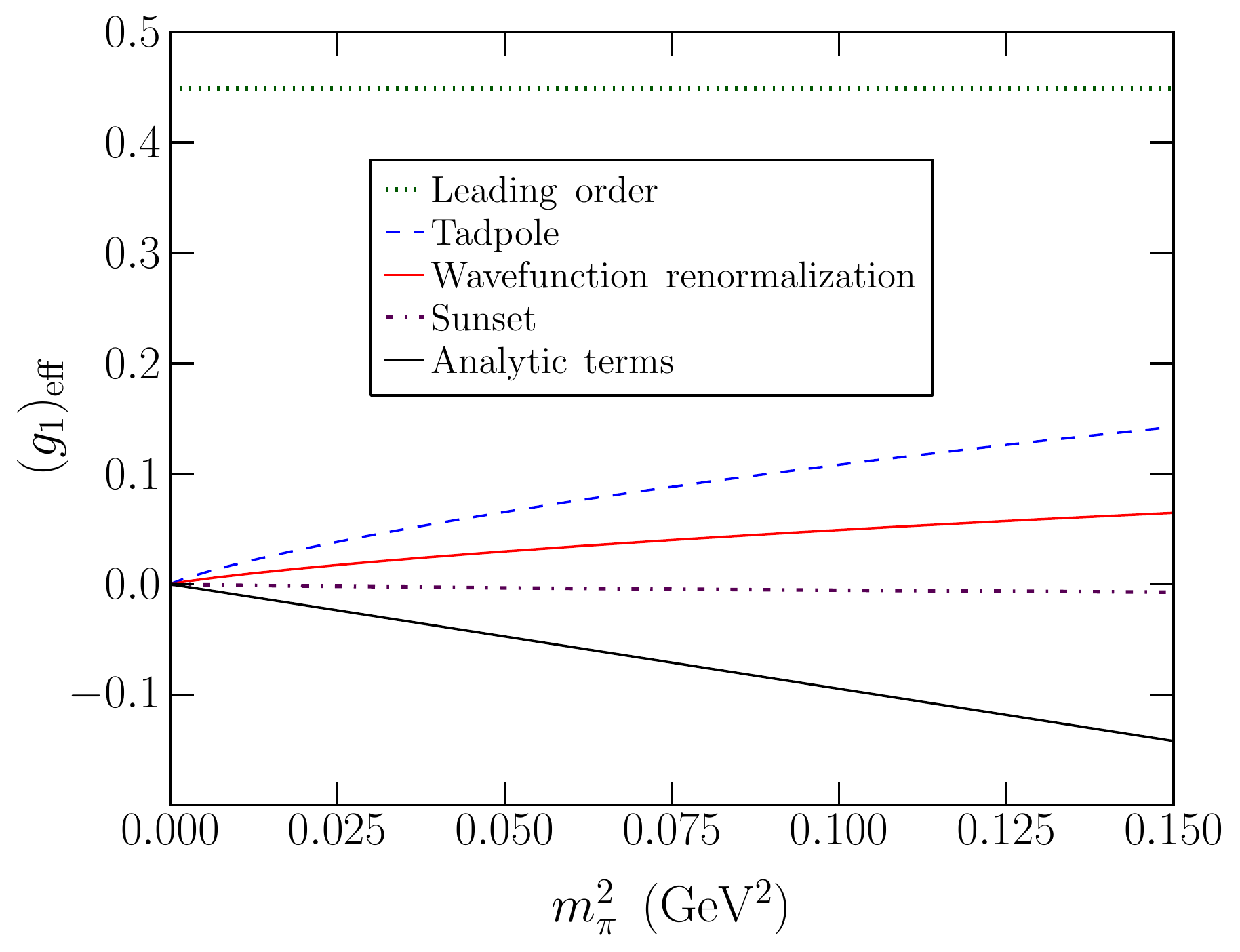}}
\caption{\label{fig:g1continuumindividual}Contributions from individual classes of Feynman diagrams in HH$\chi$PT (see Ref.~\protect{\cite{Detmold:2011rb}}) to the fitted function $(g_1)_{\rm eff}$,
evaluated in the infinite volume, continuum limit, for $m_\pi^{(\mathrm{vv})}=m_\pi^{(\mathrm{vs})}$. The renormalization scale is $\mu=4\pi f$.}
\end{figure*}

\begin{figure*}[ht!]
 \centerline{\includegraphics[width=0.49\linewidth]{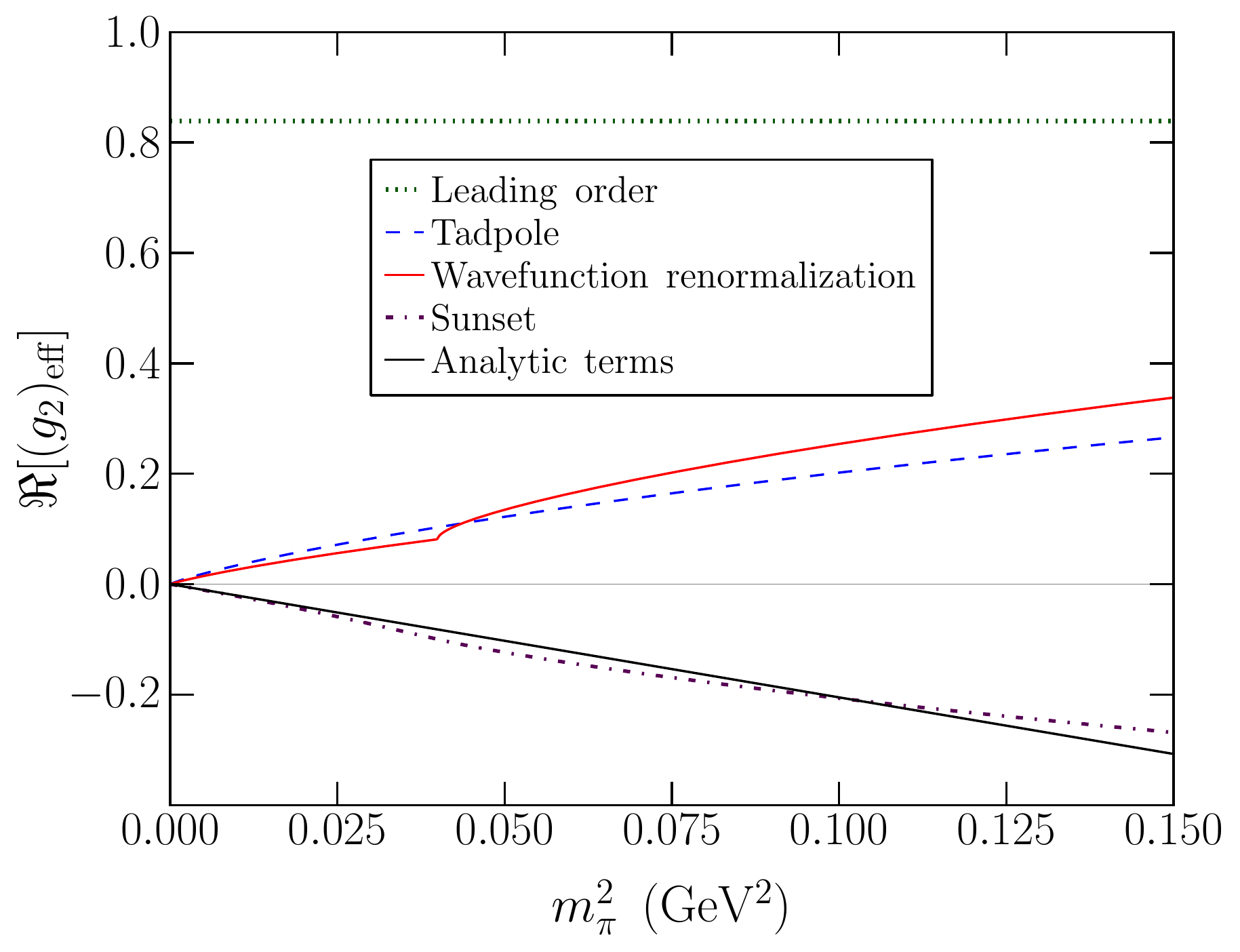}\hfill\includegraphics[width=0.49\linewidth]{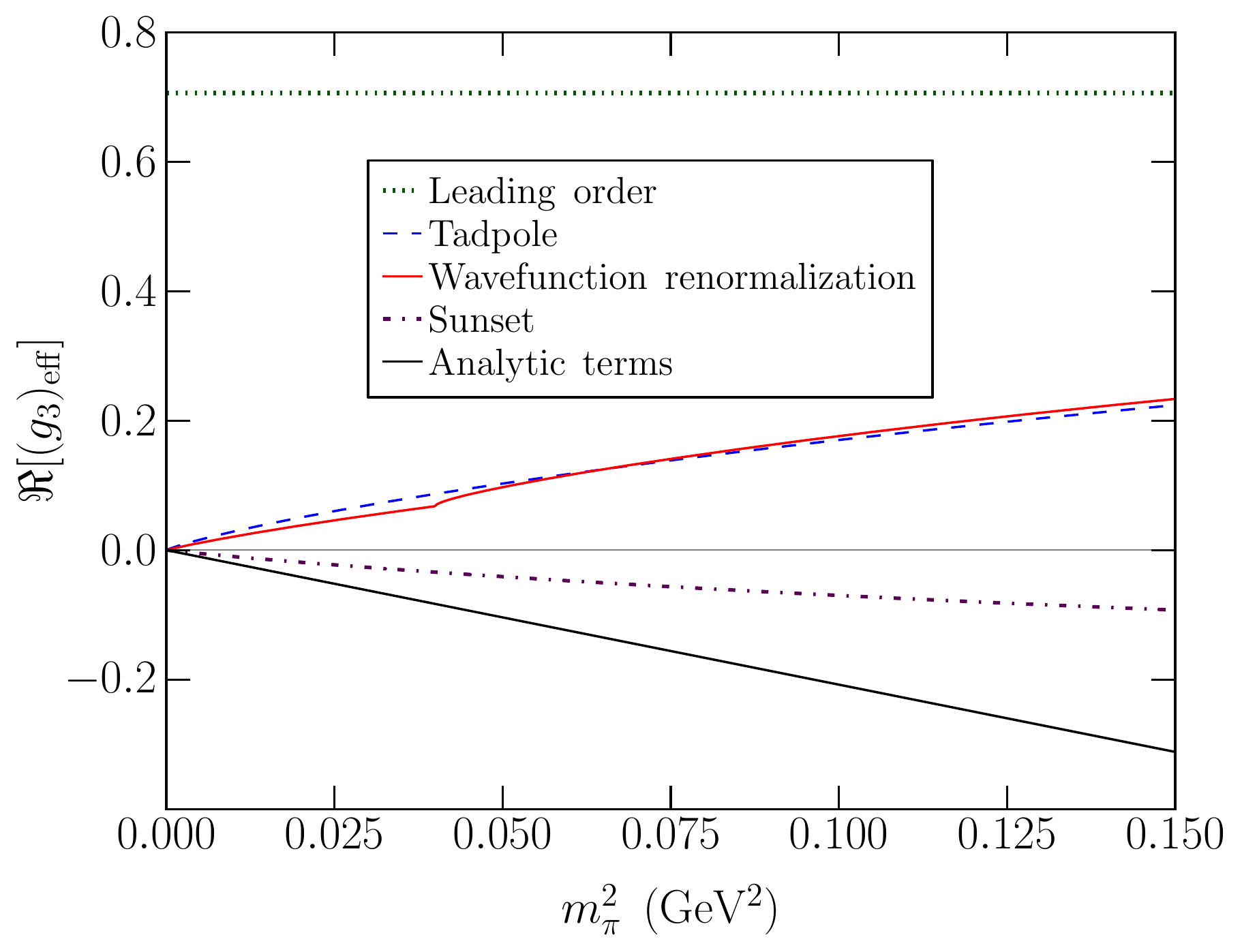}}
\caption{\label{fig:g2g3continuumindividual}Contributions from individual classes of Feynman diagrams in HH$\chi$PT (see Ref.~\protect{\cite{Detmold:2011rb}}) to the real part of the
fitted functions $(g_2)_{\rm eff}$ (left panel) and $(g_3)_{\rm eff}$ (right panel), evaluated in the infinite volume, continuum limit, for $m_\pi^{(\mathrm{vv})}=m_\pi^{(\mathrm{vs})}$.
The renormalization scale is $\mu=4\pi f$.}
\end{figure*}

The individual contributions from different classes of Feynman diagrams in HH$\chi$PT \cite{Detmold:2011rb} to the fitted functions $(g_1)_{\rm eff}$, $\Re[(g_2)_{\rm eff}]$, and $\Re[(g_3)_{\rm eff}]$ (evaluated
for $a=0$, $L=\infty$, and $m_\pi^{(\mathrm{vv})}=m_\pi^{(\mathrm{vs})}$) are shown in Figs.~\ref{fig:g1continuumindividual}, \ref{fig:g2g3continuumindividual}.
Note that while the sum of all contributions (including the analytic terms) is independent of the renormalization scale $\mu$, the individual contributions are not,
and the figures are based on the natural scale $\mu=4\pi f_\pi$. For the range of pion masses considered here, the NLO contributions are significantly smaller than the LO
contribution (which is equal to $g_i$). This, and the natural size of the fitted coefficients $c_i^{(\mathrm{vv})}$, $c_i^{(\mathrm{vs})}$, indicates that
the chiral expansion of the axial-current matrix elements is well-behaved here.

To estimate the size of systematic uncertainties caused by the missing NNLO terms in the fits to the quark-mass and lattice-spacing dependence,
we performed fits to the data using modified functions $(g_i)^{({\rm NLO+HO})}_{\rm eff}$ that include
higher-order analytic terms:

\begin{eqnarray}
\nonumber (g_i)^{({\rm NLO+HO})}_{\rm eff}(a, m, n_{\rm HYP}) &=& (g_i)^{\rm (NLO)}_{\rm eff}(a, m, n_{\rm HYP}) \\
\nonumber && + g_i \Bigg[ c_i^{(\mathrm{vv,vv})}\: [m_\pi^{(\mathrm{vv})}]^4
+ c_i^{(\mathrm{vs,vs})}\: [m_\pi^{(\mathrm{vs})}]^4
+ c_i^{(\mathrm{vv,vs})}\: [m_\pi^{(\mathrm{vv})}]^2 [m_\pi^{(\mathrm{vs})}]^2 \\
&& \phantom{+ g_i \Bigg[}+ d^{\rm(vv)}_{i,\:n_{\rm HYP}}\:a^2\:[m_\pi^{(\mathrm{vv})}]^2 + d^{\rm(vs)}_{i,\:n_{\rm HYP}}\:a^2\:[m_\pi^{(\mathrm{vs})}]^2
+ h_{i,\:n_{\rm HYP}} \:a^4  \Bigg].  \label{eq:HOfit}
\end{eqnarray}
Here, the functions $(g_i)^{\rm (NLO)}_{\rm eff}$ are as defined in Eqs.~(\ref{eq:g1chiralcontinuumfit}), (\ref{eq:g2chiralcontinuumfit}), and (\ref{eq:g3chiralcontinuumfit}).
Because we do not have enough data to fit all the parameters in Eq.~(\ref{eq:HOfit}), we constrained the parameters corresponding to the higher-order terms
using Gaussian priors centered around zero and with widths equal to some dimensionless factor $w$ times the relevant natural scales:
\begin{eqnarray}
\nonumber c_i^{(\mathrm{vv,vv})} &=& 0 \:\pm\: w / \Lambda_\chi^4, \\
\nonumber c_i^{(\mathrm{vs,vs})} &=& 0 \:\pm\: w / \Lambda_\chi^4, \\
\nonumber c_i^{(\mathrm{vv,vs})} &=& 0 \:\pm\: w / \Lambda_\chi^4, \\
\nonumber d^{(\mathrm{vv})}_{i,\:n_{\rm HYP}} &=& 0 \:\pm\: w\: \Lambda_{\rm QCD}^2/\Lambda_\chi^2, \\
\nonumber d^{(\mathrm{vs})}_{i,\:n_{\rm HYP}} &=& 0 \:\pm\: w\: \Lambda_{\rm QCD}^2/\Lambda_\chi^2, \\
 h_{i,\:n_{\rm HYP}} &=& 0 \:\pm\: w \:\Lambda_{\rm QCD}^4. \label{eq:NNLOw}
\end{eqnarray}
Here we used $\Lambda_\chi=4\pi f_\pi$ with $f_\pi=132$ MeV, and $\Lambda_{\rm QCD}=300$ MeV. The fit results for the axial couplings $g_i$
as a function of the width factor $w$ are given in Table \ref{tab:NNLO}. While the case $w=0$ corresponds to the
original NLO fits, in the limit $w\rightarrow\infty$ the new parameters would become unconstrained (because we have insufficient data,
we are unable to perform fits in this limit).

\begin{table}[ht!]
\begin{tabular}{ccccccccccccc}
\hline\hline
  $w$   & \hspace{1ex} & $g_1$ & \hspace{1ex} & $\delta\sigma(g_1)$ & \hspace{1ex} & $g_2$ & \hspace{1ex} & $\delta\sigma(g_2)$ & \hspace{1ex} & $g_3$ & \hspace{1ex} & $\delta\sigma(g_3)$ \\
\hline
  0     && 0.449(47)  && 0             && 0.84(20)  && 0             && 0.71(12)  && 0             \\
  1     && 0.449(47)  && 0.0020        && 0.84(20)  && 0.0023        && 0.71(12)  && 0.0045        \\
  5     && 0.452(48)  && 0.0089        && 0.84(20)  && 0.014\nb      && 0.70(12)  && 0.017\nb      \\
  10    && 0.455(50)  && 0.016\nb      && 0.84(20)  && 0.024\nb      && 0.70(12)  && 0.026\nb      \\
  50    && 0.464(72)  && 0.054\nb      && 0.82(22)  && 0.099\nb      && 0.68(15)  && 0.094\nb      \\
  100   && 0.452(94)  && 0.082\nb      && 0.78(26)  && 0.17\nb\nb    && 0.63(21)  && 0.17\nb\nb    \\
\hline\hline
\end{tabular}
\caption{\label{tab:NNLO}Results of higher-order fits using Eq.~(\ref{eq:HOfit}) as a function of the width factor $w$ defined in Eq.~(\ref{eq:NNLOw}).}
\end{table}

As can be seen in Table \ref{tab:NNLO}, the shifts in the central values of the axial couplings are smaller than the statistical
errors up to the very large width $w=100$. This is a consequence of the smallness of the quantities
$m_\pi^4/(4\pi f_\pi)^4$, $a^2 \Lambda_{\rm QCD}^2 m_\pi^2/(4\pi f_\pi)^2$, and $a^4 \Lambda_{\rm QCD}^4$
for the pion masses and lattice spacings where we have data. The shifts in the central values fluctuate statistically
and can be close to zero even for large $w$ (at least for $g_1$).
However, including the higher-order terms leads to a systematic increase in the uncertainties of the fit parameters $g_i$
(as calculated from the Hessian of $\chi^2$), as expected. Also shown in the table is the quantity
\begin{equation}
\delta\sigma(g_i) = \sqrt{ \sigma^2(g_i)^{({\rm NLO+HO})} - \sigma^2(g_i)^{\rm (NLO)}}, \label{eq:sigmaincrease}
\end{equation}
where $\sigma(g_i)^{\rm (NLO)}$ is the original uncertainty of $g_i$ from the NLO fit, and $\sigma(g_i)^{({\rm NLO+HO})}$ is
the new uncertainty of $g_i$ from the higher-order fit (\ref{eq:NNLOw}). To calculate (\ref{eq:sigmaincrease}) we used
more digits for $\sigma(g_i)^{\rm (NLO)}$ and $\sigma(g_i)^{({\rm NLO+HO})}$ than shown in Table \ref{tab:NNLO}.
Equation (\ref{eq:sigmaincrease}) gives the additional uncertainty in $g_i$, calculated using quadrature, that
results from the higher-order terms. This additional uncertainty $\delta\sigma(g_i)$
scales roughly linearly with the width parameter $w$. For a reasonable choice of $w$, the quantity $\delta\sigma(g_i)$
can be considered to be the systematic uncertainty in $g_i$ from the NLO fit due to the missing NNLO terms.
Here we choose the conservative value of $w=10$ for this purpose. The resulting estimates of relative
systematic uncertainties can be found in Table \ref{tab:systerrs}. There, we also show the estimates
of the other relevant sources of uncertainties: effects of higher excited states in the fits to $R_i(t)$ as discussed
in Sec.~\ref{sec:sourcesinkextrap}, and the effects the sea-strange-quark mass being about 10\% above the physical value,
as discussed in Sec.~\ref{sec:latticeparams}. Including the estimates of the total systematic uncertainties, our final results for
the axial couplings, based on (\ref{eq:g1finalresult}) and (\ref{eq:g2g3finalresult}), are then
\begin{eqnarray}
  g_1 \:\:=&\:\:\: 0.449   \pm 0.047_{\:\rm stat}   \pm 0.019_{\:\rm syst}    \:\:\:&=\:\:0.449   \pm 0.051, \nonumber \\
  g_2 \:\:=&\:\:\: 0.84\nb \pm 0.20_{\:\rm stat}\nb \pm 0.04_{\:\rm syst}\nb  \:\:\:&=\:\:0.84\nb \pm 0.20,  \nonumber \\
  g_3 \:\:=&\:\:\: 0.71\nb \pm 0.12_{\:\rm stat}\nb \pm 0.04_{\:\rm syst}\nb  \:\:\:&=\:\:0.71\nb \pm 0.13. \label{eq:finalresults}
\end{eqnarray}

\begin{table}[ht!]
\begin{tabular}{lcccccc}
\hline\hline
  Source   & \hspace{1ex} & $g_1$ & \hspace{1ex} & $g_2$ & \hspace{1ex} & $g_3$  \\
\hline
  NNLO terms in fits of $m_\pi$- and $a$-dependence  && 3.6\% && 2.8\% && 3.7\% \\
  Higher excited states in fits to $R_i(t)$          && 1.7\% && 2.8\% && 4.9\% \\
  Unphysical value of $m_s^{(\rm sea)}$              && 1.5\% && 1.5\% && 1.5\% \\
\\[-2ex]
  Total                                              && 4.2\% && 4.3\% && 6.3\% \\
\hline\hline
\end{tabular}
\caption{\label{tab:systerrs} Estimates of systematic uncertainties in the axial couplings $g_i$.}
\end{table}

\clearpage

\clearpage

\section{Comparison with previous results for the axial couplings}
\label{sec:comparison}

We begin this section by discussing previous lattice calculations of the heavy-meson axial-current matrix elements
and the corresponding extractions of $g_1$. A summary of results is shown in Table \ref{tab:g1comparison}. All of the
past works used an order-$a$ improved Wilson action \cite{Sheikholeslami:1985ij} for the light quarks, and
variants of the Eichten-Hill action \cite{Eichten:1989kb, DellaMorte:2003mn} for the static heavy quark.
The first lattice estimate for $g_1$ was obtained in the pioneering work of Ref.~\cite{deDivitiis:1998kj},
using a $12^3 \times 24$ lattice and quenched gauge fields, where the fermion
determinants in the path-integral weight are set to 1, which means that the
vacuum-polarization effects of the light quarks are neglected.
In Ref.~\cite{deDivitiis:1998kj}, the average of $(g_1)_{\rm eff}$ from two different
valence pion masses (760 MeV, 900 MeV) was taken as the result for $g_1$. Quenched calculations
of $g_1$ were also reported in Refs.~\cite{Abada:2003un} and \cite{Negishi:2006sc}. The results
for $g_1$ in these works were obtained by extrapolating data for $(g_1)_{\rm eff}$, at pion
masses in the range of about 550 to 850 MeV, linearly in $[m_\pi^{(\mathrm{vv})}]^2$ to $m_\pi^{(\mathrm{vv})}=0$.

\begin{table}[ht!]
\begin{tabular}{llcccll}
 \hline\hline
Reference                                        & \hspace{2ex} &  $n_f$, action   & \hspace{2ex} & $[m_{\pi}^{(\rm vv)}]^2$ (GeV$^2$)  & \hspace{2ex} &  $g_1$ \\
\hline
De Divitiis {\it et al.}, 1998 \cite{deDivitiis:1998kj}            &  &  $0$, clover  &  & 0.58 - 0.81   &  &   $0.42 \pm 0.04 \pm 0.08$                          \\
Abada {\it et al.}, 2004 \cite{Abada:2003un}                       &  &  $0$, clover  &  & 0.30 - 0.71   &  &   $0.48 \pm 0.03 \pm 0.11$                          \\
Negishi {\it et al.}, 2007 \cite{Negishi:2006sc}                   &  &  $0$, clover  &  & 0.43 - 0.72   &  &   $0.517 \pm 0.016$                                 \\
Ohki {\it et al}., 2008 \cite{Ohki:2008py}                         &  &  $2$, clover  &  & 0.24 - 1.2    &  &   $0.516 \pm 0.005 \pm 0.033 \pm 0.028 \pm 0.028 $  \\
Be\'cirevi\'c {\it et al.}, 2009 \cite{Becirevic:2009yb}           &  &  $2$, clover  &  & 0.16 - 1.2    &  &   $0.44 \pm 0.03 ^{+0.07}_{-0.00}$                  \\
Bulava {\it et al.}, 2010 \cite{Bulava:2010ej}                     &  &  $2$, clover  &  & 0.063 - 0.49  &  &   $0.51 \pm 0.02$                                   \\
\\ [-2ex]
This work                                                          &  &  $2+1$, domain wall    &  & $0.052$ - $0.12$      &  &  $0.449\pm0.047_{\:\rm stat}\pm 0.019_{\:\rm syst}$ \\
\hline\hline
\end{tabular}
\caption{\label{tab:g1comparison} Comparison of lattice QCD results for the mesonic axial coupling $g_1$. Also shown are the numbers of dynamical light-quark flavors $n_f$,
the fermion lattice action, and the range of valence pion masses used in the calculation.}
\end{table}

Since calculations without sea quarks have uncontrolled systematic errors, more
recent lattice calculations of $g_1$ have been performed with dynamical flavors, albeit only for $n_f=2$.
The first of these was done in Ref.~\cite{Ohki:2008py}, using two different lattices of sizes $12^3\times 24$
and $16^3\times 32$, and pion masses in the range 490 - 1100 MeV. Stochastic all-to-all propagators
were used to reduce the statistical uncertainties. In Ref.~\cite{Ohki:2008py}, the data for
the axial-current matrix elements $(g_1)_{\rm eff}$
was fitted using different approaches: linear in $m_\pi^2$, linear+quadratic in $m_\pi^2$, or linear+quadratic+logarithmic
in $m_\pi^2$, using the average of $g_1$ from the linear and the linear+quadratic+logarithmic fits as the final result.
A second unquenched calculation was published in Ref.~\cite{Becirevic:2009yb}, using three different lattice spacings
and pion masses in the range from 400 to 1100 MeV. In Ref.~\cite{Becirevic:2009yb}, the coupling
$g_1$ was obtained from a linear+logarithmic fit of $(g_1)_{\rm eff}$. Recently, the
axial couplings of orbitally excited heavy-light mesons were also included \cite{Becirevic:2012zz}. Another $n_f=2$ calculation of $g_1$
was reported in Ref.~\cite{Bulava:2010ej}, with three different lattice spacings and pion masses down to 250 MeV. In Ref.~\cite{Bulava:2010ej},
the result of an extrapolation of $(g_1)_{\rm eff}$ linear in $m_\pi^2$ was given as the value of $g_1$.

The coefficient of the chiral logarithm used in the fits of the axial-current matrix elements $(g_1)_{\rm eff}$
in Refs.~\cite{Ohki:2008py} and \cite{Becirevic:2009yb} was set equal to that of the strong decay $P^* \to P\:\pi$ in $SU(2)$ HH$\chi$PT \cite{Fajfer:2006hi}, because
the corresponding loop contributions for the axial-current matrix elements were not known at that time.
The NLO expression for the strong-decay amplitude in $SU(2)$ HH$\chi$PT is proportional to
\begin{equation}
 \mathcal{M}(P^* \to P\:\pi) \propto g_1 \left[ 1 - 4g_1^2\frac{m_\pi^2}{(4\pi f_\pi)^2} \log \frac{m_\pi^2}{\mu^2} + \tilde{c}\: m_\pi^2   \right]. \label{eq:strongdecaylog}
\end{equation}
We have recently derived the NLO expressions for the axial-current matrix elements in $SU(2)$, $SU(3)$, $SU(4|2)$, and $SU(6|3)$ HH$\chi$PT \cite{Detmold:2011rb}.
As discussed in Ref.~\cite{Detmold:2011rb}, the chiral expansion of the axial-current matrix elements contains an additional
tadpole loop contribution, which modifies the coefficient of the logarithm. In the $SU(2)$ case, one has
\begin{equation}
 (g_1)_{\rm eff} = g_1 \left[ 1 - (2+4g_1^2)\frac{m_\pi^2}{(4\pi f_\pi)^2} \log \frac{m_\pi^2}{\mu^2} + c\: m_\pi^2   \right].  \label{eq:g1efflog}
\end{equation}
Because $g_1\approx 0.5$, the coefficient of the logarithm in Eq.~(\ref{eq:g1efflog}) is numerically about 3 times larger than the coefficient
of the logarithm in Eq.~(\ref{eq:strongdecaylog}). The logarithm makes $(g_1)_{\rm eff}$ as a function of $m_\pi^2$ curve downward
when $m_\pi^2$ is decreased (see Fig.~\ref{fig:g1}).
The results for $g_1$ from the previous unquenched lattice calculations, which incorrectly used Eq.~(\ref{eq:strongdecaylog}) or did not include any logarithm
in the fits, would be significantly lower if the correct HH$\chi$PT formula (\ref{eq:g1efflog}) had been used instead.
We have attempted fits to the data of \cite{Ohki:2008py, Becirevic:2009yb, Bulava:2010ej} using Eq.~(\ref{eq:g1efflog}), obtaining
values of $g_1$ that are about 10 to 20\% lower than what is published in these works. Note, however, that HH$\chi$PT
is not expected to converge in the upper range of the pion masses in \cite{Becirevic:2009yb, Ohki:2008py}. For the data
used in the present work, incorrect fits linear in $m_\pi^2$ or using Eq.~(\ref{eq:strongdecaylog}) give values
for $g_1$ that are higher than the correct result, Eq.~(\ref{eq:g1finalresult}), by 12\% and 8\%, respectively.

Next, we move to the discussion of various theoretical estimates of the axial couplings $g_1$, $g_2$, $g_3$
based on approximations, models, and experimental data. A comparison of these estimates to
our QCD results is shown in Table \ref{tab:g123comparison}. The nonrelativistic
quark model (NRQM) predicts $g_1=g_A^{ud}$, $g_2=2g_A^{ud}$ and $g_3=\sqrt{2} g_A^{ud}$ \cite{Yan:1992gz},
where $g_A^{ud}=1$ is the axial coupling of the single-quark transition $u \to d$. Interestingly, if
$g_A^{ud}$ is set to 0.75, the value needed to reproduce the experimental value of the
nucleon axial charge, one obtains $g_1=0.75$, $g_2=1.5$, $g_3=1.06$, still significantly larger
than our QCD results. The predicted \emph{ratios} of the axial couplings in the NRQM are, however,
consistent with our lattice determination. The relativistic quark models of Refs.~\cite{Colangelo:1994jc}
and \cite{Becirevic:1999fr} give $g_1=1/3$ and $g_1=0.6\pm0.1$, respectively.

\begin{table}[ht!]
\begin{ruledtabular}
\begin{tabular}{lcccc}
Reference                                             &   Method                       &   $g_1$                            &  $g_2$              &  $g_3$            \\
\hline
Yan {\it et al.}, 1992 \cite{Yan:1992gz}              &  Nonrelativistic quark model   &  1                                 &  2                  &  $\sqrt{2}$       \\
Colangelo {\it et al.}, 1994 \cite{Colangelo:1994jc}  &  Relativistic quark model      &  $1/3$                             &  $\hdots$           &  $\hdots$         \\
Be\'cirevi\'c, 1999 \cite{Becirevic:1999fr}           &  Quark model with Dirac eq.    &  $0.6 \pm 0.1$                     &  $\hdots$           &  $\hdots$         \\
Guralnik {\it et al.}, 1992 \cite{Guralnik:1992dj}    &  Skyrme model                  &  $\hdots$                          &  1.6                &  1.3              \\
Colangelo {\it et al.}, 1994 \cite{Colangelo:1994es}  &  Sum rules                     &  0.15 - 0.55                       &  $\hdots$           &  $\hdots$         \\
Belyaev {\it et al.}, 1994 \cite{Belyaev:1994zk}      &  Sum rules                     &  $0.32 \pm 0.02$                   &  $\hdots$           &  $\hdots$         \\
Dosch and Narison, 1995 \cite{Dosch:1995kw}           &  Sum rules                     &  $0.15 \pm 0.03$                   &  $\hdots$           &  $\hdots$         \\
Colangelo and Fazio, 1997 \cite{Colangelo:1997rp}     &  Sum rules                     &  0.09 - 0.44                       &  $\hdots$           &  $\hdots$         \\
Pirjol and Yan, 1997 \cite{Pirjol:1997nh}             &  Sum rules                     &  $\hdots$                          &  $<\sqrt{6-g_3^2}$  &  $<\sqrt{2}$      \\
Zhu and Dai, 1998 \cite{Zhu:1998vg}                   &  Sum rules                     &  $\hdots$                          &  $1.56\pm0.30\pm0.30$  &   $0.94\pm0.06\pm0.20$  \\
Cho and Georgi, 1992 \cite{Cho:1992nt}                &  $\mathcal{B}[D^* \to D\:\pi]$, $\mathcal{B}[D^* \to D\:\gamma]$    &  $0.34 \pm 0.48$  &  $\hdots$           & $\hdots$          \\
Arnesen {\it et al.}, 2005 \cite{Arnesen:2005ez}      &  \hspace{-4ex} $\mathcal{B}[D_{(s)}^*\!\! \to \! D_{(s)}\pi]$, $\mathcal{B}[D_{(s)}^*\!\! \to \! D_{(s)}\gamma]$, $\Gamma[D^*]$ & 0.51 &  $\hdots$           &  $\hdots$         \\ 
Li {\it et al.}, 2010 \cite{Li:2010rh}                &  $\mathrm{d}\Gamma[B\to\pi\ell \nu]$  &  $<0.87$                           &  $\hdots$           & $\hdots$          \\
Cheng, 1997 \cite{Cheng:1997rp}                       &  $\Gamma[\Sigma_c^*\to\Lambda_c\:\pi]$, NRQM   & $0.70 \pm 0.12$           &  $1.40 \pm 0.24$    &  $0.99 \pm 0.17$  \\
\\ [-2ex]
This work                                             &  Lattice QCD                    &  $0.449 \pm 0.051$                 &  $0.84 \pm 0.20$    & $0.71 \pm 0.13$   \\
\end{tabular}
\end{ruledtabular}
\caption{\label{tab:g123comparison} Comparison of our lattice QCD results for the axial couplings $g_1$, $g_2$, and $g_3$ with other determinations as reported
in the literature [all results are shown in our normalization, see Eq.~(\ref{eq:conventions})]. Here, NRQM stands for nonrelativistic quark model. Where decay widths
or branching fractions are listed under ``Method'', these are experimental inputs. As discussed in the main text, the axial couplings extracted from experimental data are
defined away from the static limit in some cases.
When a reference contained multiple results for the same coupling and did not specify which one is the most reliable,
we quote here the range from the lowest result minus its uncertainty up to the highest result plus its uncertainty.}
\end{table}

Another theoretical approach for estimating the axial couplings is the large-$N_c$ limit of QCD, where $N_c$ is the number of colors. In the limit $N_c \to \infty$,
one finds that the baryonic couplings satisfy the relation \cite{Guralnik:1992dj, Jenkins:1993af}
\begin{equation}
\left.\frac{g_2}{g_3}\right|_{N_c=\infty}=\sqrt{\frac32}\approx1.22.
\end{equation}
For comparison, our lattice QCD result for this ratio is
\begin{equation}
 \frac{g_2}{g_3} = 1.19(26),
\end{equation}
and the nonrelativistic quark model predicts $g_2/g_3=\sqrt{2}\approx1.41$.

The axial couplings have also been estimated using sum rules \cite{Colangelo:1994es, Belyaev:1994zk, Dosch:1995kw, Colangelo:1997rp, Pirjol:1997nh, Zhu:1998vg},
with results as shown in Table \ref{tab:g123comparison}. For the heavy-meson coupling $g_1$, most sum rule determinations are smaller than
our lattice QCD result, and much smaller than the NRQM value. In contrast, the values of $g_2$ and $g_3$ obtained using sum rules in Ref.~\cite{Zhu:1998vg}
are larger than our lattice results.

Experimental data for various heavy-hadron decay processes has also been used to determine the axial couplings. In Ref.~\cite{Cho:1992nt}, electromagnetic interactions
were included in HH$\chi$PT, and the coupling $g_1$ was extracted from the measured branching fractions $\mathcal{B}[D^* \to D\:\pi]$ and
$\mathcal{B}[D^* \to D\:\gamma]$ at tree level, finding $g_1=0.43\pm0.61$ for $m_c=1.5$ GeV and $g_1=0.34\pm0.48$ for $m_c=1.7$ GeV. Note that these
values for $g_1$ are not defined in the static limit; they are effective values corresponding to the $D^* D \pi$ coupling. A similar calculation, which
additionally included the leading nonanalytic effects in the radiative decays, is reported in Ref.~\cite{Amundson:1992yp}. The complete $1/m_Q$ and loop corrections
in both the strong and radiative decays were included in the analysis of Ref.~\cite{Stewart:1998ke}. There, the fit to experimental data for the
branching fractions $\mathcal{B}[D_{(s)}^* \to D_{(s)}\:\pi]$ and $\mathcal{B}[D_{(s)}^* \to D_{(s)}\:\gamma]$ gave two possible
solutions for $g_1$. The fit of Ref.~\cite{Stewart:1998ke} was updated later by including experimental results for $\Gamma[D^*]$ \cite{Anastassov:2001cw},
leading to $g_1 \simeq 0.51$ \cite{Arnesen:2005ez}, where (unlike in Ref.~\cite{Stewart:1998ke}) $g_1$ is defined in the static limit.

Recently, $g_1$ was also extracted from data for the $B\to\pi\ell \nu$ form factors, giving results for $g_1$ in the range from $0.02 \pm 0.32$ up to $0.73^{+0.14}_{-0.12}$
depending on the parametrizations of the form factor shape \cite{Li:2010rh}.
The measured widths of the baryonic decays $\Sigma_c^*\to\Lambda_c\:\pi$
were used in Refs.~\cite{Pirjol:1997nh} and \cite{Cheng:1997rp} to estimate $g_3$, with the result $0.99 \pm 0.17$. The NRQM relations
then give $g_1=g_3/\sqrt{2}=0.70 \pm 0.12$ and $g_2 = g_3/2 = 1.40 \pm 0.24$ \cite{Cheng:1997rp}. However, as discussed in Sec.~\ref{sec:decays},
the value of ``$g_3$'' extracted directly from $\Gamma[\Sigma_c^*\to\Lambda_c\:\pi]$ 
should really be considered as an effective value of the decay coupling constant at $m_Q=m_c$, deviating from the static-limit
axial coupling by corrections of order $\Lambda_{\rm QCD}/m_c \sim$ 30\%.

\section{Calculation of decay widths}
\label{sec:decays}

In this section, we use our lattice QCD results for the axial couplings $g_2$ and $g_3$ to calculate various decay widths
of heavy baryons. At leading order in the chiral expansion, the widths for the strong decays $S\to T\:\pi$ are
\begin{equation}
  \Gamma[S \to T\: \pi]=c_{\rm f}^2\:\frac{1}{6\pi f_\pi^2}
  \left(g_3+\frac{\kappa_J}{m_Q}\right)^2\frac{M_T}{M_S} \: |\mathbf{p}_\pi|^3,
  \label{eq:decaywidth}
\end{equation}
where $S$ and $T$ now denote physical $s_l=1$ and $s_l=0$ heavy-baryon states such as $\Sigma_b$
and $\Lambda_b$, $|\mathbf{p}_\pi|$ is the magnitude of the pion momentum in the $S$ rest frame,
\begin{equation}
|\mathbf{p}_\pi| = \frac{\sqrt{[(M_S-M_T)^2-m_\pi^2][(M_S+M_T)^2-m_\pi^2]}}{2 M_S},
\end{equation}
and $c_{\rm f}$ is a flavor factor \cite{Ivanov:1999bk},
\begin{equation}
 c_{\rm f} = \left\{
\begin{array}{ll}
1          &{\rm for}\hspace{2ex} \Sigma_Q^{(*)} \to \Lambda_Q \:\pi^\pm, \\
1          &{\rm for}\hspace{2ex} \Sigma_Q^{(*)} \to \Lambda_Q \:\pi^0, \\
1/\sqrt{2} &{\rm for}\hspace{2ex} \Xi_Q^{\prime(*)} \to \Xi_Q \: \pi^\pm, \\
1/2        &{\rm for}\hspace{2ex} \Xi_Q^{\prime(*)} \to \Xi_Q \: \pi^0.
\end{array}\right.
\end{equation}
The $m_Q=\infty$ expression for $\Gamma$ can be found for example in \cite{Pirjol:1997nh}.
In Eq.~(\ref{eq:decaywidth}), we included the term $\kappa_J/m_Q$
to account for the first-order corrections for a finite heavy-quark mass. The parameters $\kappa_J$
are related to the additional couplings in the order-$1/m_Q$ HH$\chi$PT Lagrangian \cite{Cheng:1993gc}.
Terms suppressed by $(m_\pi/\Lambda_\chi)^2$ and $(\Lambda_{\rm QCD}/m_Q)^2$, which are omitted from
(\ref{eq:decaywidth}), lead to small systematic uncertainties in $\Gamma$.

\begin{figure}[ht!]
\includegraphics[width=0.5\linewidth]{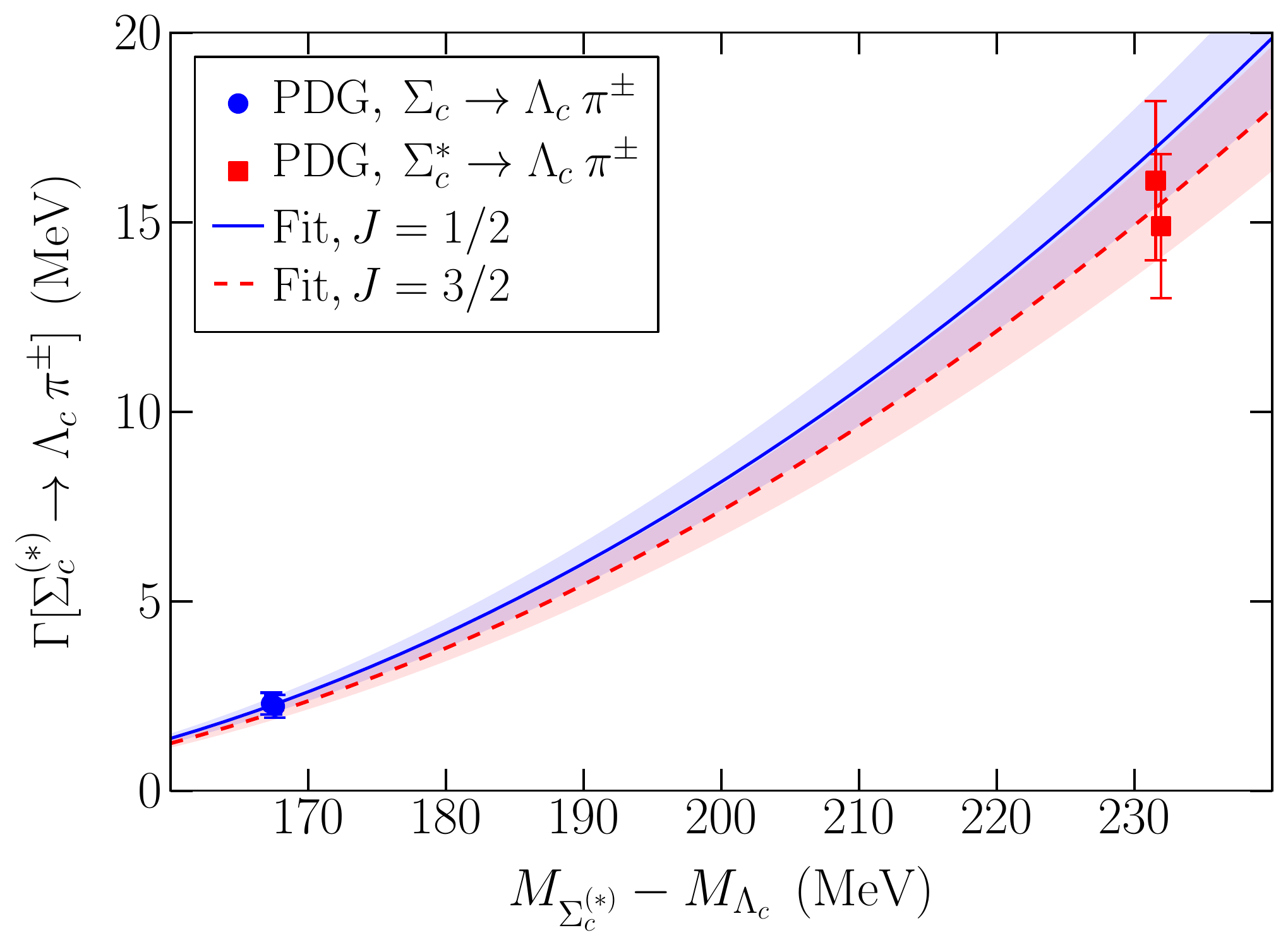}
  \caption{Experimental data for $\Gamma[\Sigma_c^{(*)} \to \Lambda_c\:\pi^{\pm}]$ from Ref.~\protect{\cite{Nakamura:2010zzi}},
along with fits using Eq.~(\ref{eq:decaywidth}), for $J=1/2$ (solid curve) and $J=3/2$ (dashed curve).}
  \label{fig:Sigmacwidthfit}
\end{figure}

To determine $\kappa_{1/2}$ and $\kappa_{3/2}$, we performed fits of experimental data \cite{Nakamura:2010zzi}
for the widths of the $\Sigma_c^{++}$, $\Sigma_c^0$ ($J=1/2$)
and the $\Sigma_c^{*++}$, $\Sigma_c^{*0}$ ($J=3/2$) using (\ref{eq:decaywidth}),
where we constrained $g_3$ to our lattice QCD result (\ref{eq:finalresults}) and set $m_Q=\frac12 M_{J/\psi}$. These fits
are shown in Fig.~\ref{fig:Sigmacwidthfit} and gave the results
\begin{eqnarray}
\nonumber \kappa_{1/2}&=&0.55(21)\:\:{\rm GeV},\:\:\:\:{\rm Cov}(\kappa_{1/2},\:g_3) = -0.025 \:\:{\rm GeV}, \\
\kappa_{3/2}&=&0.47(21)\:\:{\rm GeV},\:\:\:\:{\rm Cov}(\kappa_{3/2},\:g_3) = -0.025 \:\:{\rm GeV}. \label{eq:kappa}
\end{eqnarray}

\begin{figure}[ht!]
\includegraphics[width=0.5\linewidth]{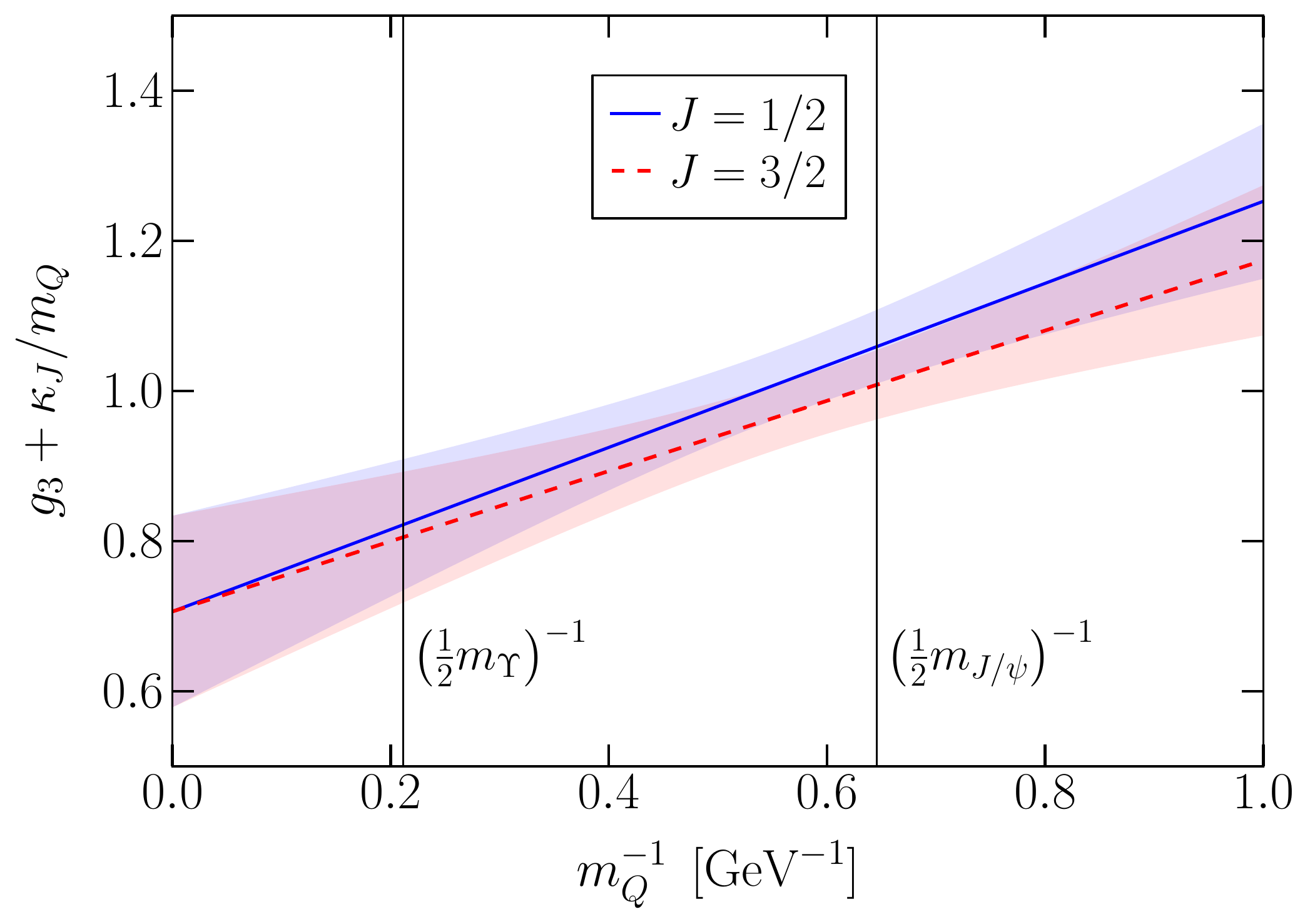}
  \caption{Value and uncertainty of the quantity $(g_3+\kappa_J/m_Q)$, which enters in the strong-decay width (\ref{eq:decaywidth}),
as a function of the inverse heavy-quark mass $m_Q^{-1}$, for $J=1/2$ (solid curve) and $J=3/2$ (dashed curve).
At $m_Q^{-1}=0$ the function is equal to $g_3$, which is given by our lattice QCD result (\ref{eq:finalresults}). The vertical lines
indicate our choices for the inverse bottom and charm quark masses.}
  \label{fig:gpluskappaoverm}
\end{figure}

The fit parameters $\kappa_J$ are correlated with $g_3$, and therefore we also show the covariances in Eq.~(\ref{eq:kappa}).
The value of the sum $g_3+\frac{\kappa_J}{m_Q}$ in Eq.~(\ref{eq:decaywidth}) is plotted as a function of $1/m_Q$
in Fig.~\ref{fig:gpluskappaoverm}. For $m_Q=\frac12 M_{J/\psi}$, the values
of $g_3+\frac{\kappa_J}{m_Q}$ are determined dominantly by the experimental input used to fit $\kappa_J$:
\begin{eqnarray}
\nonumber g_3+\frac{\kappa_{1/2}}{\frac12 m_{J/\psi}} &=& 1.059(49), \\
g_3+\frac{\kappa_{3/2}}{\frac12 m_{J/\psi}} &=& 1.008(46).
\end{eqnarray}
Using the masses of the $\Xi_c^{*}$ and $\Xi_c$ baryons from Ref.~\cite{Nakamura:2010zzi},
we obtain predictions for $\Gamma[\Xi_c^{*+} \to \Xi_c^+\pi^0,\: \Xi_c^0\,\pi^+]$
and $\Gamma[\Xi_c^{*0} \to \Xi_c^0\pi^0,\: \Xi_c^+\,\pi^-]$ as shown in Table \ref{tab:strongdecays}. There,
we also show other predictions from the literature, as well as upper limits from experiments
\cite{Avery:1995ps, Gibbons:1996yv}. Our results for $\Gamma[\Xi_c^{*+}]$
and $\Gamma[\Xi_c^{*0}]$ are compatible with these limits.

\begin{figure}[ht!]
\includegraphics[width=0.5\linewidth]{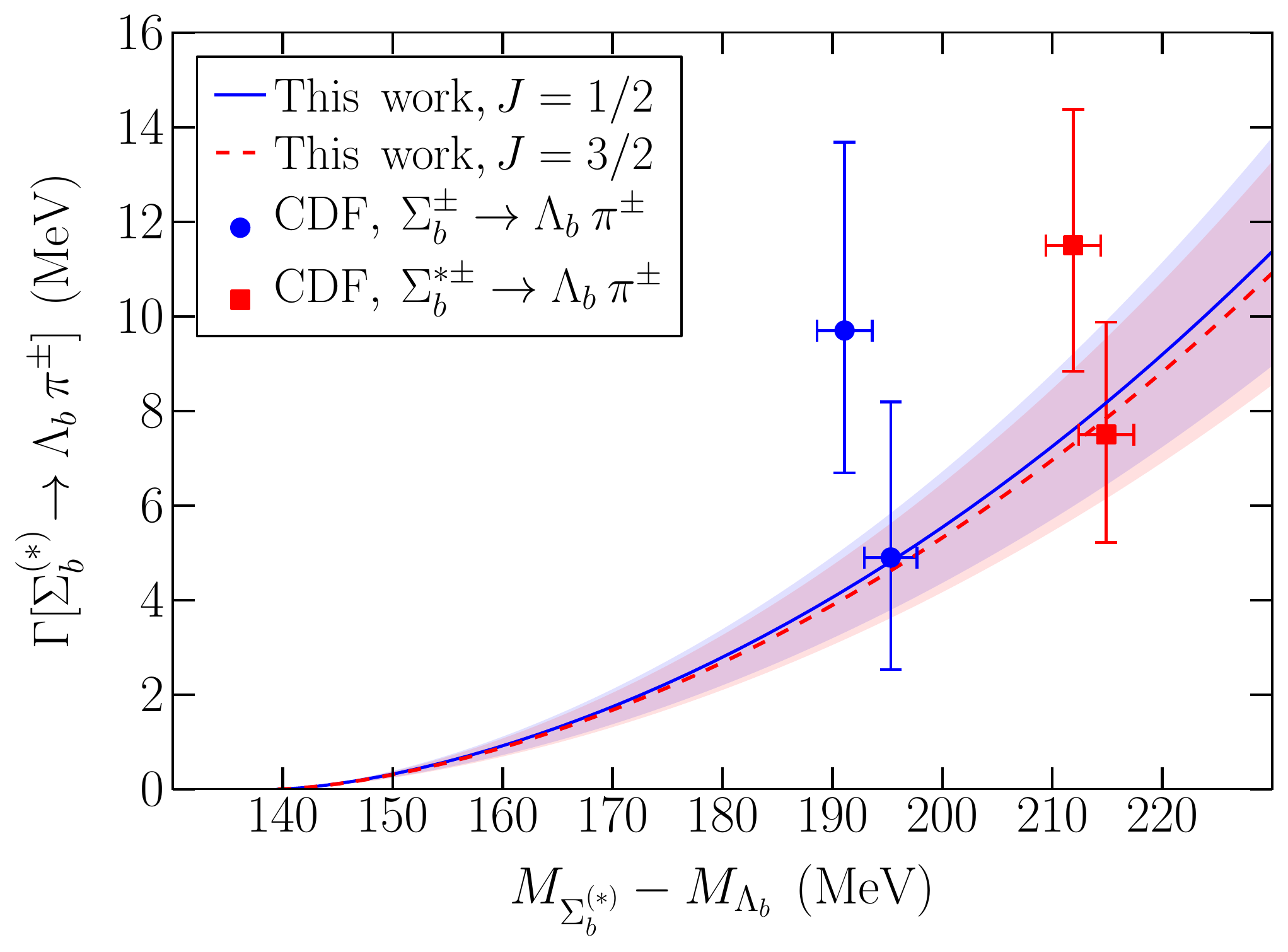}
  \caption{Widths of the decays $\Sigma_b^{(*)\pm}\to \Lambda_b\:\pi^\pm$ as functions of the $\Sigma_b^{(*)}-\Lambda_b$ mass difference.
The curves (solid: $\Sigma_b$, dashed: $\Sigma_b^*$) and shaded regions show our predictions and their uncertainties.
The experimental data points are from CDF \protect{\cite{CDF:2011ac}}.}
  \label{fig:Sigmabwidth}
\end{figure}

We can also make predictions for the radiative decay $\Xi_c^{*0} \rightarrow \Xi_c^0\: \gamma$, which
is forbidden at tree level but can be mediated by loops because of flavor-$SU(3)$ breaking. Using HH$\chi$PT, it
has been shown that the branching fraction of this decay is related
to the axial coupling $g_2$ as follows \cite{Lu:1995te}:
\begin{equation}
 \mathcal{B}[ \Xi_c^{*0} \rightarrow \Xi_c^0 \gamma ] = (1.0\pm0.3) \times 10^{-3}\:g_2^2.
\end{equation}
Combining this with our lattice QCD result for $g_2$, Eq.~(\ref{eq:finalresults}),
and our calculated strong-decay width $\Gamma[\Xi_c^{*0} \to \Xi_c^0\pi^0,\: \Xi_c^+\,\pi^-]=2.78(29)$ MeV,
we obtain
\begin{eqnarray}
\nonumber \mathcal{B}[ \Xi_c^{*0} \rightarrow \Xi_c^0 \gamma ] &=& (7 \pm 4) \times 10^{-4}, \\
\Gamma[ \Xi_c^{*0} \rightarrow \Xi_c^0 \gamma ] &=& (2.0\pm1.1)\:\:{\rm k\hspace{0.1ex}eV}.
\end{eqnarray}
Next, we discuss the strong decays of bottom baryons. To calculate these widths, we evaluated
(\ref{eq:decaywidth}) for $m_Q=\frac12 M_\Upsilon$. In this case the values of $g_3+\frac{\kappa_J}{m_Q}$
are determined dominantly by the lattice result (\ref{eq:finalresults}) for $g_3$:
\begin{eqnarray}
\nonumber g_3+\frac{\kappa_{1/2}}{\frac12 m_{\Upsilon}} &=& 0.822(87), \\
g_3+\frac{\kappa_{3/2}}{\frac12 m_{\Upsilon}} &=& 0.805(87). \label{eq:kappaUpsilon}
\end{eqnarray}
Our calculated widths $\Gamma[\Sigma_b^{(*)}\! \to \Lambda_b\:\pi^\pm]$ as functions of the
$\Sigma_b^{(*)}-\Lambda_b$ mass difference are shown as the curves in Fig.~\ref{fig:Sigmabwidth}.
Using the experimental values of the baryon masses \cite{CDF:2011ac, Nakamura:2010zzi},
we obtain the results for $\Gamma[\Sigma_b^{(*)}\! \to \Lambda_b\:\pi^\pm]$ shown in Table \ref{tab:strongdecays},
in agreement with the widths measured by the CDF collaboration \cite{CDF:2011ac}.

\begin{table}[ht!]
\begin{ruledtabular}
\begin{tabular}{cccccccccc}
Hadron   &  Ref.~\cite{Pirjol:1997nh} & Ref.~\cite{Tawfiq:1998nk} & Ref.~\cite{Ivanov:1999bk} & Ref.~\cite{Albertus:2005zy,Hernandez:2011tx} &  Ref.~\cite{Cheng:2006dk} & Ref.~\cite{Hwang:2006df} & Ref.~\cite{Chen:2007xf}  &  This work  &  Experiment \\
\hline
$\Sigma_b^+$       & $\hdots$  & $\hdots$ & $\hdots$   & 6.0        & $\hdots$ &  4.35    & 3.5  & $4.2(1.0)$       & $9.7^{+3.8+1.2}_{-2.8-1.1}$  \cite{CDF:2011ac}   \\
$\Sigma_b^-$       & $\hdots$  & $\hdots$ & $\hdots$   & 7.7        & $\hdots$ &  5.77    & 4.7  & $4.8(1.1)$       & $4.9^{+3.1}_{-2.1}\pm1.1$  \cite{CDF:2011ac}   \\
$\Sigma_b^{*+}$    & $\hdots$  & $\hdots$ & $\hdots$   & 11.0       & $\hdots$ &  8.50    & 7.5  & $7.3(1.6)$       & $11.5^{+2.7+1.0}_{-2.2-1.5}$ \cite{CDF:2011ac}   \\
$\Sigma_b^{*-}$    & $\hdots$  & $\hdots$ & $\hdots$   & 13.2       & $\hdots$ &  10.44   & 9.2  & $7.8(1.8)$       & $7.5^{+2.2+0.9}_{-1.8-1.4}$  \cite{CDF:2011ac}   \\
$\Xi_b^{*0}$       & $\hdots$  & $\hdots$ & $\hdots$   & $\hdots$   & $\hdots$ & $\hdots$ & 0.85 & $0.51(16)$       & $2.1\pm1.7$ \cite{Chatrchyan:2012ni}                    \\
$\Xi_c^{*+}$       & 1.2 - 4.1 & 1.81     & $3.04(37)$ & $3.18(10)$ & $2.7(2)$ & $\hdots$ & 1.13 & $2.44(26)$       & $<3.1$ (CL=90\%)  \cite{Gibbons:1996yv}           \\
$\Xi_c^{*0}$       & 1.2 - 4.0 & 1.88     & $3.12(33)$ & $3.03(10)$ & $2.8(2)$ & $\hdots$ & 1.08 & $2.78(29)$       & $<5.5$ (CL=90\%)  \cite{Avery:1995ps}             \\
\end{tabular}
\end{ruledtabular}
\caption{\label{tab:strongdecays}Results in MeV for the total strong-decay widths of charm and bottom baryons.}
\end{table}

In our previous work \cite{Detmold:2011bp} we predicted that the widths of the 
$\Xi_b^\prime$ and $\Xi_b^*$ are less than 1.1 and 2.8 MeV, respectively. Very recently, the CMS collaboration
has observed the $\Xi_b^{*0}$, finding a width of $2.1\pm 1.7$ MeV \cite{Chatrchyan:2012ni}\footnote{Without a spin identification,
there is a small possibility that the state observed by CMS is the $\Xi_b^{\prime 0}$ instead. We do not consider this further.}.
The mass difference to the $\Xi_b^-$ was measured to be
\begin{equation}
M_{\Xi_b^{*0}}-M_{\Xi_b^-} = 154.41 \pm 0.79\:\:{\rm MeV}. \label{eq:XibstarzeroXibminus}
\end{equation}
The $\Xi_b^{*0}$ can decay into $\Xi_b^- \pi^+$ (seen by CMS) and into $\Xi_b^0\: \pi^0$. Taking $M_{\Xi_b^-}-M_{\Xi_b^0}=3.1\pm5.6\pm1.3$ MeV
from the CDF measurement reported in Ref.~\cite{Aaltonen:2011wd}, we have
\begin{eqnarray}
M_{\Xi_b^{*0}}-M_{\Xi_b^0} &=& 157.5 \pm 5.8\:\:{\rm MeV}. \label{eq:XibstarzeroXibzero}
\end{eqnarray}
Using the results (\ref{eq:XibstarzeroXibminus}) and (\ref{eq:XibstarzeroXibzero}),
we can update our calculation of the $\Xi_b^{*0}$ width and find
\begin{equation}
\Gamma[\Xi_b^{*0} \rightarrow \Xi_b^- \pi^+,\: \Xi_b^0\:\pi^0] =  0.51 \pm 0.16\:\:{\rm MeV}.
\end{equation}
Given the observed mass difference (\ref{eq:XibstarzeroXibminus}), and assuming that
$M_{\Xi_b^{*}}-M_{\Xi_b^{\prime}}\approx M_{\Sigma_b^{\ast}} - M_{\Sigma_b} = 21 \pm 2$ MeV
\cite{Aaltonen:2007rw}, it is likely that the decay $\Xi_b^{\prime0} \to \Xi_b^- \pi^+$ is kinematically forbidden.

\section{Conclusions}
\label{sec:conclusions}

The chiral dynamics of mesons and baryons containing a heavy quark is controlled at leading order by three axial couplings $g_1$, $g_2$, and $g_3$.
Knowledge of the values of these couplings is an essential ingredient for precision QCD calculations in flavor physics. In this paper, we
have discussed in detail the first complete lattice QCD determination of $g_1$, $g_2$, and $g_3$. We have extracted
the axial couplings by fitting numerical data for matrix elements of the axial current using the quark-mass and volume dependence
calculated in $SU(4|2)$ heavy-hadron chiral perturbation theory. Our final results are
\begin{eqnarray}
  g_1 \:\:&=&\:\:\: 0.449   \pm 0.047_{\:\rm stat}   \pm 0.019_{\:\rm syst}, \nonumber \\
  g_2 \:\:&=&\:\:\: 0.84\nb \pm 0.20_{\:\rm stat}\nb \pm 0.04_{\:\rm syst}, \nonumber \\
  g_3 \:\:&=&\:\:\: 0.71\nb \pm 0.12_{\:\rm stat}\nb \pm 0.04_{\:\rm syst}. \label{eq:finalresults2}
\end{eqnarray}
The systematic uncertainties in (\ref{eq:finalresults2}) are very small, because our analysis is based on
data at low pion masses, with a large volume, and at two different lattice spacings. We have also carefully
removed the excited-state contamination in the matrix elements by extrapolating the ratios of correlation
functions to infinite source-sink separation.

Previous lattice calculations of heavy-hadron axial couplings had only considered the mesonic coupling $g_1$. The early calculations of $g_1$
did not include dynamical quarks and hence are contaminated by uncontrolled systematic errors.
The $n_f=2$ calculations typically used large quark masses and the fits to the quark-mass-dependence were performed either linearly
in $m_\pi^2$ or with an incorrect coefficient of the chiral logarithm. Had the correct coefficient
been used, significantly lower values of $g_1$ would have been obtained in these previous studies.

For the range of pion masses considered in our work ($230 \hspace{1ex}{\rm MeV} \lesssim m_\pi \lesssim 350$ MeV), the chiral expansion of
the axial-current matrix elements between heavy-light hadron states is found to be well-behaved. The
next-to-leading-order contributions are small compared to the leading-order contributions, and NNLO contributions
are negligible.
The rapid convergence of the chiral expansion is also a consequence of the smallness of the static-light axial couplings (\ref{eq:finalresults2}).
It is interesting to compare the chiral dynamics of hadrons containing a heavy quark with that of
light baryons. Being particularly light, the interactions of virtual pions (and other pseudo-Goldstone bosons) 
produce significant contributions to many properties of baryons, and generically these effects scale quadratically 
with the strength with which a given baryon sources pions. This, in turn, is determined by the relevant axial coupling, 
$g_{1,2,3}$ in the case of heavy hadrons, and $g_A\approx 1.26$, $|g_{N\Delta}|\sim1.6$ and $g_{\Delta\Delta}\sim-1.9$  
in the case of light baryons \cite{Jenkins:1991ne, Jenkins:1990jv}.
From the numerical values of these couplings, it is apparent that chiral dynamics is more perturbative
for heavy-light hadrons than that for light baryons.

Our results for the heavy-light axial couplings, Eq.~(\ref{eq:finalresults2}), are significantly smaller than the values
one obtains in the nonrelativistic quark model, $g_1=g_A^{ud}$, $g_2=2g_A^{ud}$ and $g_3=\sqrt{2} g_A^{ud}$,
where $g_A^{ud}=1$ is the axial coupling of the single-quark transition $u \to d$. Even if
$g_A^{ud}$ is set to 0.75, as needed to reproduce the experimental value of the
nucleon axial charge, the corresponding quark-model values of $g_{1,2,3}$ are still significantly smaller than
the results (\ref{eq:finalresults2}) from first-principles lattice QCD.

We have used our results for $g_2$ and $g_3$ to calculate strong and radiative decay widths of charm
and bottom baryons. For the strong decays, we have taken into account the order-$1/m_Q$ corrections, which we have constrained
by combining experimental data for charmed baryon decay rates with our lattice determination of $g_3$.
We found that the $1/m_Q$ corrections are significant (their 
effect on the amplitudes for $\Sigma_{Q}^{(*)}\to\Lambda_Q\:\pi$ decays is about 40\% at $m_Q=m_c$ and about 13\%
at $m_Q=m_b$). As a consequence, the coupling $g_3$ cannot be reliably extracted from experimental data for 
charmed baryon decays alone, and our lattice calculation in the static limit is crucial to
calculate the widths of bottom baryons. Our results for the widths of the $\Sigma_b^{(*)}$ baryons
are in agreement with recent measurements at Fermilab.

Our determination of the axial couplings can also improve
the precision of future lattice QCD calculations of other heavy-hadron properties such as masses,
decay constants, and form factors, because
the axial couplings control the dependence of these properties on the light-quark masses. Therefore, the calculation
of the axial couplings from first principles also has an impact on searches for beyond-the-Standard-Model
physics at the LHC and the planned SuperB experiment. Importantly, our results include the baryonic couplings
$g_2$ and $g_3$. Heavy baryons may offer additional opportunities for probing the structure of new physics as a consequence
of the different spin quantum numbers.

\section{Acknowledgments}

We would like to thank Hai-Yang Cheng, Kostas Orginos, Martin Savage, Brian Tiburzi, Andr\'e Walker-Loud, and Matt Wingate
for helpful discussions. We are indebted to the RBC and UKQCD collaborations for access to the
gauge field configurations used in this work and to Robert Edwards and Balint Jo\'o
for the development of the {\tt chroma} library \cite{Edwards:2004sx} with which some of the
calculations were performed. The work of WD is supported in part by
Jefferson Science Associates, LLC under U.S.~Department of Energy Contract
No.~DE-AC05-06OR-23177 and by the Jeffress Memorial Trust, J-968. WD
and SM were supported by DOE Outstanding Junior Investigator Award
DE-S{C0}0{0-17}84 and DOE Grant No.~DE-FG02-04ER41302.
CJDL is supported by NSC Taiwan Grant No.~99-2112-M-009-004-MY3.
We acknowledge the hospitality of Academia Sinica Taipei, National Chiao-Tung University, National Centre
for Theoretical Sciences Taiwan, The College of William and Mary, and Thomas Jefferson 
National Accelerator Facility. This research
made use of computational resources provided by NERSC and the NSF Teragrid (NCSA, TACC and NICS).

\clearpage

\appendix

\section{Plots of raw data}
\label{sec:rawdata}

\begin{figure*}[ht!]
  \includegraphics[width=0.44\linewidth]{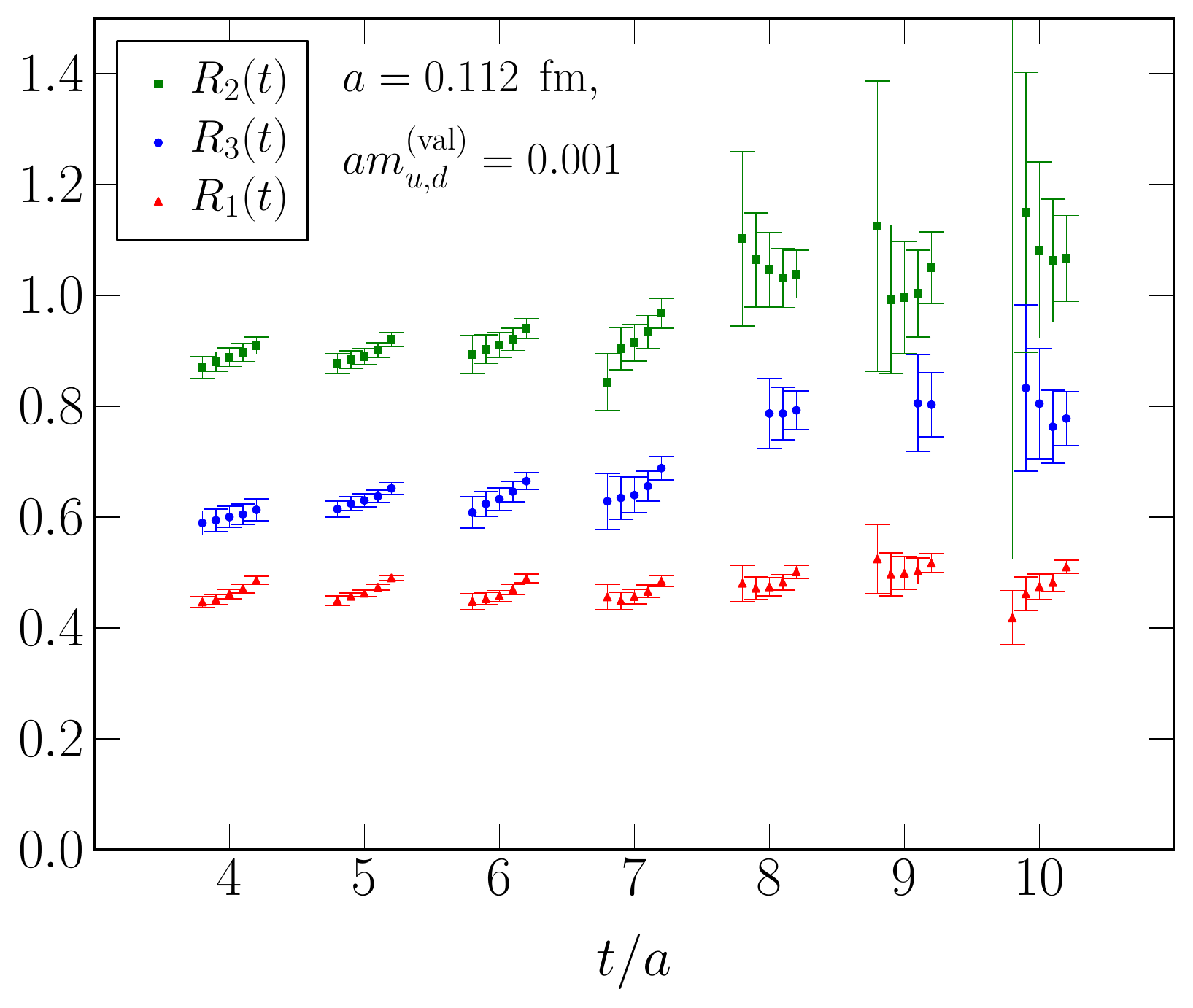}  \hfill \includegraphics[width=0.44\linewidth]{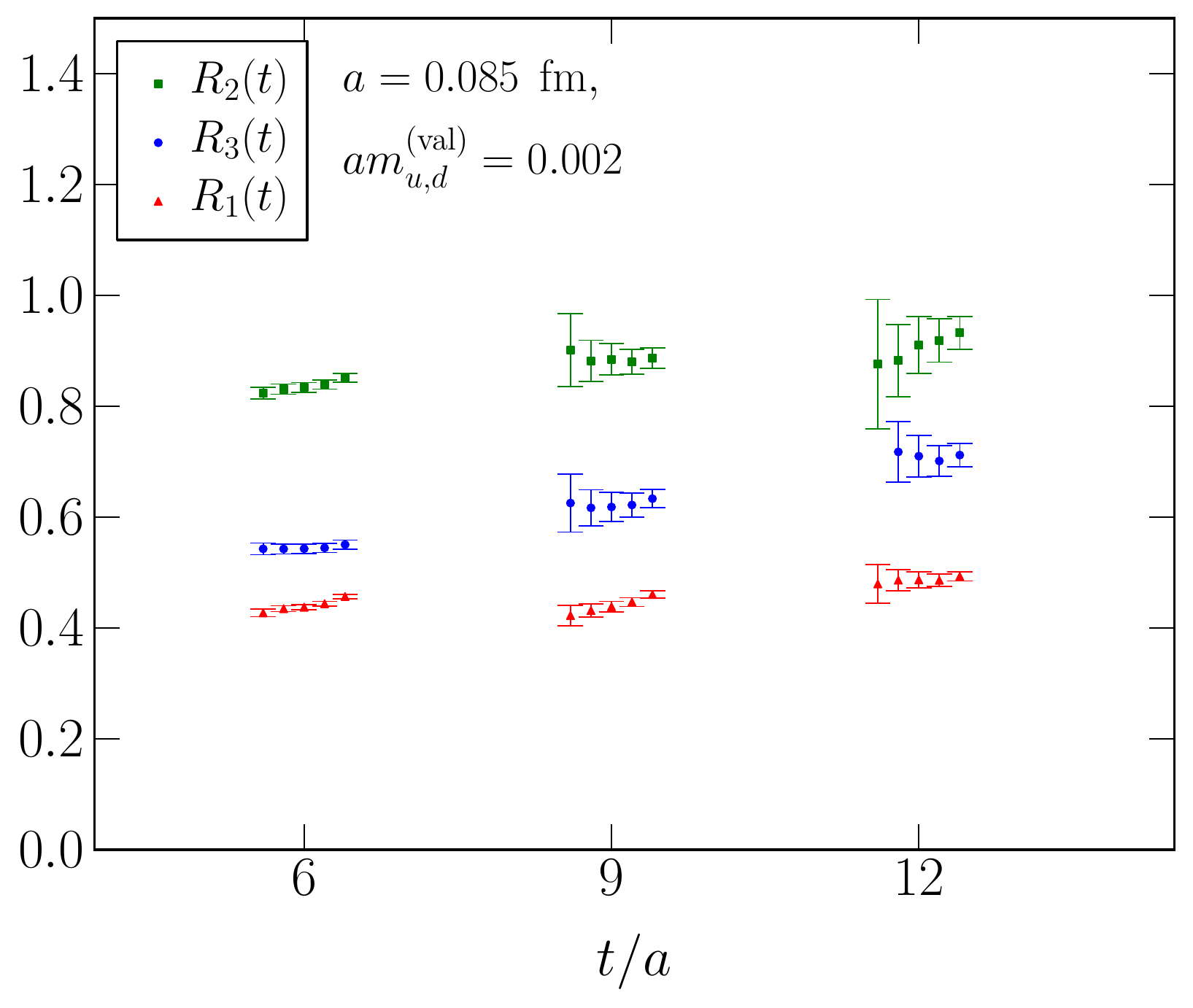}
  \includegraphics[width=0.44\linewidth]{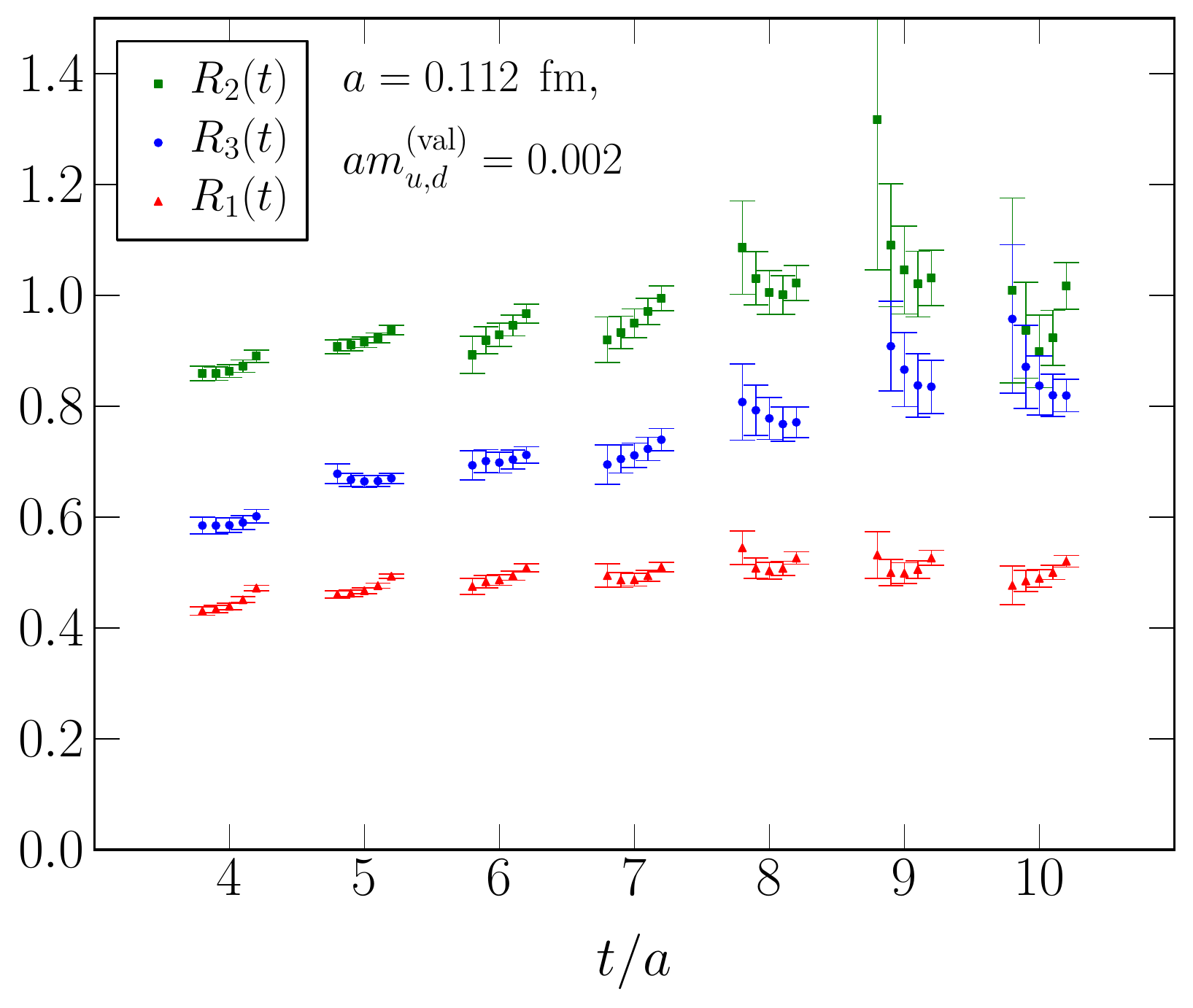}  \hfill \includegraphics[width=0.44\linewidth]{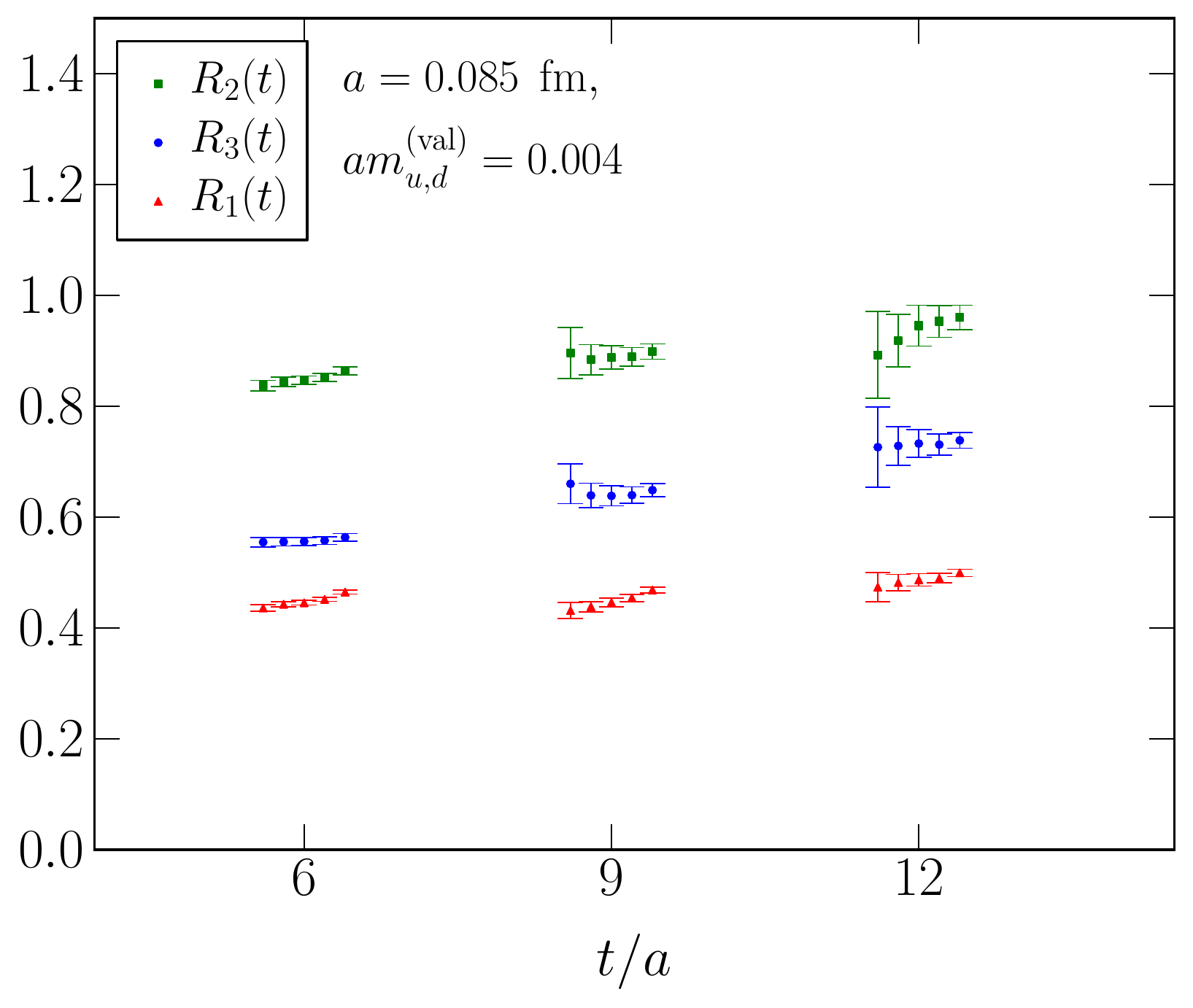}
  \includegraphics[width=0.44\linewidth]{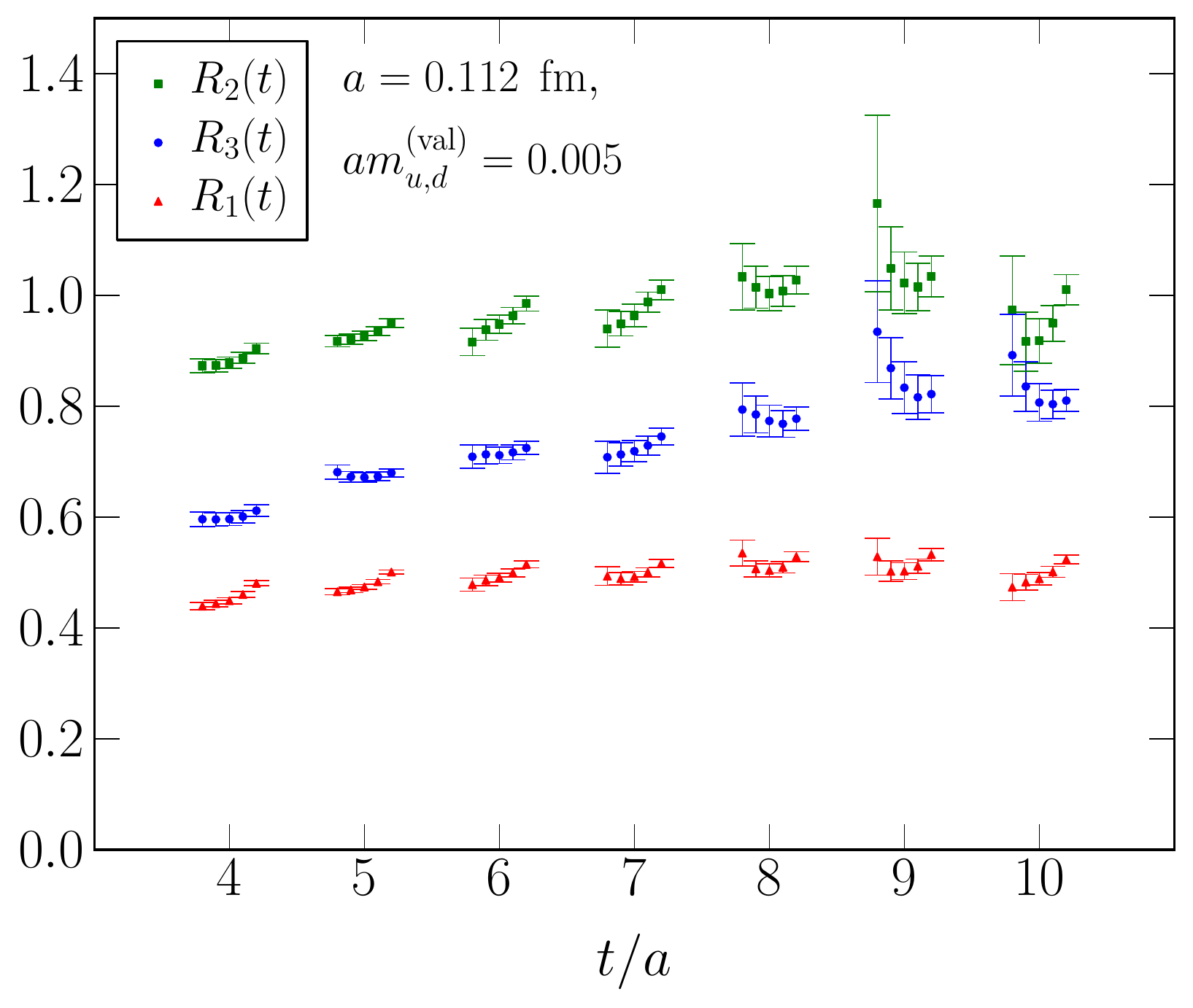}  \hfill \includegraphics[width=0.44\linewidth]{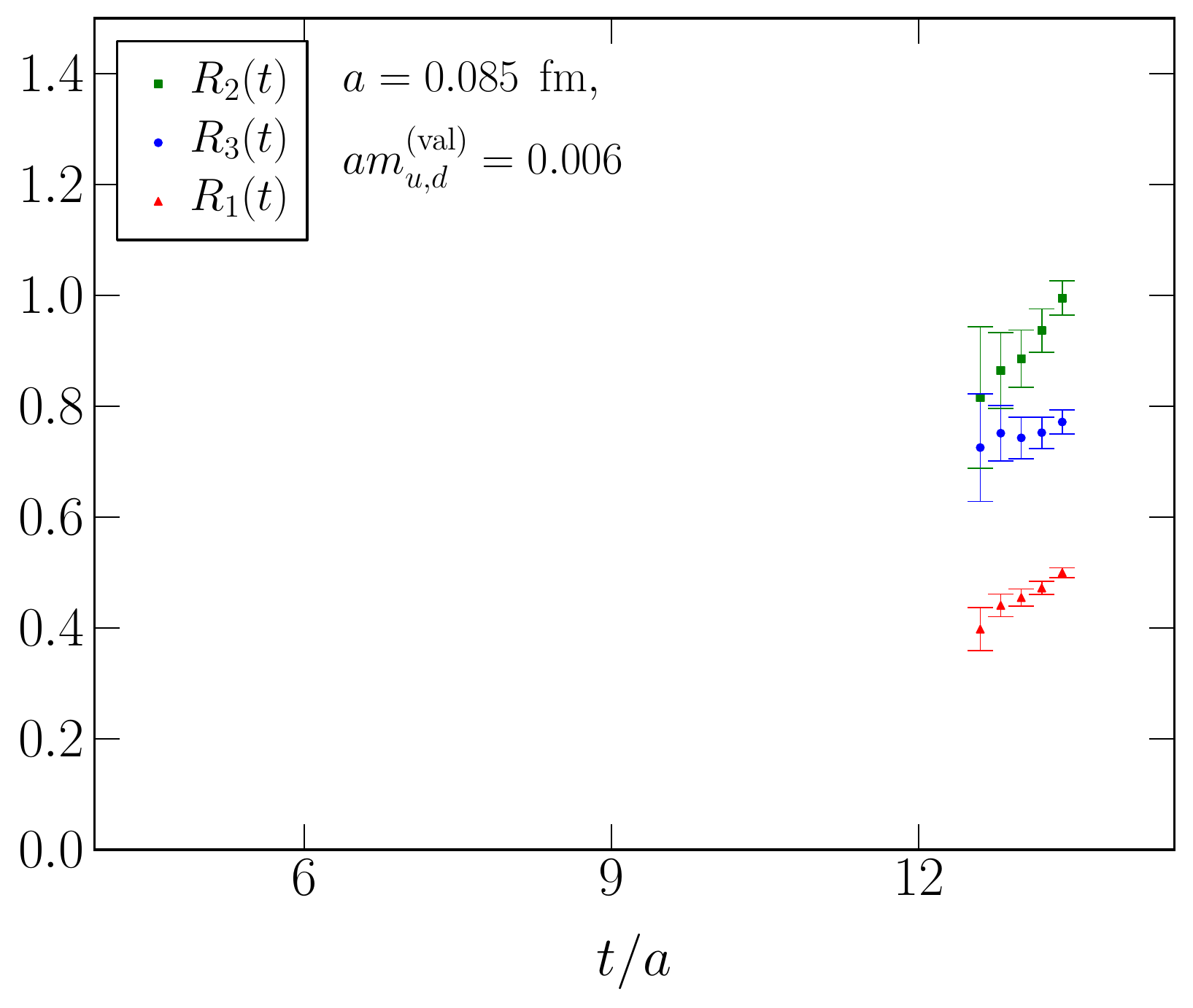}
\caption{\label{fig:rawdatag2g3} Summary of all data points for $R_1(t)$, $R_2(t)$ and $R_3(t)$.
At each value of $t/a$, results from up to five different values of $n_{\rm HYP}$ are shown (from left to right: $n_{\rm HYP}=1,2,3,5,10$; points offset horizontally for legibility;
in some cases there are no results for $R_3$ for the lowest values of $n_{\rm HYP}$, because the statistical fluctuations were too large to calculate the square root of the double ratio).
In physical units, the range of the horizontal axis in all plots is from $t=0.336$ fm to $t=1.23$ fm.}
\end{figure*}

\clearpage

\section{Comparison of standard ratio method and summation method}
\label{sec:summationmethod}

To extract the effective axial couplings from the ratios $R_i(t, \:t')$ defined in equations (\ref{eq:R1}), (\ref{eq:R2}), and (\ref{eq:R3}),
we defined $R_i(t)$ to be the average of $R_i(t,\: t')$ over a symmetric range of $t'$ values around $t/2$ in a region where there was no discernible $t'$ dependence, which essentially amounts to using
\begin{equation}
 R_i(t, \:t/2). \label{eq:standardratio2}
\end{equation}
An alternative approach for extracting $g_{\rm eff}$ is the summation method \cite{Maiani:1987by, Gusken:1989ad, Capitani:2010sg, Bulava:2010ej, Bulava:2011yz}. In the following, we only consider
the case of the simple ratios (\ref{eq:R1}) and (\ref{eq:R2}) for degenerate spectra. One defines the summed ratio $S_i(t)$ by summing $R_i(t,\: t')$ over all values of $t'$,
\begin{equation}
 S_i(t)=\:a\sum_{t'=0}^t R_i(t,\: t').  \label{eq:Si}
\end{equation}
For large $t$, one expects \cite{Maiani:1987by, Gusken:1989ad}
\begin{equation}
S_i(t) \approx c_i + (g_i)_{\rm eff}\: t,   \label{eq:Silarget}
\end{equation}
with some constant $c_i$. Thus, the coupling $(g_i)_{\rm eff}$ can be extracted by taking the derivative \cite{Bulava:2010ej, Bulava:2011yz},
\begin{equation}
 R_i^{\rm sum}(t)=\frac{\mathrm{d}}{\mathrm{d} t}  S_i(t),  \label{eq:summedratio}
\end{equation}
which is approximated by a finite difference on the lattice.
Assuming that there is a nonvanishing off-diagonal matrix element of the axial current between the ground-state hadron and an excited state with an energy gap $\delta$
(for our data, contamination from off-diagonal matrix elements actually appears to be very small, as discussed in Secs.~\ref{sec:resultsratios} and \ref{sec:sourcesinkextrap}),
one expects that the systematic uncertainties of (\ref{eq:standardratio2}) and (\ref{eq:summedratio}) due to this excited state are of order \cite{Bulava:2011yz}
\begin{eqnarray}
\nonumber R_i(t) - (g_i)_{\rm eff} &=& O(e^{-\frac12 \delta_i \:t}), \\
 R_i^{\rm sum}(t) - (g_i)_{\rm eff} &=& O(t\:e^{-\delta_i \:t})
\end{eqnarray}
[see Eq. (\ref{eq:R1simplespectral}) for the spectral decomposition of $R_i(t)$]. Thus, the excited-state contamination in $R_i^{\rm summed}(t)$ decays effectively with twice the energy gap relevant for $(g_i)_{\rm eff}(t)$,
but at the cost of an additional factor of $t$ in front of the exponential, which may be important at intermediate values of $t$.

\begin{figure*}[ht!]
 \includegraphics[width=0.47\linewidth]{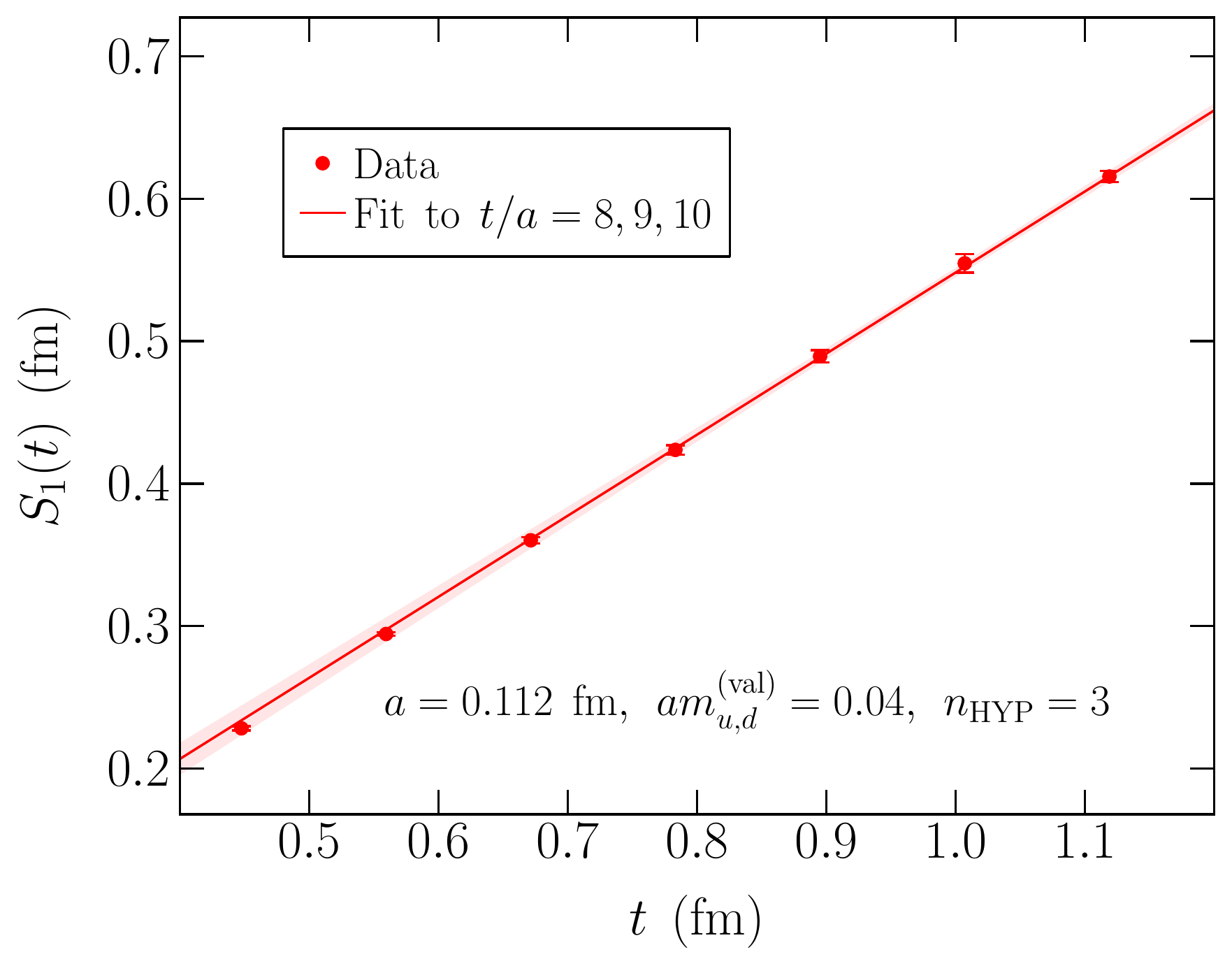}  \hfill \includegraphics[width=0.47\linewidth]{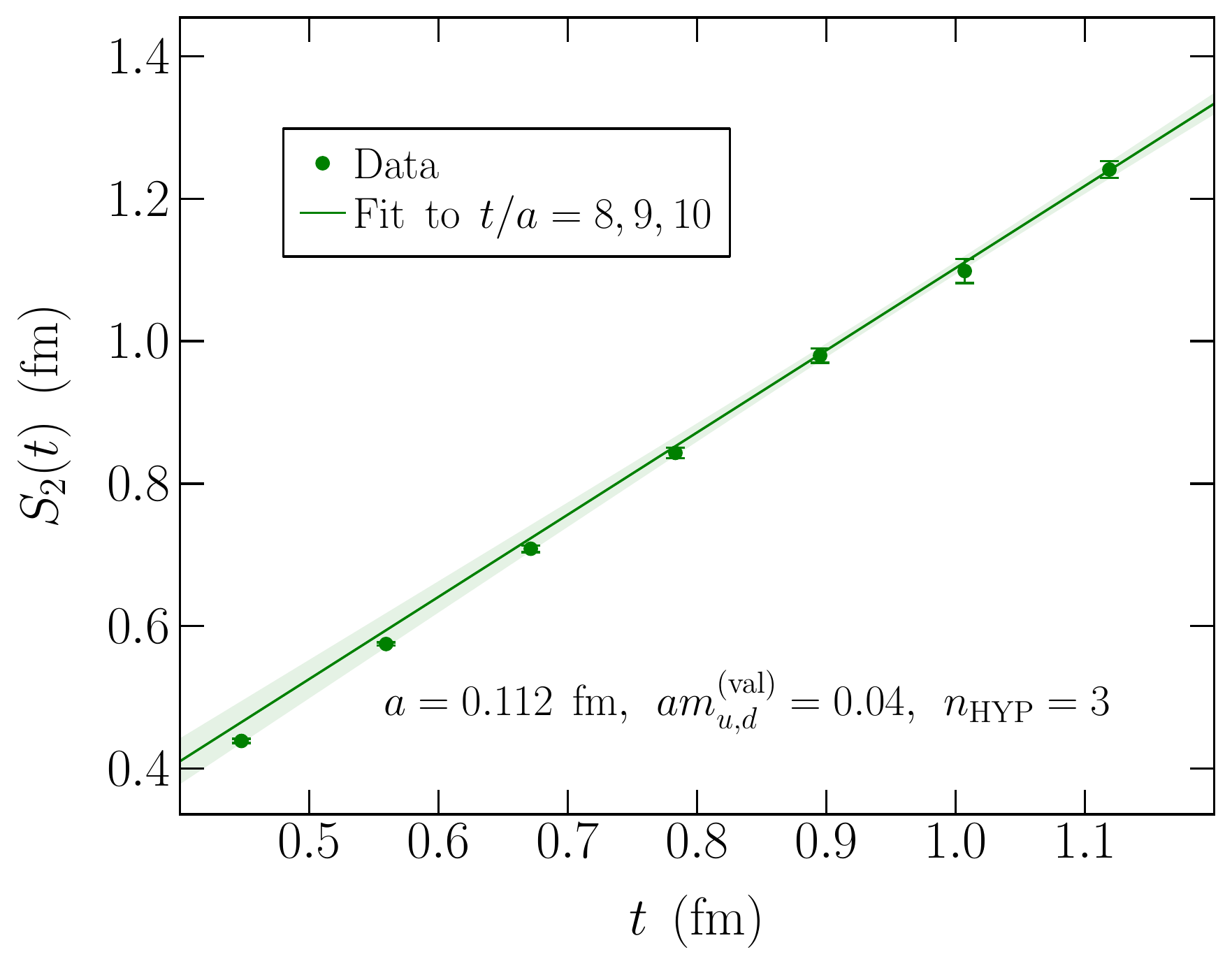}
\caption{\label{fig:summedratio} Fits to the summed ratios $S_1(t)$ and $S_2(t)$, in the range $t/a=8,9,10$. The data are for $a=0.112$ fm, and a heavy quark mass of $m_{u,d}^{(\mathrm{val})}=0.04$ (close to the physical strange quark mass;
the large mass was chosen here for the smaller statistical uncertainties) and $n_{\rm HYP}=3$.}
\end{figure*}

\begin{figure*}[ht!]
 \includegraphics[width=0.47\linewidth]{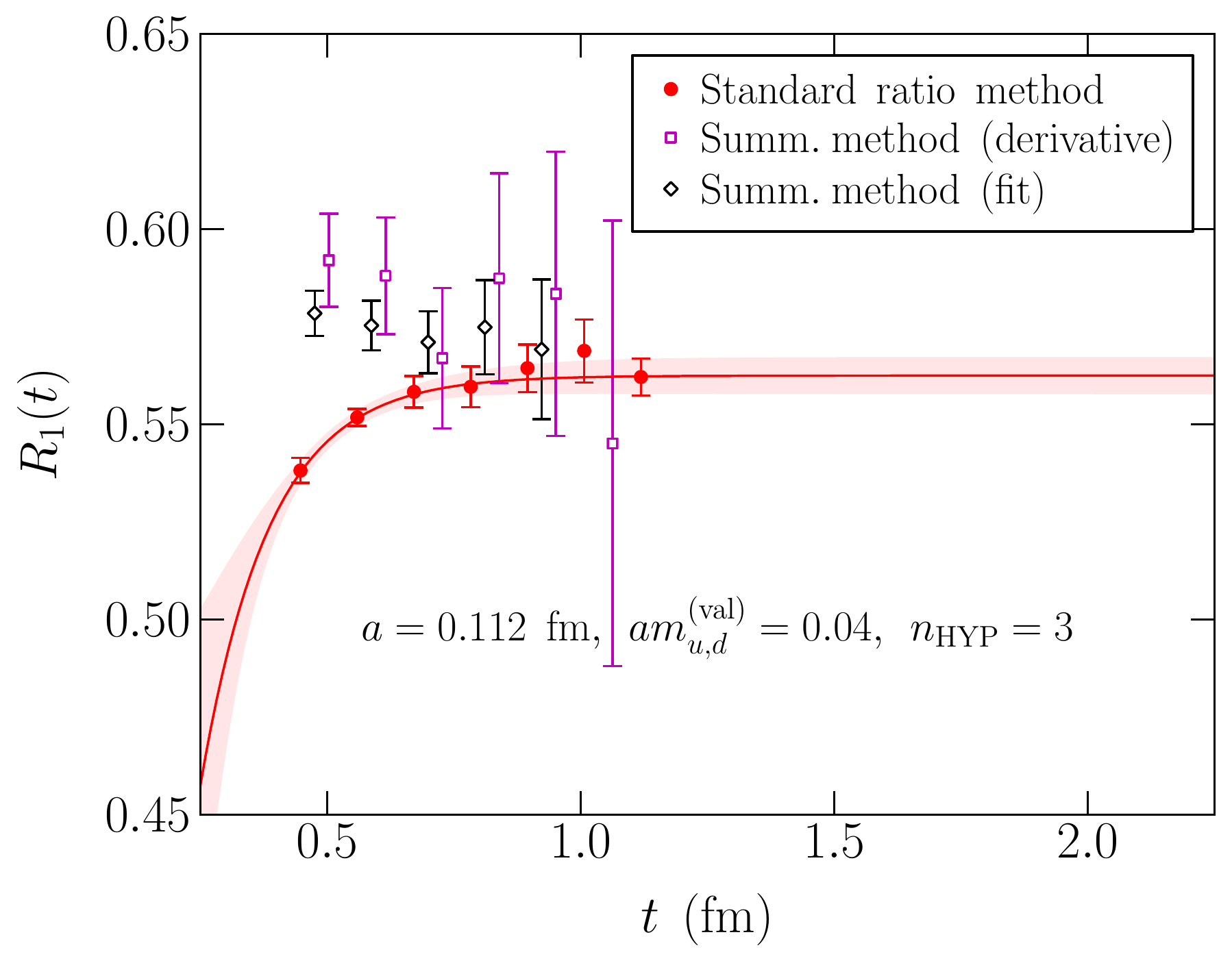}  \hfill \includegraphics[width=0.47\linewidth]{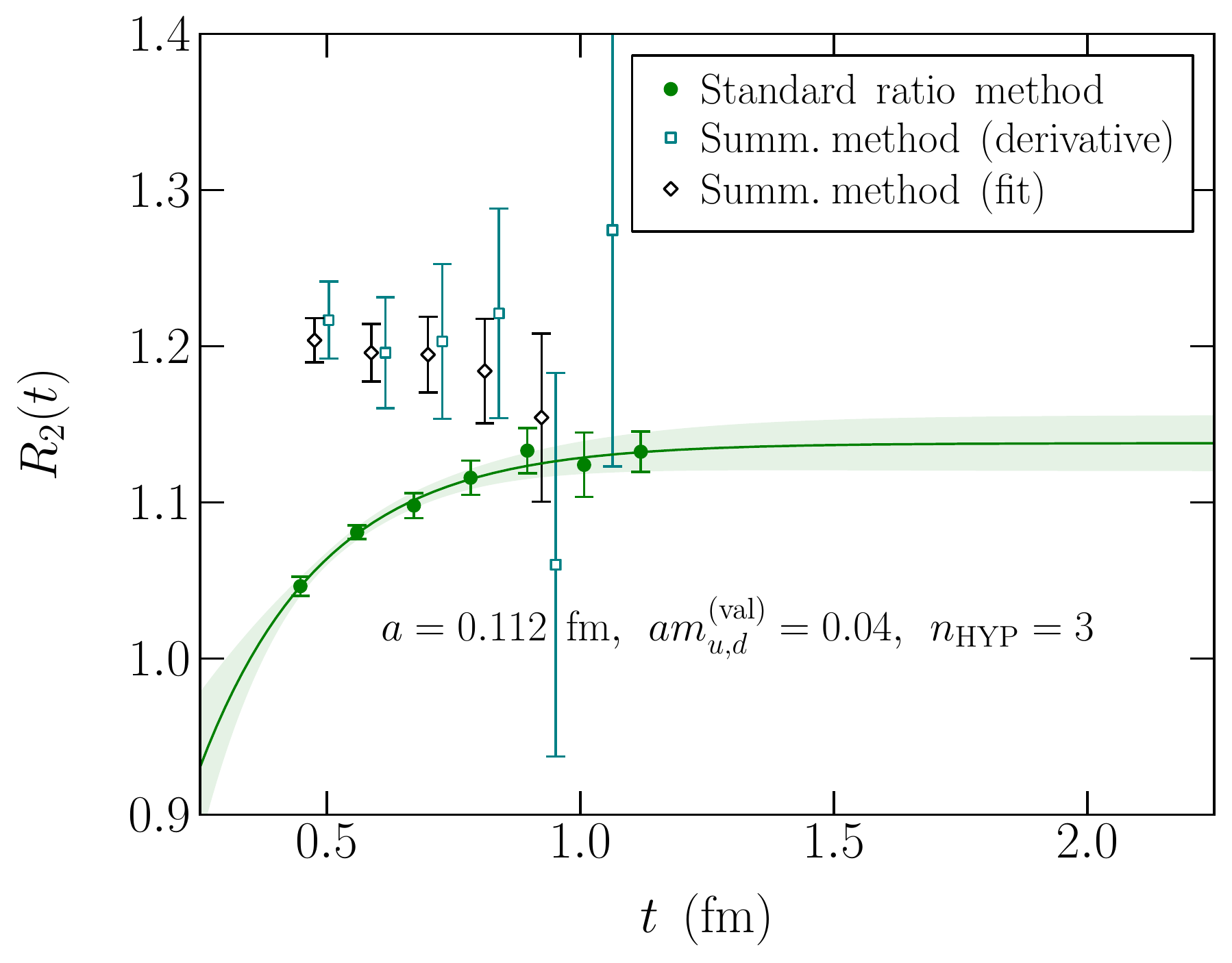}
\caption{\label{fig:derivativemethod} Comparison of the standard ratio method and the summation method for $R_1$ and $R_2$ at $a=0.112$ fm, for a heavy quark mass of $a m_{u,d}^{(\mathrm{val})}=0.04$ (close to the physical strange quark mass; the large mass was chosen here for the smaller statistical uncertainties) and $n_{\rm HYP}=3$.
Shown is the dependence on the source-sink separation $t$. The derivative of the summed ratio, Eq. (\ref{eq:summedratio}), is approximated using $R_i^{\rm sum}=[S_i(t)+S_i(t+a)]/a$ and plotted
with square symbols at $t+\frac12a$. The results for $(g_i)_{\rm eff}$ from linear fits to $S_i(t)$ using Eq.~(\ref{eq:Silarget}) are plotted with diamond symbols. We performed fits in the range $t=t_{\rm min}\: ...\: 10\: a$, and the
results are plotted at $t=t_{\rm min}+\frac 14 a$. For the data from the standard ratio method (circles), the curves indicate the results of fits with the form $R_i(t)=(g_i)_{\rm eff}-A_i\:e^{-\delta_i \: t}$ with free parameters $(g_i)_{\rm eff}$, $A_i$, and $\delta_i$ [see Sec.~\ref{sec:sourcesinkextrap}].}
\end{figure*}

Alternatively to taking the derivative as in Eq.~(\ref{eq:summedratio}), one may fit $S_i(t)$ using the linear function (\ref{eq:Silarget}) with parameters $c_i$ and $(g_i)_{\rm eff}$.
In Fig.~\ref{fig:summedratio}, we show numerical results for $S_i(t)$, along with such fits.
In Fig.~\ref{fig:derivativemethod}, we compare numerical results for the standard ratio (\ref{eq:standardratio2}), the derivative of the summed ratio (\ref{eq:summedratio}),
and the results for $(g_i)_{\rm eff}$ from linear fits to $S_i(t)$ using Eq.~(\ref{eq:Silarget}).
For our data, the results from the summation method, especially for the derivative of the summed ratio, are seen to suffer from much larger statistical uncertainties than the standard ratio.
This was also found in Ref.~\cite{Bulava:2011yz} and is not unexpected, because the relative statistical uncertainty in the difference of two
similarly-sized observables (the discrete derivative used here) is much larger than the relative statistical uncertainty in the individual observables. Of course there are correlations
which can improve the situation, and we did take these into account when calculating (\ref{eq:summedratio}), but because of the way that our lattice calculation was set up (data at successive values of $t$
did not always have neighboring source locations), the correlations were not optimal.

It appears that the systematic errors of the results from the summation method
at short $t$ are similar in magnitude to the systematic errors of the results from the
standard ratio method at the same $t$, but the deviations from $(g_i){\rm eff}$ have the opposite sign.
This shows that valuable information about systematic errors can be obtained by comparing both methods.
For the present data, our process of extrapolating the results from the standard ratio to infinite $t$ is superior
because of the much smaller statistical uncertainty.

Similarly to the work done in Ref.~\cite{Bulava:2011yz}, we also studied models for the three-point and two-point functions
with excited states. We found that at intermediate values of $t$, the systematic uncertainties of (\ref{eq:standardratio2}) and (\ref{eq:summedratio}) were strongly dependent on the assumptions made in the model.
For some models, the standard ratio showed an adantage while for others the summation method showed an advantage, so that again we were not able to draw definitive conclusions.

Further methods for the calculation of hadron-to-hadron matrix elements are based on the Generalized Eigenvalue Problem \cite{Bulava:2011yz} and the
``Generalized Pencil-of-Function'' \cite{Aubin:2010jc}. These techniques use matrices of correlation functions with multiple interpolating fields to
reduce the excited-state contamination at finite $t$. Because we only have data from one interpolating field for each hadron, we
cannot test these methods here.

\end{document}